\pdfoutput=1  % Instructs arXiv to run this with pdflatex rather than latex.

\documentclass[aps,prd,superscriptaddress,floatfix,nofootinbib,preprintnumbers,eqsecnum,twocolumn]{revtex4-1}

\usepackage{amsmath,float,graphicx,placeins,amssymb,multirow,pgfplots,pgfplotstable}
\usepackage{tikz}
\usepackage{soul}
\setstcolor{red}

\usepackage{comment}
\usepackage{enumerate,slashed,cancel,comment,color,bbm,graphicx,rotating,placeins,mathtools,multirow,hhline}
\usepackage[normalem]{ulem}
\usepackage{array}
\usepackage{enumitem}
\usepackage{hyperref}
\usepackage{color}
 \hypersetup
 { 
   colorlinks,
   linkcolor={blue!80!black},
   citecolor={blue!70!black},
   urlcolor={blue!70!black}
 }

\usepackage{lipsum}
\usepackage{diagbox}
\usetikzlibrary{shapes.geometric,arrows}
\graphicspath{}

\def\msmall#1 {\mbox{\small{$#1$}}}
\def\qbar{{\overline{q}}}
\def\pfZ{{\cal Z}}

\def\beq{\begin{equation}}
\def\eeq{\end{equation}}
\def\beqn{\begin{eqnarray}}
\def\eeqn{\end{eqnarray}}
\def\apo{\mbox{\small ${\frac{\alpha'}{2}}$}}
\def\half{\mbox{\small ${\frac{1}{2}}$}}
\def\threehalf{\mbox{\small ${\frac{3}{2}}$}}

\def\quarter{\mbox{\small ${\frac{1}{4}}$}}
\def\sqapo{\mbox{\tiny $\sqrt{\frac{\alpha'}{2}}$}}
\def\sqap{\mbox{\tiny $\sqrt{{\alpha'}}$}}
\def\sqapxtwo{\mbox{\tiny $\sqrt{2{\alpha'}}$}}
\def\aptwo{\mbox{\tiny ${\frac{\alpha'}{2}}$}}
\def\apofour{\mbox{\tiny ${\frac{\alpha'}{4}}$}}
\def\bosqtwo{\mbox{\tiny ${\frac{\beta}{\sqrt{2}}}$}}
\def\btosqtwo{\mbox{\tiny ${\frac{\tilde{\beta}}{\sqrt{2}}}$}}
\def\apofour{\mbox{\tiny ${\frac{\alpha'}{4}}$}}
\def\sqaptwo{\mbox{\tiny $\sqrt{\frac{\alpha'}{2}}$}  }
\def\apoeight{\mbox{\tiny ${\frac{\alpha'}{8}}$}}
\def\sapoeight{\mbox{\tiny ${\frac{\sqrt{\alpha'}}{8}}$}}

\newcommand{\newc}{\newcommand}
\def\calZ{{\cal Z}}
\def\calM{{\cal M}}
\def\calV{{\cal V}}
\def\calF{{\cal F}}
\def\calG{{\cal G}}

\def\calX{{\cal X}}
\def\calK{{\cal K}}
\def\calA{{\cal A}}
\def\calB{{\cal B}}

\def\mathbbX{{\mathbbm{X}}}

\def\mathbbK{{\mathbbm{K}}}
\def\mathbbZ{{\mathbbm{Z}}}
\def\mathbbT{{\mathbbm{T}}}
\def\mathbbA{{\mathbbm{A}}}
\def\mathbbB{{\mathbbm{B}}}

\def\bQ{{\bf Q}}
\def\bT{{\bf T}}

\def\Qs{{\bf q}}

\def\Thetabar{{\overline{\Theta}}}
\def\etabar{{\overline{\eta}}}

\def\ie{{\it i.e.}\/}
\def\eg{{\it e.g.}\/}
\def\etc{{\it etc}.\/}
\def\inbar{\,\vrule height1.5ex width.4pt depth0pt}
\def\IR{\relax{\rm I\kern-.18em R}}
 \font\cmss=cmss10 \font\cmsss=cmss10 at 7pt
\def\IQ{\relax{\rm I\kern-.18em Q}}
\def\IZ{\relax\ifmmode\mathchoice
 {\hbox{\cmss Z\kern-.4em Z}}{\hbox{\cmss Z\kern-.4em Z}}
 {\lower.9pt\hbox{\cmsss Z\kern-.4em Z}}
 {\lower1.2pt\hbox{\cmsss Z\kern-.4em Z}}\else{\cmss Z\kern-.4em Z}\fi}

\def\oneRes{{       %  residue at s=1
     \underset{s=1}{\rm Res}
  }}
\def\Str{{\rm Str}}
\def\zStr{{         %  zero-mass supertrace
      \underset{M=0}{\rm Str}
       \,}}

\def\zLStrE{{         %  zero-mass supertrace
      \underset{M_L=0}{{\rm Str}_E}
       \,}}

\def\pStr{{         %  positive-mass supertrace
      \underset{M>0}{\rm Str}
       \,}}
\def\effStr{{         %  EFT supertrace
      \underset{\small 0<M\lesssim \mu}{\rm Str}
       \,}}

\def\pStrE{{         %  positive-mass supertrace
      \underset{M_L>0}{{\rm Str}_E}
       \,}}
\def\pLStrE{{         %  positive-mass supertrace
      \underset{M_L>0}{{\rm Str}_E}
       \,}}

\begin{document}

\title{On the Running of Gauge Couplings in String Theory} 

\def\andname{\hspace*{-0.5em}} % gets rid of "and" in author list
\author{Steven Abel}
\email[Email address: ]{s.a.abel@durham.ac.uk}
\affiliation{IPPP and Department of Mathematical Sciences, Durham University, Durham, DH1 3LE, United Kingdom}
\affiliation{CERN, Theoretical Physics Department, CH 1211 Geneva 23  Switzerland}
\author{Keith R. Dienes}
\email[Email address: ]{dienes@arizona.edu}
\affiliation{Department of Physics, University of Arizona, Tucson, AZ 85721 USA}
\affiliation{Department of Physics, University of Maryland, College Park, MD 20742 USA}
\author{Luca A. Nutricati}
\email[Email address: ]{luca.a.nutricati@durham.ac.uk}
\affiliation{IPPP and Department of Mathematical Sciences, Durham University, Durham, DH1 3LE, United Kingdom}

\begin{abstract}
 In this paper we conduct
 a general, model-independent analysis of the running of gauge couplings within closed string theories.  Unlike previous discussions in the literature, our calculations fully respect the underlying  modular invariance of the string and include the contributions from the infinite towers of string states which are ultimately responsible for many of the properties for which string theory is famous, including an enhanced degree of finiteness and UV/IR mixing.  In order to perform our calculations, we adopt a formalism that was recently developed for calculations of the Higgs mass within such theories, and demonstrate that this formalism can also be applied to calculations of gauge couplings.
In general, this formalism gives rise to an ``on-shell'' effective field theory (EFT) description in which the final results are expressed in terms of supertraces over the physical string states, and in which these quantities exhibit an EFT-like ``running'' as a function of an effective spacetime mass scale.
We find, however, that the calculation of the gauge couplings differs in one deep way from that of the Higgs mass:  while the latter results depend on purely on-shell supertraces,
the former results have a different modular structure which causes them to depend on {\it off-shell}\/ supertraces as well.
In some regions of parameter space, our results demonstrate how certain expected field-theoretic behaviors can emerge from the highly UV/IR-mixed environment.  In other situations, by contrast,
our results give rise to a number of intrinsically stringy behaviors that transcend what might be expected within an effective field theory approach.
\end{abstract}

\maketitle
  \tableofcontents

\vspace{0.3cm}
%  {\flushleft{Preprints: CERN-TH-2023-044, IPPP/23/16}}
%   This information goes in the metadata.

\def\ie{{\it i.e.}\/}
\def\eg{{\it e.g.}\/}
\def\etc{{\it etc}.\/}
\def\taubar{{\overline{\tau}}}
\def\qbar{{\overline{q}}}
\def\kbar{{\overline{k}}}
\def\bQ{{\bf Q}}
\def\calT{{\cal T}}
\def\calN{{\cal N}}
\def\calF{{\cal F}}
\def\calM{{\cal M}}
\def\calZ{{\cal Z}}

\def\beq{\begin{equation}}
\def\eeq{\end{equation}}
\def\beqn{\begin{eqnarray}}
\def\eeqn{\end{eqnarray}}
\def\apo{\mbox{\small ${\frac{\alpha'}{2}}$}}
\def\half{\mbox{\small ${\frac{1}{2}}$}}
\def\sqapo{\mbox{\tiny $\sqrt{\frac{\alpha'}{2}}$}}
\def\sqap{\mbox{\tiny $\sqrt{{\alpha'}}$}}
\def\sqapxtwo{\mbox{\tiny $\sqrt{2{\alpha'}}$}}
\def\aptwo{\mbox{\tiny ${\frac{\alpha'}{2}}$}}
\def\apofour{\mbox{\tiny ${\frac{\alpha'}{4}}$}}
\def\bosqtwo{\mbox{\tiny ${\frac{\beta}{\sqrt{2}}}$}}
\def\btosqtwo{\mbox{\tiny ${\frac{\tilde{\beta}}{\sqrt{2}}}$}}
\def\apofour{\mbox{\tiny ${\frac{\alpha'}{4}}$}}
\def\sqaptwo{\mbox{\tiny $\sqrt{\frac{\alpha'}{2}}$}  }
\def\apoeight{\mbox{\tiny ${\frac{\alpha'}{8}}$}}
\def\sapoeight{\mbox{\tiny ${\frac{\sqrt{\alpha'}}{8}}$}}

\newc{\gsim}{\lower.7ex\hbox{$\;\stackrel{\textstyle>}{\sim}\;$}}
\newc{\lsim}{\lower.7ex\hbox{$\;\stackrel{\textstyle<}{\sim}\;$}}
\def\calM{{\cal M}}
\def\calV{{\cal V}}
\def\calF{{\cal F}}
\def\bQ{{\bf Q}}
\def\bT{{\bf T}}
\def\Qs{{\bf q}}

\def\ie{{\it i.e.}\/}
\def\eg{{\it e.g.}\/}
\def\etc{{\it etc}.\/}
\def\inbar{\,\vrule height1.5ex width.4pt depth0pt}
\def\IR{\relax{\rm I\kern-.18em R}}
 \font\cmss=cmss10 \font\cmsss=cmss10 at 7pt
\def\IQ{\relax{\rm I\kern-.18em Q}}
\def\IZ{\relax\ifmmode\mathchoice
 {\hbox{\cmss Z\kern-.4em Z}}{\hbox{\cmss Z\kern-.4em Z}}
 {\lower.9pt\hbox{\cmsss Z\kern-.4em Z}}
 {\lower1.2pt\hbox{\cmsss Z\kern-.4em Z}}\else{\cmss Z\kern-.4em Z}\fi}

\section{Introduction and motivation}

String theory is widely regarded as providing the ultimate ``UV completion'' of theories which successfully describe experimental phenomena at lower energy scales.   
Such theories include  the Standard Model as well as its various extensions.   
However, it is not always clear how one might draw an explicit map between these full string theories on the one hand and observable
low-energy phenomena on the other.
Because the fundamental scale of string theory is normally considered to be unreachably remote, and
because the particle spectrum of the string is
generally quantized in units of this scale,
one traditionally attempts to extract low-energy
phenomenological predictions from string theory by focusing on the effects associated with only the lightest of the string modes. 

Unfortunately, this approach towards string phenomenology robs us of the full power of string theory to provide new insights into low-energy phenomena. 
String theory, as a theory of extended objects, does not merely produce light states --- it also gives rise to infinite towers of massive states which are also an intrinsic part of the string spectrum.  Indeed, the ``stringiness''
of string theory --- \ie, the fundamental features of string theory that transcend our field-theoretic expectations and therefore have the power to suggest new solutions to old puzzles --- lie within these states.  By disregarding these states and their accumulated contributions to low-energy physics, we are severing the link between the UV-complete theory and its low-energy phenomenology.  This reduces us to working within an effective field theory (EFT) whose relevant operators are very hard to explain. 

For this reason, it may be argued that a proper approach to understanding many of the low-energy phenomenological implications of string theory is one in which these infinite towers of states are retained and their effects are incorporated in a natural way throughout our calculations.  Indeed, the
effects of such states are likely to be the most relevant for fundamental phenomenological questions --- such as {\it hierarchy}\/ problems --- which focus on the difficulties of maintaining a peaceful coexistence of  both light and heavy scales within a quantum-mechanical universe.

One clue as to the power of these infinite towers of states is that string theories generally have finiteness properties that transcend what can be expected in field theory.   One normally attributes these finiteness properties to the extended nature of the string --- a feature lacking in theories based on point particles ---  but this extended nature of the string is precisely what gives rise to these infinite towers of states.   For perturbative closed strings (which will be our main focus throughout this paper), worldsheet modular invariance is the exact fundamental symmetry which governs these states and their interactions.   Thus, 
modular invariance holds
the key to much of the stringiness of string theory and the finiteness (or softened divergences) associated with its low-energy phenomenological predictions.   However, modular invariance also leads to much more, including a unique and surprising form of UV/IR mixing that can severely distort the validity of effective field theories (EFTs), even at low energies where one might have assumed EFT-based approaches to hold.

For this reason, it is important to develop fully modular-invariant methods of extracting low-energy phenomenological predictions from string theory.   By their very nature, these are methods in which the full towers of string states play an important role and cannot be neglected. It is then hoped that the inclusion of these infinite towers of states and the preservation of the underlying modular symmetry can lead to new ways of approaching long-standing phenomenological puzzles.
Indeed, as originally advocated in Ref.~\cite{Dienes:2001se},
this might be one route towards developing non-traditional approaches towards addressing hierarchy problems.  

In this paper, we shall calculate the running of the one-loop gauge couplings within string theory.   This is an old and classic topic within string phenomenology, but we shall employ a formalism for doing this calculation which fully respects modular invariance and which thereby incorporates all of the ``magic'' to which string theory gives rise.   We shall begin in Sect.~II by reviewing the framework~\cite{Abel:2021tyt} within which we shall perform this calculation.   We shall also summarize the prior results in this field and highlight the ways in which our approach (and our eventual results) will be different.
Sect.~III then forms the main body of this paper.
Within this section, we shall systematically perform our calculations, ultimately developing a completely general picture of how gauge couplings run within four-dimensional closed string theories.    Along the way we shall also discuss several new results which may have wider applicability beyond our specific gauge-coupling calculation.     These include new theorems concerning the cancellations of various supertraces of modular-invariant operators.   
We shall also discuss the effects of {\it entwinement}\/, a phenomenon which emerges within the context of our gauge-coupling calculation and which shifts the meaning of ``physicality'' when characterizing different states in the string spectrum.
We shall then summarize our main results and possible directions for future research in Sect.~\ref{sec:discussion}.

%============================================

\section{Preliminaries:  Our framework and connection to prior literature  \label{sec:preliminaries} }

In Ref.~\cite{Abel:2021tyt}, a framework was developed for performing calculations of the Higgs mass in a fully modular-invariant way.    As discussed there, 
this framework is completely general and can be applied to any string model (vacuum state).
Moreover, although the focus within Ref.~\cite{Abel:2021tyt} centered around calculations of the Higgs mass, this framework can be applied to numerous quantities of phenomenological interest, including the running of the gauge couplings. Explicitly performing such a calculation is thus the primary goal of this paper.

In this section, we shall begin by reviewing the salient features of this framework and the various steps that are involved.   With these steps explicitly elucidated, we shall then discuss prior calculations of the running gauge couplings that exist in the literature --- including the classic calculation of Kaplunovsky~\cite{Kaplunovsky:1987rp} --- and discuss precisely which parts of those prior calculation preserve modular invariance and which parts do not.   We shall then outline the primary goals of this paper within this language.

\subsection{Our analysis framework} 

\begin{figure*}[t]
\centering
\includegraphics[keepaspectratio, width=1.02\textwidth]{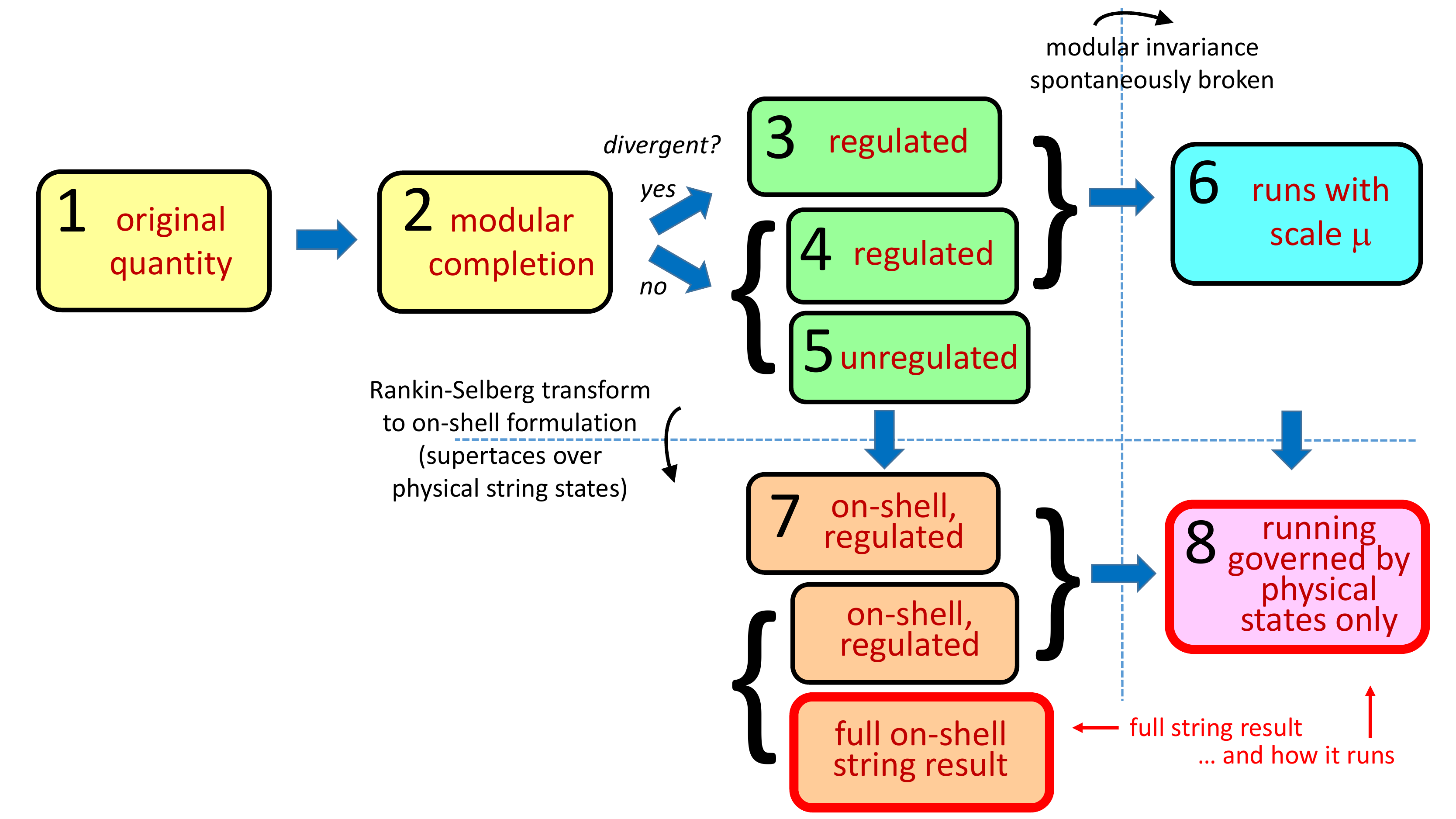}
\caption{Distinct steps associated with our analysis of an arbitrary physical quantity, as discussed in the text.   The particular sequence of steps to be followed depends on the goal of the analysis.  Particularly relevant results are those in the lower portions of this sketch, in which our physical quantity is expressed purely in terms of contributions from on-shell, physical string states and also in a form which runs as a function of a spacetime mass scale $\mu$.}
\label{fig:process}
\end{figure*}

Within the framework developed in Ref.~\cite{Abel:2021tyt}, 
the calculation of a given low-energy quantity $\zeta$  
proceeds through a number of distinct steps.  These steps are illustrated schematically in Fig.~\ref{fig:process}.  For clarity we shall now enumerate these steps individually although many of them are deeply connected to each other and may be performed simultaneously.
Explicit examples of each step will be given later in this section.

%======================================
%======================================

\begin{enumerate}
    \item \label{step1} As a starting point, one constructs what may be considered to be the ``string-theoretic'' equivalent of the one-loop field-theoretic
    contributions to the relevant quantity $\zeta$ coming from each of the string states.  In doing this, one must sum over the contributions from the infinite towers of string states, regardless of their masses.   This is a sum of the contributions from the entire tower of states as they propagate around the worldsheet torus, with these contributions weighted appropriately by the naive vertex factors corresponding to their charges and couplings.  However, even though we have summed over the entire string spectrum, the resulting expression may not be modular invariant.
    \item \label{modularcompletion} Second, if needed, one then performs a ``modular completion'' of the above expression for $\zeta$.   This will generally require the introduction of additional terms which may be interpreted as coming from extra intrinsically stringy effects such as gravitational backreactions.  In such situations, the tight constraints of modular invariance render these modular completions fairly unique.  Thus, after this step, one has obtained a general, fully modular-invariant, string-theoretic expression for the quantity $\zeta$ under study. 
    \end{enumerate}
    
  One could, in principle, stop here.    However, one natural question that arises is whether this expression for $\zeta$ is finite, or whether it might diverge in certain string backgrounds.    Because this quantity is fully modular invariant, this expression will already exhibit the elimination or softening of the divergences that would have otherwise been expected in field theory.   Thus, the divergence structure of $\zeta$ might be very different from what one would expect in ordinary quantum field theory.  
  
  We then have different options, depending on whether $\zeta$ is finite or divergent.
  \begin{enumerate}[resume]
  \item \label{divergent} If the quantity resulting from Step~\ref{modularcompletion}  is potentially divergent, one must regulate this quantity {\it in a manner which is consistent with modular invariance}.  
  (Indeed, any regulator which breaks modular invariance is likely to introduce precisely the sorts of spurious effects we are hoping to avoid.)
  This passage from the divergent quantity $\zeta$ to the regulated quantity $\widehat \zeta$ is indicated as Step~3 within Fig.~\ref{fig:process}.
  Thus, after this step, one has a fully string-theoretic (and hence modular-invariant) expression for $\zeta$ which is also {\it finite}.  We shall let $\widehat \zeta$ denote this finite, regulated quantity.
    In general, $\widehat\zeta$ will depend on a regulator parameter $a$ or collections of parameters $\lbrace a_i\rbrace$. 
    The resulting regulated quantity $\widehat \zeta(a_i)$ will then be finite for all $\lbrace a_i\rbrace$ {\it except}\/ those limiting values of $a_i$ which correspond to removing the regulator.  
    \item \label{finite1} Alternatively, if the quantity resulting from Step~\ref{modularcompletion}  is finite, we have two possibilities.  One possibility is to nevertheless choose to regulate this quantity in the same way as in Step~\ref{divergent}.   This then deforms $\zeta$ into another finite quantity $\widehat \zeta(a_i)$ which remains finite even in the limit when the regulator is removed. This is shown as Step~4 within Fig.~\ref{fig:process}.
    \item \label{finite2}  Finally, if the quantity resulting from Step~\ref{modularcompletion}  is finite, an alternative possibility is to simply recognize that no regulation is needed.   In that case, one can forego the regulator entirely and simply retain the expression obtained in Step~\ref{modularcompletion}.  We indicate this choice as Step~5 within Fig.~\ref{fig:process}.
    \end{enumerate}
    
At this stage (green boxes in Fig.~\ref{fig:process}), we have a quantity $\widehat \zeta$ which is fully modular invariant and finite.   This quantity will depend on regulator parameters $a_i$ if we have employed Steps~\ref{divergent} or \ref{finite1}, but will be independent of $a_i$ if we have followed Step~\ref{finite2}.  

There are now several different options for how one might proceed.   These different paths ultimately correspond to recasting the finite expression $\widehat \zeta$ obtained in Steps~\ref{divergent} through \ref{finite2} in different forms that are useful for different purposes.
\begin{enumerate}[resume]
    \item \label{step6} If we are interested in extracting an EFT-like ``running'' for $\widehat\zeta(a_i)$, we can start from Step~\ref{divergent} or Step~\ref{finite1} and proceed to identify an appropriate combination $f(a_i)$ of the $a_i$-parameters with a spacetime scale $\mu$.   As discussed in Ref.~\cite{Abel:2021tyt}, such an identification breaks modular invariance by adopting a particular EFT-like direction for spacetime ``UV'' versus ``IR'' physics (\ie, a particular UV/IR direction for $\mu$)  in terms of the otherwise UV/IR-blind worldsheet combination $f(a_i)$.   This step nevertheless respects all other aspects of the modular symmetry, and can be viewed as merely breaking the modular symmetry {\it spontaneously}\/.   One then obtains a running quantity $\widehat \zeta(\mu)$.   Indeed, this is the  step at which we first introduce the notion of a spacetime energy scale into the theory.
    \end{enumerate}

All expressions up to this point receive explicit contributions from the full towers of string states.   These include not only physical, ``on-shell'' level-matched states (whose left- and right-moving mass contributions are equal), but also unphysical ``off-shell'' states (whose left- and right-moving mass contributions are unequal).  Note that the off-shell states can only appear within loops, and thus cannot serve as in-states or out-states in any string amplitude.   Indeed, it is the on-shell states which have field-theory analogues, while the off-shell states are intrinsically stringy.   Thus, if our goal for comparison purposes is to recast our string results into an on-shell form which is as close as possible to what might arise in field theory, we would like to rewrite $\widehat \zeta$ in terms of the contributions from only the physical, on-shell states as fully as possible.

To do this, we can utilize certain methods derived from modular-function theory which involve the so-called ``Rankin-Selberg'' transform~\cite{rankin1a,rankin1b,selberg1}.
The mathematics behind this transform is reviewed in Ref.~\cite{Abel:2021tyt} and ultimately allows us to express a one-loop {\it string-theoretic}\/ amplitude as the {\it residue}\/ of a deformed {\it field-theoretic}\/ amplitude, evaluated at a location in the complex plane associated with the deformation parameter where the field-theoretic amplitude has a  pole.  This relation between a string amplitude and a (deformed) field-theory amplitude then enables us to obtain an expression for the string amplitude which involves supertraces over the contributions from only the physical string states.
    \begin{enumerate}[resume]
    \item \label{step7} If we perform a Rankin-Selberg transform starting from the results of Step~\ref{divergent}, \ref{finite1}, or \ref{finite2} (green boxes in Fig.~\ref{fig:process}), we then obtain corresponding results (orange boxes in Fig.~\ref{fig:process}) which involve supertraces over only the physical string states.   Such results preserve modular invariance fully and represent an alternative --- and often more transparent --- formulation for $\widehat \zeta$ which enables a direct comparison with what might have been expected in field theory.   
    In particular, if we apply the Rankin-Selberg transform to the results of Steps~\ref{divergent} or \ref{finite1}, we obtain results which also depend on our regulator parameters $a_i$.   However, if we apply the Rankin-Selberg transform to the results of Step~\ref{finite2}, our result depends on the physical supertraces only and does not involve any regulator parameters (orange box with red border in Fig.~\ref{fig:process}).  {\it This may then be viewed as our final result for the string quantity $\widehat \zeta$ --- one which is fully modular invariant and involves only the supertraces over physical string states.}
    \item \label{step8} Alternatively, if we calculate the Rankin-Selberg transform of the results of Step~\ref{step6} (blue box in Fig.~\ref{fig:process}) --- or equivalently identify $f(a_i)$ with $\mu$ within the results of Step~\ref{step7} (upper two orange boxes in Fig.~\ref{fig:process}) --- we obtain an expression for $\widehat \zeta(\mu)$ in which the supertraces over the physical string states govern the running of
    $\widehat \zeta(\mu)$.   This is indicated by the purple box with the red border in Fig.~\ref{fig:process}.  As discussed in Ref.~\cite{Abel:2021tyt}, these results preserve modular invariance as fully as possible and yet resemble as closely as possible the running of physical quantities in field theory.   {\it This result thus describes $\widehat\zeta$ as a running quantity, where the running is now governed purely by the supertraces of the physical string states.}  This formulation for $\widehat \zeta(\mu)$ is particularly useful for studying the maximal extent to which an EFT description of $\widehat \zeta(\mu)$ at low energies emerges and remains valid within the full modular-invariant string theory.
    \end{enumerate}
    
The most important results of this analysis are those which are indicated in the red-bordered boxes in the lower right portion of Fig.~\ref{fig:process}.  As discussed above, these results respectively express our original quantity in terms of the supertraces over only the physical states in the string spectrum, and also describe how this quantity runs as a function of a spacetime energy scale $\mu$.  Indeed, the limit in which the regulator is removed will typically correspond to taking the deep-IR limit $\mu\to 0$ (or equivalently the deep-UV limit $\mu\to \infty$, given that modular invariance requires an invariance under the scale duality $\mu\to M_s^2/\mu$, as originally pointed out in Ref.~\cite{Abel:2021tyt}).
In this limit, the result of Step~\ref{step8} reduces to the regulator-independent result of Step~\ref{step7}.

Even though we have broken this analysis procedure into distinct steps, we stress that many of these steps are deeply connected and can be performed simultaneously.  For example, as discussed in Ref.~\cite{Abel:2021tyt}, it is possible to proceed directly from the results of Step~\ref{modularcompletion} to those of Step~\ref{step7} through a so-called ``{\it regulated}\/ Rankin-Selberg'' transform.    Likewise, if we are not interested in interpreting our physical quantities as ``running'' with respect to a spacetime scale $\mu$, we need never be concerned with Step~\ref{step6} or Step~\ref{step8}.

\subsection{Prior literature:   Results to date}

To date, this procedure has been applied to two different quantities $\zeta$ of phenomenological interest:
the one-loop cosmological constant $\Lambda$, and the one-loop Higgs mass $m_\phi^2$.   Here the Higgs field $\phi$ is identified as any scalar field $\phi$ whose fluctuations can affect the masses of other string states throughout the string spectrum.   We shall now present some of the main results of these prior analyses.  These results will not only serve to  illustrate the different steps of this procedure but will also be relevant later in this paper.

For the one-loop closed-string cosmological constant (vacuum energy) $\Lambda$, Step~\ref{step1} requires that 
we begin with the standard expression~\cite{Polchinski:1985zf} which is nothing other than the one-loop string partition function $\pfZ(\tau)$ integrated over the fundamental domain $\calF$ of the modular group.  Indeed, if we define the standard four-dimensional one-loop string amplitude for any operator insertion $A$ as
\begin{equation}
    \langle A \rangle ~\equiv~
      \int_\calF \frac{d^2\tau}{\tau_2^2}\, \tau_2^{-1}\, \sum_{m,n}\, (-1)^F  \,A_{mn} ~ \qbar^m q^n~,
\label{amplitudeA}
\end{equation}
then the corresponding one-loop vacuum energy $\Lambda$ is nothing but
\begin{equation}
    \Lambda~=~ - \frac{\calM^4}{2} \langle {\bf 1} \rangle~.
\label{Lambdanoinsertion}
\end{equation}
Here $\tau$ is the one-loop torus modular parameter with real and imaginary parts $\tau_{1,2}$ respectively;   $q\equiv e^{2\pi i \tau}$;  $\calF$ is the fundamental domain of the modular group; 
the sum $\sum_{m,n}$ is over all discrete string states with 
 right- and left-moving worldsheet energies $(m,n)$, normalized so that the corresponding spacetime mass $M^2$ is given by
 $M^2= \half(M_R^2+ M_L^2) = 
 \frac{2}{{\alpha'}} (m+n)$ where the string scale $M_s$ and reduced string scale $\calM$ are given by $M_s \equiv 2\pi \calM = 1/\sqrt{\alpha'}$;
$F$ is the spacetime fermion number of each state contributing in the sum;  and $A_{mn}$ are the eigenvalues of the operator $A$ when acting on each $(m,n)$ string state.  
Note that $d^2\tau/\tau_2^2$ is the modular-invariant measure for the $\tau$-integration, while the extra prefactor
$\tau_2^{-1}$ within the integrand of Eq.~(\ref{amplitudeA}) emerges from the integration over the continuum of modes associated with the uncompactified spacetime coordinates and reflects the fact that the four-dimensional string partition function, prior to insertions, has modular weight $k=-1$. The curved shape of the lower portion of the fundamental domain $\calF$ implies that the amplitude in Eq.~(\ref{amplitudeA}) receives contributions from not only the physical (level-matched) ``on-shell'' string states with $m=n$ but also the unphysical (intrinsically stringy) ``off-shell'' states with $m\not=n$.
 Indeed, we see from Eq.~(\ref{Lambdanoinsertion}) that for the cosmological constant $\Lambda$ the only ``insertion'' into the partition-function in Eq.~(\ref{amplitudeA}) is given by $A={\bf 1}$, the identity operator.  This makes sense for a vacuum energy, since all states contribute equally and independently of their possible charges or other characteristics.
 
The result in Eq.~(\ref{Lambdanoinsertion}) thus represents Step~\ref{step1}.   Given that $A={\bf 1}$, this expression is fully modular invariant and no modular completion is needed.  This result then carries over to Step~2.   Proceeding to Step~3, we ask whether this quantity is divergent. In principle, there are indeed certain states within the string spectrum which could cause divergences:   these are physical tachyons for which $m=n<0$.   Since the presence of such tachyons destabilizes the theory, we shall restrict our attention to string theories in which such states are absent.   It then follows that $\Lambda$ is finite.  According to the procedure we have sketched in Fig.~\ref{fig:process}, we then have two options which amount to whether or not we wish to impose a regulator.  For considerations of $\Lambda$ alone, there is no need to do so, since $\Lambda$ is already finite.   We shall
therefore carry this expression for $\Lambda$ unchanged into Step~5.    

Our final step (Step 7 within Fig.~\ref{fig:process}) is to evaluate the Rankin-Selberg transform of the expression in Eq.~(\ref{Lambdanoinsertion}).  This is not difficult, and leads immediately to a result first derived in Ref.~\cite{Dienes:1995pm}:
\beq
           \Lambda ~=~ \frac{1}{24} \calM^2 \,
            \Str M^2
\label{Lamresult}
\eeq  
where our supertrace `Str' notation indicates a statistics-weighted trace over the spectrum of only physical string states~\cite{Dienes:1995pm}:
\beq
              \Str \, A ~\equiv~   
       \lim_{y\to 0} \, \sum_{{\rm states}~ i}  (-1)^{F_i} \, A_i \,  e^{- y \alpha' M_i^2}~
\label{supertracedef2}
\eeq
with the index $i$ labeling the different physical states in the spectrum.
This definition of the supertrace will be discussed further in Sect.~\ref{sec:string-comp}.~
Intimately connected with the result in Eq.~(\ref{Lamresult}) and emerging from the same analysis is also an additional constraint~\cite{Dienes:1995pm}
\beq
   \Str\, {\bf 1}~=~ 0~.
\label{strone}
\eeq
The results in 
Eqs.~(\ref{Lamresult}) and (\ref{strone}) hold for any tachyon-free closed string theory in four spacetime dimensions and even generalize~\cite{Dienes:1995pm} to other dimensionalities as well, with $\Str\,M^{2\beta}=0$ for all $0\leq \beta\leq \half(D-4)$ in general and with $\Lambda_D\sim \calM^2 \,\Str\, M^{D-2}$ where $\Lambda_D$ is the corresponding one-loop cosmological constant in $D$ spacetime dimensions.

These results are truly remarkable.   In ordinary four-dimensional quantum field theory, we would expect that $\Lambda$ would be a divergent quantity for which 
$\Str \,{\bf 1}$ governs the quartic divergence
and $\Str\,M^2$ governs the quadratic divergence.
However, we now see that in a four-dimensional tachyon-free modular-invariant string theory  $\Lambda$ is actually {\it finite} and moreover that $\Str\,M^2$ gives its {\it value}.
Likewise, $\Str \,{\bf 1}$ actually
{\it vanishes}.

These results are the consequence of a governing ``misaligned supersymmetry''~\cite{Dienes:1994np, Dienes:1995pm} which has been proven to exist within the spectra
of all tachyon-free modular-invariant string theories.   Indeed, this symmetry indicates that  bosonic and fermionic string states must be distributed across the infinite string spectrum in such a way that the spectrum is either exactly spacetime supersymmetric (a ``degenerate'' form of misaligned supersymmetry) or configured in a
precise mathematical way wherein any surplus of bosonic states at a given mass level triggers the existence of an even greater surplus of fermionic states at an even higher mass level, which in turn triggers the existence of an even larger surplus of bosonic states at an even higher mass level, and so forth.  The sizes of these alternating surpluses grow exponentially as a function of mass, thereby explaining how even a non-supersymmetric string can remain consistent not only with the Hagedorn transition but also with finite supertrace results such as that in Eq.~(\ref{Lamresult}). Misaligned supersymmetry thus lies at the heart of the remarkable finiteness properties of closed strings~\cite{Dienes:1994np,Dienes:1995pm, Dienes:2001se} and will ultimately underpin the results of this paper as well.

To date, the only other physical quantity which has been studied within the full framework sketched in Fig.~\ref{fig:process} is the Higgs mass $m_\phi^2$.  This analysis was performed in Ref.~\cite{Abel:2021tyt}, and we shall outline the salient results here.  In general, as stated above, the Higgs will be viewed as any state whose VEV affects the masses of at least some of the corresponding string states.  We shall work within the Higgsed phase of the theory and accordingly assume that the Higgs field has a non-zero VEV and is already sitting at the minimum of its potential.   Clearly the Higgs mass then corresponds to the curvature of this potential at that minimum.  In order to calculate this curvature, we can regard the masses $M^2$ of all string states in the Higgsed phase to be functions of $\phi$, where $\phi$ parametrizes the fluctuations of the Higgs field around this minimum.
In complete analogy to Eq.~(\ref{Lambdanoinsertion}), it then turns out that~\cite{Abel:2021tyt}
the one-loop Higgs mass can then be written as
\beq
        m_\phi^2  ~=~  -\frac{\calM^2}{2}
              \biggl\langle \tau_2 \mathbbX_1 + \tau_2^2\, \mathbbX_2 \biggr\rangle + ...
\label{Higgsstep1}
\eeq
where the insertions $\mathbbX_1$ and 
$\mathbbX_2$ into the partition function sum
are given by
\beqn
  \mathbbX_1 ~&\equiv&~  -\frac{1}{4\pi} \partial_\phi^2 M^2  \biggl |_{\phi=0}\nonumber\\
  \mathbbX_2 ~&\equiv&~ \frac{1}{16\pi^2 \calM^2} ( \partial_\phi M^2)^2 \biggl |_{\phi=0}~.
\eeqn
These insertions thus tally the effective Higgs ``charges'' (or equivalently the contributions to the curvature of the effective Higgs potential) from each state.  Indeed, these are the strengths with which each state couples to the Higgs, as measured by the degree to which the mass of the state responds to fluctuations of the Higgs VEV.~
The result in Eq.~(\ref{Higgsstep1}) can thus serve as the starting point (Step~\ref{step1}) for our analysis.

As shown in Ref.~\cite{Abel:2021tyt}, the insertion of these non-trivial $\mathbbX_i$ operators breaks the modular invariance due to a subtle modular anomaly.   As a result, a modular completion is needed.  It turns out~\cite{Abel:2021tyt} that the appropriate completion in this case can be achieved by introducing an additional constant into the operator insertions, so that Eq.~(\ref{Higgsstep1}) now takes the fully modular-invariant (completed) form
\beq
        m_\phi^2  ~=~  -\frac{\calM^2}{2}
              \left\langle  \frac{\xi}{4\pi^2} +  \tau_2 \mathbbX_1 + \tau_2^2\, \mathbbX_2 \right\rangle 
\label{higgscompletion1}
\eeq
where $\xi$ is an ${\cal O}(1)$ parameter which describes the way in which the particular Higgs field under study is embedded within the corresponding string model.   Indeed, this extra term can be interpreted as arising from the universal gravitational backreactions associated with the direct Higgs couplings to the individual string states.
Eq.~(\ref{higgscompletion1}) is therefore fully modular invariant and serves as the result of Step~\ref{modularcompletion}.   Note that use of our result in Eq.~(\ref{Lambdanoinsertion}) enables us to rewrite Eq.~(\ref{higgscompletion1}) in the form
\beq
        m_\phi^2  ~=~    \frac{\xi}{4\pi^2} \frac{\Lambda}{\calM^2} 
             -\frac{\calM^2}{2}
              \biggl\langle \tau_2 \mathbbX_1 + \tau_2^2\, \mathbbX_2 \biggr\rangle ~,
\label{higgscompletion2}
\eeq
thereby indicating the existence of a surprising string-theoretic connection between the Higgs mass and the cosmological constant~\cite{Abel:2021tyt}.  It is intriguing that such relations join together precisely the two quantities whose values lie at the heart of the two most pressing hierarchy
problems in modern physics.

In general, the quantity in Eq.~(\ref{higgscompletion2}) can diverge at most logarithmically.   This is also a striking result, indicating that modular invariance has significantly softened what would otherwise have been a field-theoretic {\it quadratic}\/ divergence of the Higgs mass.   Moreover, we see that this quantity is actually {\it finite}\/ unless the string model in question happens to contain a net number of massless $\mathbbX_2$-charged string states.   
For simplicity, we shall therefore proceed under the assumption that the net number of massless $\mathbbX_2$-charged string states
vanishes in the string model under discussion,
and merely note that the analysis presented in Ref.~\cite{Abel:2021tyt} is completely general and considers all possible cases, including those in which the net number of such states is non-zero.

Given these assumptions, we can now continue to express these results in different forms.   One possibility is to proceed directly through Step~5 towards Step~7 by taking the Rankin-Selberg transform of our modular-complete result in Eq.~(\ref{higgscompletion2}).  In this way, one finds that the Higgs mass can generally be expressed in terms of the contributions from only the physical string states~\cite{Abel:2021tyt}:
\beq
  m_\phi^2 ~=~ \left. \frac{1}{24} \calM^2 \,
            \Str \left[ D_\phi^2  M^2(\phi)\right]\,\right|_{\phi=0}~
\label{higgssupertrace}
\eeq    
where we have defined the modular-covariant double-$\phi$ derivative
\beq
   D_\phi^2   ~\equiv~ \partial_\phi^2 +  \frac{\xi}{4\pi^2 \calM^2}~.
\label{Dphi2}
\eeq
The result in Eq.~(\ref{higgssupertrace}) is thus the Higgs-mass analogue of the $\Lambda$-result in Eq.~(\ref{Lamresult}).

Another possibility is to analyze how our string-theoretic Higgs mass runs as a function of a spacetime mass scale $\mu$.   For this purpose we start from the result in Eq.~(\ref{higgscompletion2}) and proceed towards Step~\ref{finite1} by introducing a suitable regulator.  As discussed in Ref.~\cite{Abel:2021tyt}, there are many requirements on such regulators, chief among them that they be completely modular invariant.   
One compelling class of such regulators can be formulated by deforming our one-loop amplitudes
\beq
   \langle A \rangle  ~\to~ \langle A \rangle_{\widehat{\cal G}}
\label{amplitudeAG}
\eeq
where $\langle A \rangle_{\widehat\calG}$ is defined exactly as in 
Eq.~(\ref{amplitudeA}) except that the integrand is now multiplied by 
an appropriate modular-invariant regulator function $\widehat{\calG}(a_i, \tau)$, with $a_i$ denoting the internal regulator parameters.
We then must demand  that $\widehat{\calG}(a_i,\tau)$ exhibit certain properties in order to ensure that we have a sensible regulator.    In particular, for such a regulator, we demand that there exist a combination or function $f(a_i)$  of regulator parameters 
such that taking $f(a_i)\to 0$ effectively removes the regulator while taking any non-zero value of $f(a_i)$ allows the regulator to suppress the unwanted divergences but otherwise leave the theory intact as far as possible.  Given that all such divergences must come from those portions of the integration region in which $\tau\to \tau_{\rm cusp}$ (where $\tau_{\rm cusp}$ are the so-called ``cusp'' points $\tau_{\rm cusp}= i\infty$ or $\tau_{\rm cusp} = p/q$, where $p,q\in\mathbbZ$), we thus have three requirements for suitable modular-invariant regulator functions $\widehat{\calG}(a_i,\tau)$:
\begin{itemize}
    \item  For all $f(a_i)>0$, we require that  $\widehat{\calG}(a_i,\tau) \to 0$ sufficiently rapidly as $\tau\to \tau_{\rm cusp}$.   This enables our regulator to suppress divergences and yield a finite one-loop string amplitude.
    \item For all $f(a_i)>0$, we also require that
    $\widehat{\calG}(a_i,\tau)\approx 1$ when $\tau$ is sufficiently far away from the cusp points.  This ensures that our regulator, while suppressing divergences near the cusp points, leaves the remainder of the theory intact as much as possible.
    \item Finally, as $f(a_i)\to 0$, we require that $\widehat{\calG}(a_i,\tau)\to 1$ for all $\tau$.   This ensures the existence of a limit in which our regulator is effectively  removed and our original theory is obtained.
\end{itemize}
We shall also need to require for consistency that $\widehat{\calG}(\tau)$ satisfy an additional algebraic identity~\cite{Abel:2021tyt} whose significance will be discussed shortly.

In Ref.~\cite{Abel:2021tyt}, a suitable modular-invariant regulator function $\widehat{\calG}(a_i,\tau)$ meeting all of these criteria was developed.  This regulator function will be discussed in detail in Sect.~\ref{sec:string-comp}.~  However, using this regulator, we can then take Step~\ref{finite1} by evaluating 
\beq
 \widehat m_\phi^2(\rho,a)  ~=~  -\frac{\calM^2}{2}
              \left\langle  \frac{\xi}{4\pi^2} +  \tau_2 \mathbbX_1 + \tau_2^2\, \mathbbX_2 \right\rangle_{\widehat{\calG}} ~.
\label{higgscompletion3}
\eeq
We then follow Step~\ref{step6} by mapping to a spacetime mass scale $\mu$ via the identification~\cite{Abel:2021tyt}
\beq
        \mu^2 ~=~  f(a_i) \,M_s^2~,
\label{muscale}
\eeq
after which we follow Step~\ref{step8} by evaluating the Rankin-Selberg transform.   
A detailed discussion of the Rankin-Selberg procedure is provided in Ref.~\cite{Abel:2021tyt}.
The end result of this analysis yields our final on-shell result for the running Higgs mass, expressed completely in terms of supertraces over only physical string states.  Indeed, this result takes the form
\beq
   \widehat m_\phi^2(\mu) ~=~ \widehat m^2_\phi(\mu)\biggl|_\mathbbX +\, \frac{\xi}{4\pi^2 \calM^2} \widehat \Lambda(\mu)~
\label{twoterms}
\eeq
where the two different terms on the right side represent the contributions ultimately stemming from the different terms in Eq.~(\ref{higgscompletion1}).

The algebraic forms of these final results~\cite{Abel:2021tyt} are fairly complicated (involving infinite sums of Bessel functions) and thus not particularly illuminating. 
However, the total result for the running Higgs mass 
$\widehat m_\phi^2 (\mu)$
is plotted in Fig.~3 of Ref.~\cite{Abel:2021tyt}.
One important feature of this running is a ``scale-duality'' invariance~\cite{Abel:2021tyt} under
$\mu \to M_s^2/\mu$.
As discussed in Ref.~\cite{Abel:2021tyt},
the emergence of scale duality is a general phenomenon, an unavoidable consequence of modular invariance and its corresponding UV/IR symmetries.   

The existence of scale duality nevertheless places an additional constraint on potential regulator functions $\widehat{\calG}(a_i, \tau)$.
Specifically, scale-duality symmetry in conjunction with the identification in Eq.~(\ref{muscale}) together require that our regulator function $\widehat{\calG}(a_i,\tau)$ also exhibit an invariance under any transformations on the parameters $a_i$ for which $f(a_i)\to 1/f(a_i)$.   Phrased slightly differently, the transformation $f(a_i)\to 1/f(a_i)$ must be a symmetry of the regulator.   
Otherwise, it would not be possible to identify a spacetime mass scale $\mu$ consistent with scale duality.   
Thus, while a regular function without this additional symmetry might have been sufficient if our only goal were to tame divergences, this extra symmetry is required if we wish to further identify some combination $f(a_i)$ of  regulator parameters with a spacetime mass scale $\mu$ and thereby express our results as quantities that run with $\mu$.

Given the explicit expressions for $\widehat m_\phi^2 (\mu)\bigl|_\mathbbX$ and $\widehat \Lambda(\mu)$ in Ref.~\cite{Abel:2021tyt}, it is possible to verify that $\lim_{\mu\to 0}\widehat \Lambda(\mu)= \Lambda$, as expected when the regulator is removed.
Moreover, it turns out that
\beq
   \widehat m_\phi^2 (\mu) \biggl|_\mathbbX ~=~ \frac{\partial^2}{\partial\phi^2} \widehat\Lambda(\mu,\phi)\biggl |_{\phi=0}~.
\eeq
Indeed, as discussed in Ref.~\cite{Abel:2021tyt}, this result holds independently of the choice  of regulator function $\widehat{\calG}(a_i,\mu)$.
Given Eq.~(\ref{twoterms}), we then
have 
\beqn
          \widehat m_\phi^2(\mu) ~&=&~  
             \left.\left( \partial_\phi^2 +  \frac{\xi}{4\pi^2 \calM^2} \right)  
               \, \widehat\Lambda( \mu,\phi) \right|_{\phi=0} ~~~~~\nonumber\\
             &=&~  \left. D_\phi^2   ~\widehat\Lambda( \mu,\phi) \right|_{\phi=0}~,
\eeqn
whereupon taking the $\mu\to 0$ limit we find
\beqn
  \lim_{\mu\to 0}  \widehat m_\phi^2(\mu)  ~&=&~ D_\phi^2 \,\Lambda(\phi) \biggl|_{\phi=0}
      \nonumber\\
    &=&~ \left. \frac{1}{24} \calM^2 \,
            \Str \left[ D_\phi^2  M^2(\phi)\right]\,\right|_{\phi=0}~,~~~~~~~
\label{asymplimit}
\eeqn
thereby matching the result for $m_\phi^2$ from Step~\ref{step7} in Eq.~(\ref{higgssupertrace}).
This matching is an important cross-check, since taking $\mu\to 0$ corresponds to the removal of our regulator.  
Indeed, pushing this further, we see that
$\Lambda$ and $m_\phi^2$ are related through the algebraic structure~\cite{Abel:2021tyt}
\beq
   \begin{cases}
      ~\Lambda &=~~ \Lambda(\phi)\bigl|_{\phi=0} \\
      ~m_\phi^2 &=~~ D_\phi^2 \,  \Lambda(\phi)\bigl|_{\phi=0}~,
   \end{cases}
\eeq
with this structure remaining intact even if we extend these quantities to run as functions of $\mu$.
Finally, the second of these relations suggests that we may view $\Lambda(\phi)$ as a Higgs Coleman-Weinberg potential for $\phi$ (at least locally).   This is discussed further in Ref.~\cite{Abel:2021tyt}.
  
\subsection{Goals and results of this paper}

As reviewed above, the one-loop cosmological constant $\Lambda$ and one-loop Higgs mass $m_\phi^2$ have already been analyzed within the formalism we have presented, with the central results outlined above.   In this paper, by contrast, our goal is to
analyze a third quantity:  the one-loop contributions to the gauge couplings $\alpha_i\equiv g_i^2/(4\pi)$ associated with the various gauge groups that might be present in a given string model.  

For the gauge couplings, it turns out that certain steps in the above procedure have already been performed.   
In a seminal early paper~\cite{Kaplunovsky:1987rp}, Kaplunovsky considered the so-called ``threshold corrections'' that are required to match the 
full string gauge couplings to an EFT at one loop
 and constructed an expression for such threshold corrections which we may regard as completing Step~\ref{step1}.  He recognized that this quantity generally diverges due to the contributions from certain massless states, and provided a procedure for removing this divergence.  Unfortunately, although sufficient for certain purposes, this procedure was not modular invariant.  Indeed, we shall see that even the starting point --- the notion of a ``threshold correction'' --- is not modular invariant, as it artificially separates the contributions of massless states from those of massive states.   This will be discussed further in Sect.~\ref{sec:string-comp}.

Later, in an important series of papers~\cite{Kiritsis:1994ta,Kiritsis:1996dn,Kiritsis:1998en},
Kiritsis and Kounnas revisited this issue and developed a properly modular-invariant regulator for this calculation.  
In so doing, Kiritsis and Kounnas implicitly completed Steps~2 and 3.  
Indeed, the regulator which we shall employ in this paper (and which was employed in Ref.~\cite{Abel:2021tyt})  is built upon the regulator they constructed.  However, our regulator has been generalized and modified in a certain critical way which allows us to proceed to identify a corresponding spacetime mass scale $\mu$ for all values of the regulator parameters $a_i$ and thereby express the gauge couplings as running quantities~\cite{Abel:2021tyt}.   Specifically, the regulator function used in Refs.~\cite{Kiritsis:1994ta,Kiritsis:1996dn,Kiritsis:1998en} 
satisfied the two bulleted requirements above but did not exhibit the required symmetry under $f(a_i)\to 1/f(a_i)$ which is critical for properly identifying a running spacetime mass scale $\mu$.
This will be corrected in our analysis in Sect.~\ref{sec:string-comp}.  

More importantly, however, the primary purpose of this paper is to bring this analysis of the gauge couplings to its natural conclusion.   In particular, we shall complete the remaining steps in our procedure outlined above, and seek to obtain an expression for the gauge couplings in terms of the supertraces of the contributions from only the physical string states.  Interestingly, we shall find that this cannot be done for all terms in our expressions because of the unique modular structure of the gauge couplings.    We shall therefore spend considerable time discussing this issue, and we shall develop a procedure through which these contributions can nevertheless be written as supertraces over certain string states.
We will also study the running of the gauge couplings as  functions of a spacetime mass scale $\mu$.  This will enable us to determine the properties --- and also the limits of validity --- for any associated EFT describing the behavior of the gauge couplings in closed string theories.   In particular, we will see how the running EFT emerges from our prescription and evolves as various mass thresholds are crossed.

 \section{Gauge couplings in string theory:   General treatment}
 \label{sec:string-comp}
 
\label{sec:genericpicture}

We now turn to the principal goal of this paper:   to utilize the methods outlined above in order to study the behavior of the one-loop contribution to the gauge coupling $g_G$ corresponding to any spacetime gauge group $G$ in closed string theory.  We shall normalize these couplings such that the corresponding gauge-kinetic terms are given by
\beq 
      {\cal L}~=~ -\frac{1}{4g_G^2} F^{(G)}_{\mu\nu }
      F^{(G)\mu\nu }~,
\eeq 
and we shall isolate the one-loop contributions to $g_G$ by evaluating these couplings $g_G$ to one-loop order and then separating out the tree-level contributions.   In general, these  quantities are related through
\beq 
    \left. \frac{16\pi^2}{g_G^2} \right|_{\substack{{\rm total~thru}\\ {\rm one\hbox{-}loop}\\ {\rm order}}}
    ~=~
    \left. \frac{16\pi^2}{g_G^2} \right|_{\rm tree} ~+~\Delta_G
\label{oneloopcontribution}
\eeq
where $\Delta_G$ denotes the one-loop contribution to $16\pi^2/g_G^2$.
Indeed, in string theory we know that 
$g_G|_{\rm tree}\sim e^{-\langle \phi\rangle}$ where $\langle \phi\rangle$ denotes the VEV of the dilaton $\phi$.
Our goal in this paper is thus to study the properties of $\Delta_G$.  

\subsection{Operator insertions}

In field theory, we know that $\Delta_G$ receives contributions from all of the states in our theory which transform in non-trivial representations $R$ of $G$.  Indeed, for each such state in the theory, the corresponding one-loop contribution to $\Delta_G$ is given by $b\cdot {\rm tr}_R(Q_G^2)$, where
\begin{itemize}
    \item $Q_G^2$ is the sum of the squares of the charges in the Cartan subalgebra of $G$;
    \item the trace tallies the values of $Q_G^2$ over all the states within the representation $R$ (following the convention that each CPT-conjugate particle/anti-particle pair of states is counted only once);  and
    \item the numerical coefficient $b$ encapsulates the Lorentz helicity properties of the state, with $b=\lbrace 1/3,2/3,-11/3\rbrace$ for Lorentz scalars, spinors, and vectors respectively.
\end{itemize}
Indeed, we note that these $b$-coefficients are nothing but $b = -4 (-1)^F ( S^2 -1/12)$
where $S= \lbrace 0, 1/2, 1\rbrace$ is the Lorentz spin of the corresponding state and where $F$ is the spacetime fermion number.  

Given these observations, it is straightforward to generate an analogous expression in string theory. Of course, in string theory, our traces count {\it all}\/ states in the theory independently and thus tally each member of a CPT-conjugate particle/anti-particle pair separately.  With this effective doubling of the conventions for our traces, our field-theoretic $b$-coefficients are effectively rescaled to become $b=\lbrace 1/6, 1/3, -11/6\rbrace$ for Lorentz scalars, spinors, and vectors respectively, or equivalently $b=-2 (-1)^F (S^2 - 1/12)$.  At this stage, then, our QFT-motivated expression for $\Delta_G$ in string theory can be expected to take the form
\beq 
  \Delta_G ~=~ -2 \Bigl\langle  \left(S^2 - 1/12\right)\, Q_G^2 \Bigr\rangle
\label{Gstep1}
\eeq 
where the brackets signify the full one-loop amplitude of the form given in Eq.~(\ref{amplitudeA}).  Indeed, we note from Eq.~(\ref{amplitudeA}) that these brackets already include the factor of $(-1)^F$ as well as the double sum $\sum_{m,n}$ which effects the sum over gauge-group representations $R$ and the traces over $Q_G^2$ within each $R$.
Furthermore, without loss of generality, the presence of a gauge symmetry implies that our string states populate a corresponding lattice of gauge charges.  We can then decompose
\beq
       Q_G^2 ~=~ \sum_{\ell,\ell'} c_{\ell\ell'}^{(G)} Q_\ell Q_{\ell'}
\label{Gembedding}
\eeq
where the $Q_\ell$ component is the charge operator in the $\ell^{\rm th}$ lattice direction and where  the coefficients $c_{\ell\ell'}^{(G)}$ describe how the string gauge group $G$ is {\it embedded} within the charge lattice.

Eq.~(\ref{Gstep1}) thus represents our Step~1 starting point for our study of the one-loop contributions to the gauge couplings.  Indeed, we see that this quantity is written in terms of the product of two insertions,  $Q_G^2$ and $S^2-1/12$, and thus  
resembles as closely as possible the field-theory result, only expressed in terms of a full one-loop string amplitude.  
Note that if our theory is spacetime-supersymmetric, then we are free to drop the factor of $-1/12$, since the contributions from this term will be proportional to ${\rm Tr}\,(-1)^F$ for each representation of the gauge group and thus vanish.   We shall nevertheless keep this factor for generality.  

According to the procedure outlined in Sect.~II, our next step is to perform a {\it modular completion} of this expression.   Clearly, there are two separate insertions in play:  $Q_G^2$ and $S^2-1/12$.   We shall discuss each of these in turn, since neither insertion preserves the modular invariance of the full string amplitude.

Let us first discuss the modular completion of 
$Q_G^2$.  
In general, it was shown in 
Ref.~\cite{Abel:2021tyt} that the product of any two charge bilinears can be modular completed
by substituting
\beq
      Q_\ell Q_\ell' ~\to ~   Q_\ell Q_\ell' -\frac{1}{4\pi \tau_2} \delta_{\ell,\ell'}~.
\label{QQcompletion}
\eeq 
Given the embedding in Eq.~(\ref{Gembedding}),
we thus find the modular completion of $Q_G^2$ is given by
\beq
    Q_G^2 ~\to~ Q_G^2 - \frac{\xi}{4\pi \tau_2}
\eeq
where $\xi \equiv \sum_\ell c_{\ell\ell}^{(G)}$.
Indeed, with this result,
we see that $\xi$ is ultimately related to the {\it affine level}\/ $k_G$ at which the gauge group $G$ is realized.

We now turn to the modular completion of the helicity factor $S^2-1/12$ in Eq.~(\ref{Gstep1}).
In general, a given string theory gives rise to infinite towers of states with higher and higher spins.  However, in the heterotic string, these states can ultimately be organized in terms of the CFT sector from which they arise, where the CFT in question is that associated with the transverse right-moving Lorentz group $SO(D-2)$.  In the heterotic string, there are only three such sectors:  the identity (or scalar) sector, the spinor sector, and the vector sector.  The ground states of these sectors have spins $S=\lbrace 0, 1/2, 1\rbrace$ respectively.    Loosely speaking, every other string state  can be viewed as a member of one of these sectors in the sense that it can be realized through tensor products of this vacuum state (or one of its CFT descendants) with additional vector representations arising from excitations of the left-moving coordinate bosons.  In this way, states with arbitrarily high spins can be generated.

Disregarding the contributions from the purely internal degrees of freedom and the two transverse spacetime-coordinate bosons,  
the contribution to the total partition function from the states in each of these three sectors takes the form $\Thetabar/\etabar$, where $\eta$ is the Dedekind eta-function and where $\Theta$ is given by
\beqn
   {\rm scalar:} ~~~\Theta ~&=&~ \half\left(
      \vartheta_3 + \vartheta_4 \right)\nonumber\\
{\rm spinor:} ~~~\Theta ~&=&~ \half\vartheta_2\nonumber\\
     {\rm vector:} ~~~\Theta ~&=&~ \half\left( 
      \vartheta_3 - \vartheta_4 \right)~.
\eeqn
Here $\vartheta_i$ are the three Jacobi theta-functions.   Indeed, in each of these
cases we find that
\beq 
       \Theta~\sim ~e^{\pi i \tau S^2}(1+ ...)~,
\eeq
thereby already suggesting a relationship
between $S$ and a modular derivative.

Given this, we now seek to understand how to incorporate the helicity factor $S^2 -1/12$ 
in a fully modular-invariant way into  the sum over 
string states.  A direct string calculation~\cite{Kaplunovsky:1987rp} tells us that the proper procedure to generate the helicity part is to modify the total partition function of the string theory in question, 
replacing
\beq
  \frac{\Thetabar}{\etabar} ~\to~ \frac{\partial}{\partial\taubar} 
  \left( \frac{\Thetabar}{\etabar}\right)~.
\label{helicity_replacement}
\eeq
This is the result of a full string calculation, and thus this replacement does not disturb the modular invariance of the total partition function.   In particular,
the $\tau$-derivative $d/d\taubar$ is modular-covariant when acting on a modular-covariant function of modular weight $k=0$ such as $\Thetabar/\etabar$.
Thus, no further modular completion is required after this 
replacement is implemented.
Or, to phrase this another way, the simple insertion $S^2-1/12$ has been ``modular completed'' by instead implementing the replacement in Eq.~(\ref{helicity_replacement}).

The issue that remains for us, however, is to express 
the replacement in Eq.~(\ref{helicity_replacement}) as an {\it insertion}\/ into the numerator of the partition-function trace.   We wish to do this in order to eventually express our results in terms of (weighted) traces over our original string spectrum.
To accomplish this, we observe that 
\beqn
\frac{\partial}{\partial \taubar}
\left( \frac{\Thetabar}{\etabar}\right)
  ~&=&~ \frac{1}{\etabar} \frac{\partial \Thetabar}{\partial\taubar} + \Thetabar 
   \frac{\partial}{\partial \taubar}  \frac{1}{\etabar}
    \nonumber\\
 &=&~ \frac{1}{\etabar} \left[ 
  \frac{\partial \Thetabar}{\partial\taubar}
  - \Thetabar \frac{\partial}{\partial \taubar} \log \etabar\right]~\nonumber\\
 &=&~ \frac{1}{\etabar} \left[ 
      \frac{\partial}{\partial\taubar} + \frac{\pi i}{12} \,\overline{E}_2(\taubar) \right] \Thetabar~~~
\label{derividentity}
\eeqn
where in passing to the final line we have utilized the identity
\beq 
    E_2(\tau) ~=~ \frac{1}{2\pi i} \frac{\partial}{\partial\tau} \log \eta^{24}(\tau)
\eeq
where $E_2(\tau)$ is the normalized weight-two holomorphic Eisenstein function
\beqn
E_2(\tau) ~&\equiv&~ 1 -24 \sum_{n=1}^{\infty}\sigma(n)e^{2\pi i n \tau}~\nonumber\\
      &=&~ 1 - 24 q - 72 q^2 -96 q^3 - 168 q^4 - ...~~~~~~~~
\label{defG2here}
\eeqn
with $ \sigma(n)\equiv \sum_{d|n} d$. 
We can shall find it convenient to simplify this notation slightly by writing $E_2(\tau) = \sum_{n=0}^\infty \chi_n q^n$ where
\beq 
     \chi_n ~=~ \begin{cases}
        ~~~~~~ 1 &  n=0 \\
          -24 \sigma(n) & n>0 ~.
     \end{cases}
\label{cr0}
\eeq
We thus see that the replacement in
Eq.~(\ref{helicity_replacement}) is tantamount to the insertion of the modular-covariant derivative  $\overline{D}_\taubar$ into that portion of the total partition-function trace corresponding to the spacetime Lorentz group, where
\beq
   D_\tau ~\equiv~ \frac{\partial}{\partial\tau} - 
   \frac{i \pi }{12} E_2(\tau) ~.
\label{Eisen_derivative}
\eeq
In this sense $D_\tau$ is the operator that represents $S^2-1/12$ in string theory.

As evident from this discussion, the operator $d/d\taubar$ acting purely on $\Thetabar$ represents the spin $S^2$.  Indeed, we can identify the spin $S$ as the ``helicity charge'' $\overline{Q}_H$ of the state relative to the spacetime Lorentz symmetry, where the subscript $H$ can be identified as that right-moving lattice direction $\ell$ whose trace yields $\Thetabar$.
We can therefore identify $\overline{Q}_H^2 = \frac{i}{\pi} \partial/\partial\taubar $, allowing us to express our
modular completion in the form
\beq
     \overline{Q}_H^2 - \frac{1}{12} ~~\to~ ~\overline{Q}_H^2 - \frac{1}{12} \overline{E}_2(\taubar)~.
\label{Eisen_helicity}
\eeq  

At first glance, it might have seemed from Eq.~(\ref{QQcompletion})
that the modular completion of $\overline{Q}_H^2$ would simply be $\overline{Q}_H^2\to \overline{Q}_H^2 - 1/(4\pi\tau_2)$, just as occurred for the gauge charges.   However, the critical difference here is that we are not seeking the modular completion of $Q_H^2$;   we are seeking the modular completion of $Q_H^2-1/12$.    It is the presence of the extra term $-1/12$ which induces the subtlety and ultimately requires the Eisenstein function in Eq.~(\ref{Eisen_derivative}).
Although it might have seemed that the extra shift $-1/12$ is only a pure number and thus should be completely harmless, this neglects the fact that we must preserve modular invariance.  While the insertion of $S^2$ raises the modular weight of the corresponding portion of the partition function by two, the insertion of a pure number such as $-1/12$ does not affect the modular weight at all.   {\it We thus cannot subtract $1/12$ directly from $S^2$ or $Q_H^2$ in a modular-invariant theory;  rather, the $-1/12$ must first be ``modular completed'' into a modular function (or in this case, a quasi-modular function) of weight two}\/.  As it turns out, a theorem in modular-function theory asserts that there is only one (quasi-)modular function of weight $k=2$:  this is the Eisenstein function $E_2(\tau)$.  It is thus natural and expected that the modular completion in Eq.~(\ref{Eisen_helicity}) would involve the Eisenstein function.  Indeed, in this sense we may regard $E_2/12$ as the properly normalized modular completion of $1/12$, with $E_2/12= 1/12 + {\cal O}(q)$.

As noted above, the Eisenstein series $E_2$ (unlike the Eisenstein series $E_{2k}$ for $k>1$)
is not a strict modular function.  Instead, $E_2$ is only {\it quasi}\/-modular, transforming under modular transformations as
\beq
~~~E_2\left(\frac{a\tau+b}{c\tau+d}\right) ~=~
(c\tau+d)^2 E_2(\tau) - \frac{6}{\pi} ic (c\tau+d)~.~~~
\label{E2anomaly}
\eeq 
It is the latter ``anomaly'' term in this result which spoils the true modular covariance for $E_2$.
However, this is precisely what is needed because the derivative $\partial/\partial \tau$ in Eq.~(\ref{Eisen_derivative}) also fails to be modular invariant in exactly the opposite way.  Thus, it is precisely the combination in Eq.~(\ref{Eisen_derivative}) that yields a fully modular-invariant result.

Given these results, we see that our modular-completed expression for $\Delta_G$ now takes the form
\beq 
  \Delta_G ~=~ -2 \left\langle \tau_2^2\, 
    \left( \overline{Q}_H^2 - \frac{1}{12} \overline{E}_2\right) \,
  \left( Q_G^2 - \frac{\xi}{4\pi \tau_2}\right)
  \right\rangle~.
\label{gstep2}
\eeq 
Note that the extra factor of $\tau_2^2$ that has been inserted into Eq.~(\ref{gstep2}) is another element of our modular completion.
This reflects the fact that the insertions of the helicity and gauge factors --- although preserving modular 
invariance --- also together raise the modular weight of the resulting integrand in Eq.~(\ref{amplitudeA}) by two units (from $k=-1$ to $k=+1$) for any four-dimensional string theory.  Modular invariance then dictates that such an increase in the modular weight of the integrand be accompanied by a corresponding increase in the number of leading $\tau_2$ prefactors. 
The result in Eq.~(\ref{gstep2}) then completes Step~2 of the procedure outlined in Sect.~II.

At this stage, it may be worthwhile to compare with the classic results of Kaplunovsky in  Ref.~\cite{Kaplunovsky:1987rp}.
First, we emphasize that in this paper we are simply calculating the one-loop contributions to the gauge coupling.   In particular, despite the algebraic resemblance of Eq.~(\ref{oneloopcontribution}) 
to a renormalization-group equation (RGE) for a running gauge coupling, 
at this stage we have not introduced any notion of running or scale.  Second, this conceptual difference notwithstanding, there is a further critical difference in that the contributions from the  massless states were explicitly removed within the calculation of the $\Delta_G$-term in  Ref.~\cite{Kaplunovsky:1987rp}.  This was done because a separate field-theoretic logarithmic running (assumed to be contributed from the massless states) was explicitly introduced into the renormalization-group version of Eq.~(\ref{oneloopcontribution}) in Ref.~\cite{Kaplunovsky:1987rp}.  This rendered $\Delta_G$ in Ref.~\cite{Kaplunovsky:1987rp} a mere tally of the contributions from only the massive modes.  Thus, in this sense,  the version of $\Delta_G$ in Ref.~\cite{Kaplunovsky:1987rp} became a mere threshold correction, one which is devoid of its own running.   

{\it By contrast, in this paper $\Delta_G$  will always represent the full one-loop contribution to the gauge coupling, with the contributions from both massless and massive states included together in a unified way}\/.  Indeed, it is only in such a manner that we can ever hope to preserve modular invariance throughout our calculations.   Moreover, once we proceed to introduce a scale dependence into our eventual results and consider how these quantities run, we shall even find that the contributions from the massless string states are not strictly logarithmic, but instead take a more complex form which is dictated by modular invariance and which only reduces to a logarithmic running in a certain EFT-like limit.

Certain aspects of the modular completions we have discussed here also appear in Ref.~\cite{Kaplunovsky:1987rp} and in the subsequent work reported in  Refs.~\cite{Kiritsis:1994ta,Kiritsis:1996dn,Kiritsis:1998en}.  In particular, our modular completion of $Q_G^2$ is already implicit in
Refs.~\cite{Kiritsis:1994ta,Kiritsis:1996dn,Kiritsis:1998en} and further discussed/reviewed in Refs.~\cite{Dienes:1995bx,Dienes:1996du}. Likewise, the effective ``modular completion'' of the helicity factor whereby the factor of $S^2-1/12$ is dropped in favor of the replacement in Eq.~(\ref{helicity_replacement}) already appears in Ref.~\cite{Kaplunovsky:1987rp}.   However, our subsequent reformulation of this replacement
as a partition-function insertion involving the Eisenstein function $E_2$ --- as given
 in Eqs.~(\ref{Eisen_derivative}) and (\ref{Eisen_helicity}) ---  is, as far as we are aware, new and does not appear in the prior literature.

Given our expression in Eq.~(\ref{gstep2}), we can now continue along the path outlined in Sect.~II.~  
In particular, following  Eq.~(\ref{Higgsstep1}), we see that
our total operator insertion for the gauge couplings is given by 
\beq
 \calX~\equiv~ 
 -2\tau_2^2\, 
    \left( \overline{Q}_H^2 - \frac{1}{12} \overline{E}_2\right) \,
  \left( Q_G^2 - \frac{\xi}{4\pi \tau_2}\right)~.
\label{Xdef}
\eeq
Expanding $\calX$ in leading powers of $\tau_2$ then yields
\beq
\calX ~=~ \tau_2 \mathbbX_1
    + \tau_2^2 \mathbbX_2 
\label{X1X2split}
\eeq
where we now identify
\beqn
\mathbbX_1 ~&\equiv&~ \, \,\frac{\xi}{2\pi} 
 \left( \overline{Q}_H^2  -\frac{\overline{E}_2}{12} \right) \nonumber\\
\mathbbX_2 ~&\equiv&~ -2 \left( \overline{Q}_H^2  -\frac{\overline{E}_2}{12} \right) Q_G^2~.
\label{eq:Xs}
\eeqn 
This division of the total insertion into two separate terms $\mathbbX_1$ and $\mathbbX_2$ is based on their leading powers of $\tau_2$ and will be important when we discuss how our expressions diverge and what kinds of running these quantities ultimately experience.   However, we stress that neither $\langle \tau_2 \mathbbX_1\rangle$ nor $\langle \tau_2^2 \mathbbX_2\rangle$ is modular invariant by itself.   Indeed, these two terms serve as modular completions of each other, and only their sum in Eq.~(\ref{X1X2split}) is modular invariant.
Phrased slightly differently, the splitting of the total insertion into an $\mathbbX_1$ piece and an $\mathbbX_2$ piece based on their leading powers of $\tau_2$ is not unique.   This non-uniqueness arises because modular transformations (especially the Poisson resummations that often underlie these transformations) can change the apparent powers of leading $\tau_2$ factors that appear.   Thus such resummations have the power to induce a reclassification of various terms 
as belonging to either $\mathbbX_1$ and $\mathbbX_2$.   However, once a given separation into $\mathbbX_1$ and $\mathbbX_2$ is given, it will be consistent to perform all calculations within the framework of that separation without further Poisson resummations.  This issue will be discussed further below.

Our result in Eq.~(\ref{X1X2split}) tells us that
our calculation of the gauge couplings shares the same basic algebraic structure as our calculation of the Higgs mass in Ref.~\cite{Abel:2021tyt}.
However, one important difference is the fact that $\mathbbX_{1,2}$ now depend on the worldsheet modular parameters $\tau_{1,2}$ through the Eisenstein function $E_2(\tau)$.  In other words, we now have more than simple charge insertions --- we also have the insertion of an entire modular function!  We will shortly see that this difference will have important ramifications.  

\subsection{Divergences and regulator function\label{sec:regulator}}

Our next step is to study the potential divergence structure of $\Delta_G$.   Indeed, just as in the Higgs case, it is possible for $\Delta_G$ to diverge.
For example, any level-matched massless state which carries a non-zero $\mathbbX_2$ charge will induce a divergence in $\Delta_G$ unless this state is balanced against another similar state of opposite statistics.  Indeed, in a rough sense to be clarified shortly, the divergence in $\Delta_G$ will be proportional to $\zStr \mathbbX_2$.  Likewise, we see that this divergence is at most {\it logarithmic}\/.  

The fact that  $\Delta_G$ formally diverges means that we must introduce a regulator. It is here that we pass to Step~3 within Fig.~\ref{fig:process}.  It might seem natural to proceed by simply subtracting the contributions from the masssless states (or more precisely the $\mathbbX_2$-charged massless states) from $\Delta_G$.   This is reminiscent of what was done in Ref.~\cite{Kaplunovsky:1987rp}, but introducing this sort of artificial distinction between massless and massive states necessarily breaks the modular invariance of $\Delta_G$.    

Instead, following what was done in Ref.~\cite{Abel:2021tyt} for the Higgs mass, we shall regulate our theory by deforming the one-loop amplitude as described in Eq.~(\ref{amplitudeAG}),  introducing a new regulator function $\widehat {\cal G}(\tau,\taubar)$ into the integrand:
\beqn
\Delta_G ~\to~ 
   \widehat \Delta_G ~&\equiv&~ 
   \Bigl\langle 
   \tau_2 \mathbbX_1 + \tau_2^2 \mathbbX_2
   \Bigr\rangle_{
   \widehat\calG}\nonumber\\
   ~&=&~ 
   \Bigl\langle 
   \left(\tau_2 \mathbbX_1 + \tau_2^2 \mathbbX_2\right) 
   \,\widehat\calG \Bigr \rangle~.
\label{eq:Ctilde}
\eeqn
The issue at hand is thus to choose a suitable $\widehat\calG$ regulator function.

Below Eq.~(\ref{amplitudeAG}) we have listed a number of properties that such a regulator function should exhibit.  A function satisfying all of these properties was given in Ref.~\cite{Abel:2021tyt}, adapting prior results in Ref.~\cite{Kiritsis:1994ta}, and we shall use this function here as well.  This function
$\widehat \calG_\rho(a,\tau)$ has two free regulator parameters $a_i\equiv\lbrace \rho,a\rbrace$ 
with $\rho\in \mathbb{R}^+$ and $\rho\not=1$,
and is given by 
\beq
  \widehat\calG_\rho(a,\tau) \,\equiv\,
 \frac{1}{1+\rho a^2} \,\frac{\rho}{\rho-1} a^2 \frac{\partial}{\partial a} \biggl\lbrack Z_{\rm circ}( \rho a,\tau) - Z_{\rm circ}(a,\tau)\biggr\rbrack~ 
\label{regG}
\eeq 
where
\beq
     Z_{\rm circ}(a,\tau)~\equiv~  
    \sqrt{ \tau_2}\, \sum_{k,\ell\in\mathbb{Z}} \,
   \overline{q}^{(ka-\ell/a)^2/4}  \,q^{(ka+\ell/a)^2/4}
   ~.
\label{Zcircdef}
\eeq
Note that $Z_{\rm circ}(a,\tau)$ represents the sum over the Kaluza-Klein (KK) and winding modes that would be associated with a 
bosonic worldsheet field
compactified on a circle of radius $(M_s a)^{-1}$, with $k$ and $\ell$ respectively indexing the KK and winding modes, while the leading factor of $\sqrt{\tau_2}$ is inserted into Eq.~(\ref{Zcircdef}) in order to ensure that $Z_{\rm circ}$ is modular invariant.  We stress that 
for our purposes $Z_{\rm circ}$ is merely an ingredient in the definition of our regulator and  
does not correspond to any actual physical compactification of our theory.
It turns out that $\widehat\calG_\rho(a,\tau)\to 1$ as $a\to 0$, indicating that taking $a\to 0$ removes the regulator.  Indeed, for this function we have
\beq 
            f(a_i) ~=~ \rho a^2~.
\label{ffunction}
\eeq 
It then turns out that all of the bulleted requirements below Eq.~(\ref{amplitudeAG}) are satisfied, in addition to the requirement that $\widehat \calG_\rho(a,\tau)$ exhibit an invariance under $f(a_i) \to 1/f(a_i)$, or equivalently under $a\to (\rho a)^{-1}.$   Indeed, we will eventually identify our spacetime running scale $\mu$ according to Eq.~(\ref{muscale}) with $f(a_i)$ given in Eq.~(\ref{ffunction}).

\subsection{The Rankin-Selberg transform: From amplitudes to supertraces}

\label{subsec:supertraces}

Given this choice of regulator function, 
our result for $\widehat \Delta_G$ in Eq.~(\ref{eq:Ctilde}) can then be viewed as representing Steps~3 and/or 4 in Fig.~\ref{fig:process}.   From this point, there are several options.   One possibility is to proceed directly to Step~6 by identifying a running mass scale $\mu$ according to Eq.~(\ref{muscale}), where $f(a_i)$ is given in Eq.~(\ref{ffunction}).   However, in order to extract a description of the running of the gauge couplings which is as close as possible to what we might expect from ordinary quantum field theory, we are more interested in performing the Rankin-Selberg transform of our result in 
Eq.~(\ref{eq:Ctilde}) in order to pass to Steps~7 and 8.  

Operationally, this transform can be performed in a number of different ways.  In this section, we shall describe three different approaches to evaluating this transform in order to understand their relative advantages and disadvantages.  As we shall see, the first two approaches lead to results which hold only under certain simplifying assumptions.  Indeed, we describe these approaches 
because they connect to our previous calculations in Ref.~\cite{Abel:2021tyt}.
Unfortunately, these approaches lack the complete generality that we will ultimately require for some of the later calculations in this paper.   For this reason, after describing these two approaches, we shall then proceed to outline our third and preferred approach.  As we shall see, this approach will be completely general and lead to results which have not been previously described in the literature.

To begin, let us assume that the partition function of our string theory in $D$ uncompactified spacetime dimensions takes the form
\beq
    Z(\tau) ~\equiv~ \tau_2^k\, \sum_{m,n}  a_{mn}
  \,\qbar^m q^n
  \label{partfunctbare}
\eeq
where $a_{mn}$ is the net (bosonic minus fermionic) number of string states with right- and left-moving worldsheet energies $(m,n)$ in the string spectrum and where $k\equiv 1-D/2$.
These worldsheet energies are related to the total spacetime mass $M$ of the corresponding string state via $M^2 = \half(M_L^2+ M_R^2)$ where 
$m=\alpha' M_R^2/4$
and
$n=\alpha' M_L^2/4$.
Let us further assume that we wish to consider the corresponding one-loop amplitude $\bigl\langle \sum_\ell \tau_2^\ell A^{(\ell)}\bigr\rangle$ where the operator $\sum_\ell \tau_2^\ell A^{(\ell)}$ being inserted gives the value $\sum_\ell \tau_2^\ell A^{(\ell)}_{mn}$ when acting on a state with worldsheet energies $(m,n)$.
Our one-loop amplitude is then given by
\beq
  \Bigl\langle \sum_\ell \tau_2^\ell A^{(\ell)} \Bigr\rangle ~\equiv~ \int_\calF \frac{d^2 \tau}{\tau_2^2} \,\sum_\ell\,\tau_2^{k+\ell} \,\sum_{m,n} \,a_{mn} \,A^{(\ell)}_{mn} \,\qbar^m q^n~.
 \label{RSdeff1}
\eeq
For the sake of this discussion we shall assume that this amplitude is already finite and therefore does not require any regulation.  We shall return to this issue later when we discuss what happens when we also insert our regulator.   

As long as the amplitude in Eq.~(\ref{RSdeff1}) is finite and modular invariant, the 
Rankin-Selberg transform~\cite{rankin1a,rankin1b,selberg1} then tells us that
this amplitude may equivalently be expressed as
\beq
\Bigl\langle \sum_\ell \tau_2^\ell A^{(\ell)}\Bigr\rangle ~=~ \frac{\pi}{3}\,\, \oneRes\,\, \int_0^\infty 
    d\tau_2 \,\, \tau_2^{s-2}\,\, g(\tau_2)
\label{RSdeff2}
\eeq 
where
\beqn
 g(\tau_2) ~&\equiv&~ \sum_\ell \tau_2^{k+\ell} \,\int_{-1/2}^{1/2} d\tau_1 \, \sum_{m,n}  \,a_{mn} \,A^{(\ell)}_{mn} \, \qbar^m q^n \nonumber\\
 &=&~ \sum_\ell \tau_2^{k+\ell} \,\,\sum_{n} \, a_{nn}\, A^{(\ell)}_{nn} \,
         e^{-\pi \alpha' M_n^2 \tau_2}~. 
 \label{RSdeff3}
\eeqn 
where $\alpha' M_n^2=4 n$.  Inserting Eq.~(\ref{RSdeff3}) into Eq.~(\ref{RSdeff2}) and exchanging the order of the $n$-sum and the $s$-integral/residue (an operation whose validity will be discussed below), we obtain
\beqn
 \Bigl\langle\sum_\ell \tau_2^\ell A^{(\ell)}\Bigr\rangle ~&=&~ \frac{\pi}{3} \sum_\ell
 \sum_n a_{nn} A^{(\ell)}_{nn} \nonumber\\
 && ~ \times~\oneRes\,\,
  \int_0^{\infty} d\tau_2~  
   \tau_2^{k+\ell+s-2} ~
 e^{-\pi \alpha' M_n^2 \tau_2}
 \nonumber\\
 ~&=&~  \frac{\pi}{3}
 \sum_\ell \sum_n a_{nn} A^{(\ell)}_{nn} \nonumber\\
   && ~ \times~\oneRes\,\,
  \Bigl[ (\pi \alpha' M_n^2)^{1-k-\ell-s} \nonumber\\
    && ~~~~~~~~~~~~\times~
  \Gamma(
 k+\ell+s-1)\Bigl]~.~~~~~~
 \label{RSdeff4}
\eeqn
Taking $k=-1$ (as appropriate for $D=4$)
and evaluating the residue of the Euler $\Gamma$-function we thus obtain
\beqn
\Bigl\langle \sum_\ell \tau_2^\ell A^{(\ell)} \Bigr\rangle ~&=&~ 
   \frac{\pi}{3} \sum_{\ell \leq 1}
 \frac{(-1)^{\ell+1}}{(1-\ell)!} \nonumber\\
   && ~~~\times~ 
\sum_n a_{nn} A^{(\ell)}_{nn} \,(\pi \alpha' M_n^2)^{1-\ell}~,~~~~~~~~~~
\label{othersupertrace}
\eeqn
or equivalently
\beq
\Bigl\langle \sum_\ell \tau_2^\ell A^{(\ell)} \Bigr\rangle ~=~\frac{\pi}{3}
\sum_{\ell \leq 1} 
\frac{(-1)^{\ell+1}}{(1-\ell)!} 
\, \Str  \Bigl[ A^{(\ell)} (\pi \alpha' M^2)^{1-\ell} \Bigr] ~\\
\label{finalsupertraceresult}
\eeq
where for this purpose we temporarily identify the supertrace simply as
$\Str \,A \equiv \sum_{n} a_{nn} A_{nn}$.
We thus see that the Rankin-Selberg transform procedure has allowed us to express our original amplitude 
$\bigl\langle \sum_\ell\tau_2^\ell A^{(\ell)}\bigr\rangle$ in Eq.~(\ref{RSdeff1}) as a sum of supertraces over only physical string states (\ie, states with $m=n$).  

As an example of Eq.~(\ref{finalsupertraceresult}), we can consider the amplitude with no insertions at all.   Assuming a tachyon-free theory (so that the amplitude is finite, as required), we obtain
\beq
 \langle {\bf 1} \rangle ~=~
     - \frac{1}{12} \, \Str\, M^2/\calM^2
\eeq
where $\calM \equiv M_s/(2\pi) = (2\pi\sqrt{\alpha'})^{-1}$.   We thus find~\cite{Dienes:1995pm}
\beq
 \Lambda ~\equiv~ -\frac{\calM^4}{2}
  \langle {\bf 1} \rangle ~=~
   \frac{1}{24} \calM^2 \,\Str\,M^2~,
 \eeq
 as already noted in Eq.~(\ref{Lamresult}).

Interestingly, we see that our final expression in Eq.~(\ref{finalsupertraceresult}) receives no apparent contributions from off-shell states (\ie, 
states with $m\not= n$).  Likewise, our final result receives no contributions from $A^{(\ell)}$ insertions with $\ell \geq  2$.   Because of these features, it might initially seem that the value of the amplitude $\bigl\langle \sum_\ell \tau_2^\ell A^{(\ell)} \bigr\rangle$ is independent of these states and insertions.  However, this is not correct: our result in Eq.~(\ref{finalsupertraceresult}) holds only under the assumption that this amplitude is modular-invariant, and modular invariance certainly requires these off-shell states and 
the occasional inclusion of $A^{(\ell)}$ insertions with $\ell\geq 2$, as we have seen above.   Rather, what we are learning from Eq.~(\ref{finalsupertraceresult}) that modular invariance is so powerful a symmetry that the contributions from the off-shell states and $A^{(\ell)}$ insertions with $\ell\geq 2$ are already implicitly determined by --- and thus can be written in terms of --- the contributions from the on-shell and $\ell\leq 1$ insertions.

Finally, before proceeding further, we note that our derivation of 
the result in Eq.~(\ref{finalsupertraceresult}) rests on a number of algebraic manipulations which have important qualifications and implications.  One of these assumptions is relevant for Eq.~(\ref{RSdeff4}), in which it was assumed that all $M_n$ are positive in passing to the second line.   However, if massless states are present, one can always imagine deforming our theory to give these states very small non-zero masses.  One can then perform the integration in Eq.~(\ref{RSdeff4}) rigorously and remove these masses at the end.   As discussed in Sect.~IV of Ref.~\cite{Abel:2021tyt}, this procedure is valid as long as such massless states do not cause the amplitude under study to diverge.  Indeed, we see from the final result in Eq.~(\ref{finalsupertraceresult}) that it is difficult for massless states to make contributions in any case unless $\ell= 1$.

However, a more important assumption was made in passing from Eq.~(\ref{RSdeff3}) to Eq.~(\ref{RSdeff4}), where we 
exchanged the order of the $n$-summation and
$s$-integration/residue extraction. This exchange is valid only if the $n$-summation over the spectrum does not itself introduce any new divergences.   In general, this will indeed be the case.  However, there can be limits of our theories in which the spectrum becomes so dense as to be effectively continuous.   In such cases, this procedure breaks down, additional divergences can emerge from the sums over states, and there can be non-zero contributions from the operator insertions $A^{(\ell)}$ with $\ell >1$.  However, it is easy to understand what is happening in such situations:  the theory is becoming effectively higher-dimensional.   As a result, in such cases we could equivalently shift to a higher-dimensional description from the start.  The value of $k$ would then drop below $-1$, with a corresponding change in the values of $\ell$ for which Eq.~(\ref{finalsupertraceresult}) would receive explicit contributions.
Indeed, in a limit in which our four-dimensional theory becomes 
effectively  six-dimensional, we have $k= 1-D/2 \to -2$.   Following the same derivation as above, we then find that the range of the $\ell$-summation in Eq.~(\ref{finalsupertraceresult}) becomes $\ell\leq 2$, and we now have explicit contributions from $A^{(2)}$.  This phenomenon can also be understood in terms of the discussion in the paragraph below Eq.~(\ref{eq:Xs}).  As we decompactify our theory and the string spectrum becomes dense, it becomes appropriate to perform a Poisson resummation over the corresponding Kaluza-Klein states (momentum modes).   This Poisson resummation introduces extra powers of $\tau_2$ and thus effectively reshuffles certain terms between $\mathbbX_1$ and $\mathbbX_2$.

Fortunately, it is possible to reformulate the Rankin-Selberg transform procedure in such a way as to avoid exchanging the order of the $n$-sum and the $s$-integral/residue after reaching Eq.~(\ref{RSdeff3}). This then produces a more general result which holds even when the string spectra become dense.  Indeed, rather than integrate over $\tau_2$ and then take the residue at $s=1$ as in Eq.~(\ref{RSdeff4}), we can instead proceed by recognizing from Eq.~(\ref{RSdeff2}) that the original string amplitude $\bigl\langle \sum_\ell \tau_2^\ell A^{(\ell)}\bigr\rangle$ is nothing but the Mellin transform
of $g(\tau_2)/\tau_2$.   We can therefore write $g(\tau_2)$ directly as the inverse Mellin transform of this amplitude, thereby ultimately leading to the alternative result~\cite{zag,Kutasov:1990sv}
\beq
\Bigl\langle \sum_\ell \tau_2^\ell A^{(\ell)}\Bigr\rangle ~=~ \frac{\pi}{3}\, \lim_{\tau_2\to 0}
 \, g(\tau_2)~,
\label{RSdeff2alt}
\eeq   
where $g(\tau_2)$ continues to be given by Eq.~(\ref{RSdeff3}).
This result is equivalent to Eq.~(\ref{RSdeff2}), but has the primary advantage that we can now evaluate $\bigl\langle \sum_\ell \tau_2^\ell A^{(\ell)}\bigr\rangle$ simply by taking the $\tau_2\to 0$ limit of $g(\tau_2)$ rather than by evaluating the residue of the $\tau_2$ integral of $g(\tau_2)$, as in Eq.~(\ref{RSdeff4}).  
Indeed, 
inserting Eq.~(\ref{RSdeff3}) into 
Eq.~(\ref{RSdeff2alt}), we find
\beq
\Bigl\langle\sum_\ell \tau_2^\ell A^{(\ell)}\Bigr\rangle ~=~ \frac{\pi}{3} \lim_{\tau_2\to 0} \sum_\ell
 \tau_2^{k+\ell} 
 \sum_n a_{nn} A^{(\ell)}_{nn} \,e^{-\pi \alpha' M_n^2 \tau_2}~.
 \label{RSdeff2alt2}
\eeq

The issue then boils down to how we evaluate the right side of Eq.~(\ref{RSdeff2alt2}).
Since we are taking the $\tau_2\to 0$ limit, one natural possibility would be to Taylor-expand the exponential.   Taking $k= -1$ (as appropriate for a four-dimensional theory) and recognizing that terms with $r> 1-\ell$ vanish in the $\tau_2\to 0$ limit,
we would then obtain
\beqn
&& \!\!\!\!\! \Bigl\langle\sum_\ell \tau_2^\ell A^{(\ell)}\Bigr\rangle\nonumber\\
&&=~ 
 \frac{\pi}{3} \lim_{\tau_2\to 0} \sum_\ell \sum_{r=0}^{1-\ell}
 \tau_2^{\ell+r-1} \,
 \frac{(-1)^r}{r!}  \sum_n a_{nn} A^{(\ell)}_{nn}\,   (\pi \alpha' M_n^2 )^r~\nonumber\\
&&=~ 
 \frac{\pi}{3} \lim_{\tau_2\to 0} \sum_{\ell\leq 1} \sum_{r=0}^{1-\ell}
 \tau_2^{\ell+r-1} \,
 \frac{(-1)^r}{r!} \, \,\Str\left[
 A^{(\ell)}\,   (\pi \alpha' M^2 )^r\right]
\nonumber\\
\label{RSdeffalt3}
\eeqn
where the supertrace continues to be defined as below Eq.~(\ref{finalsupertraceresult}) and where in passing to the final line we have recognized that the conditions on the $r$-sum have imposed an upper limit on the $\ell$-sum.  However,
for theories in which the original string amplitude $\bigl\langle \sum_\ell \tau_2^\ell A^{(\ell)}\bigr\rangle$ is finite, we know that the right side of Eq.~(\ref{RSdeffalt3}) cannot diverge as $\tau_2\to 0$.
We thus obtain a set of auxiliary conditions which must hold in all such theories, namely
\beq
\Str\left[
 A^{(\ell)}\,   (\pi \alpha' M^2 )^r\right]=0
 \label{auxiliary_conditions}
 \eeq
 for all $0\leq r\leq -\ell$ with $\ell\leq 1$.
As long as these auxiliary conditions are satisfied, we then have
\beq
\Bigl\langle\sum_\ell \tau_2^\ell A^{(\ell)}\Bigr\rangle
~=~ 
 \frac{\pi}{3} \, \sum_{\ell \leq 1}
 \frac{(-1)^{1-\ell}}{(1-\ell)!} \, \Str\left[ A^{(\ell)}   (\pi \alpha' M^2 )^{1-\ell} \right]~,
\label{othermethod}
\eeq
thereby exactly matching the result we obtained in Eq.~(\ref{finalsupertraceresult}). 
We thus see that the Mellin-transformed result has not only reproduced the result in Eq.~(\ref{finalsupertraceresult}), but has also furnished the explicit extra conditions in Eq.~(\ref{auxiliary_conditions})  which must be satisfied in order for this result to be valid.  Indeed, these auxiliary conditions may be viewed as the conditions under which the exchange of 
the order of $n$-summation and $s$-integration/residue extraction are valid.
For example, in the case of the vacuum amplitude $\langle {\bf 1}\rangle$, these auxiliary conditions reduce to the condition
$ \Str\, {\bf 1}=0$.
This condition, which we have already mentioned in Eq.~(\ref{strone}),
is quite remarkable, indicating that even when spacetime supersymmetry is broken in a given string model, the residual misaligned supersymmetry continues to ensure that this supertrace vanishes --- even though such theories do not permit any possible pairing of bosonic and fermionic states.

The results in Eqs.~(\ref{auxiliary_conditions}) and (\ref{othermethod})
hold for the vast majority of string theories as long as we are avoiding the edges of moduli space corresponding to  decompactification limits.   Indeed, we continue to issue this proviso because
we have still made a further critical algebraic assumption in this analysis.  This occurred when we evaluated the right side of Eq.~(\ref{RSdeff2alt2}) by Taylor-expanding the exponential and then passing the $r$-summation past the $n$-summation.   While this may be a valid step in many string theories, in this paper we shall need to consider cases in which the insertions $A^{(\ell)}$ have eigenvalues $(m,n)$-eigenvalues which are growing functions of $m$ and $n$, thereby rendering our sums over states badly divergent.   In such cases, the exponential suppression is critical for a finite result and it is therefore not possible to Taylor-expand this exponential and consider the different terms in the Taylor expansion separately.

As we have explained above, the fundamental difficulty is that the different $\ell$-terms which contribute to $g(\tau_2)$ have different {\it apparent}\/ powers of $\tau_2$, but in reality these powers of $\tau_2$ can be exchanged between these different $\ell$-terms as the result of algebraic manipulations (such as Poisson resummations) that become appropriate in certain limiting regions of moduli space.   For this reason, we should really consider $g(\tau_2)$ as a single object with its own overall $\tau_2$-dependence without attempting to draw a correspondence between this overall $\tau_2$-dependence and the different $\ell$-terms within $g(\tau_2)$.   

We can accomplish this by 
following the approach originally taken in Ref.~\cite{Dienes:1995pm}.  This approach has also been generalized in Ref.~\cite{NewSupertraceIdentities}, and in the remainder of this subsection we shall quickly derive the important results from Refs.~\cite{Dienes:1995pm,NewSupertraceIdentities} that we shall require in the rest of this paper.
In particular, returning to Eq.~(\ref{RSdeff2alt2}), we can begin by isolating
our sum over states with all modular-invariant operator insertions included, \ie,
\beq
S(\tau_2)  ~\equiv~ \sum_\ell \tau_2^\ell \sum_{n} a_{nn} A^{(\ell)}_{nn}\,
      e^{-\pi \alpha' M_n^2 \tau_2}~.
\label{SSdef}
\eeq
However, let us now assume that $S$
has an overall $\tau_2$-dependence as $\tau_2\to0$ given by
\beq
S(\tau_2) ~\sim~ \sum_{j} \, C_j \tau_2^j
\label{assumedtau2dep}
\eeq
where the coefficients $C_j$ are completely arbitrary.   Note that our assumption that this sum is finite as $\tau_2\to 0$ allows us to assume that $j\geq 0$.
Indeed, this sum is finite because we have assumed that our theory is free of physical tachyons;  because the degeneracies  grow no more rapidly than $|a_{nn}|\sim e^{\sqrt{n}}$ according to the Hagedorn phenomenon;  and because the charge-operator eigenvalues $A^{(\ell)}_{nn}$ typically grow no faster than a polynomial in $n$.  By contrast, the exponential suppression goes as $e^{-n}$ since $\alpha' M_n^2 \sim n$.   In this context, we remark that we are {\it not}\/ imposing the condition that $C_j\not=0$ only for integer values of $j$.  There indeed exist examples for which non-integer  values of $j$ contribute within the sum in Eq.~(\ref{assumedtau2dep}).   However, we shall be concerned with the lowest-lying values of $j$, and for these we can take $j\in \mathbbZ$.

Once we assume a $\tau_2$-dependence of
the form in Eq.~(\ref{assumedtau2dep}), we can take the $\tau_2$-derivative of both sides of Eq.~(\ref{assumedtau2dep}) to obtain
\beqn
&& \frac{d}{d\tau_2} \left[ 
 \sum_\ell \tau_2^\ell \sum_{n} a_{nn} A^{(\ell)}_{nn}
      \,e^{-\pi \alpha' M_n^2 \tau_2} \right] \nonumber\\
&& ~~~~~~=~ \sum_n a_{nn} \, 
 \frac{d}{d\tau_2} \left[ \sum_\ell \tau_2^\ell A^{(\ell)}_{nn}
\right]
      \,e^{-\pi \alpha' M_n^2 \tau_2} \nonumber\\
&& ~~~~~~~~~~~~+~
\sum_\ell \tau_2^\ell \sum_{n} a_{nn} A^{(\ell)}_{nn} \,(\pi \alpha' M_n^2)\,
      e^{-\pi \alpha' M_n^2 \tau_2} \nonumber\\
&& ~~~~~~\stackrel{\rm set}{=}~  \frac{d}{d\tau_2} \left(
 \sum_j \, C_j\tau_2^j \right) 
~= ~ 
 \sum_j \,(j+1) \, C_{j+1} \,\tau_2^j~.~\nonumber\\
 \label{twoforty}
 \eeqn
 Indeed, these relations hold for $\tau_2\ll 1$.   Taking the $\tau_2\to 0$ limits of Eqs.~(\ref{assumedtau2dep}) and
(\ref{twoforty})
then yields explicit expressions for the leading coefficients~\cite{Dienes:1995pm,NewSupertraceIdentities}
\beqn
C_0 &=& \lim_{\tau_2\to 0} \left[ \sum_\ell \tau_2^\ell \sum_{n} a_{nn} A^{(\ell)}_{nn} \,
      e^{-\pi \alpha' M_n^2 \tau_2} \right] \nonumber\\
C_1 &=&
\lim_{\tau_2\to 0} \left\lbrace 
\sum_n a_{nn} \frac{d}{d\tau_2}
 \left\lbrack \sum_\ell \tau_2^\ell A^{(\ell)}_{nn} \right\rbrack
      \,e^{-\pi \alpha' M_n^2 \tau_2}\right\rbrace \nonumber\\
&& \,
- \lim_{\tau_2\to 0} \left[ \sum_\ell 
 \tau_2^\ell \sum_{n} a_{nn} A^{(\ell)}_{nn} \,(\pi \alpha' M_n^2)\,
      e^{-\pi \alpha' M_n^2 \tau_2}
      \right] ~.\nonumber\\
\label{coeffsevaluated}
\eeqn
Likewise, the coefficients $C_j$ with $j\geq 2$ can be calculated in a similar fashion by taking additional $\tau_2$-derivatives (and will ultimately be useful only for theories in spacetime dimensions $D>4$).

Substituting Eq.~(\ref{assumedtau2dep}) into Eq.~(\ref{RSdeff2alt2}) and taking $k= -1$ as appropriate for four-dimensional string theories then yields
\beqn
 && \!\!\! \!\!\!\! \Bigl\langle\sum_\ell \tau_2^\ell A^{(\ell)}\Bigr\rangle ~=~ \frac{\pi}{3} \lim_{\tau_2\to 0}  \tau_2^k\, \sum_{j=0}^\infty
   C_j \tau_2^j \nonumber\\
&&=~\frac{\pi}{3}   \, C_1 \nonumber\\
&&=~ \frac{\pi}{3} 
\lim_{\tau_2\to 0} \left\lbrace 
\sum_{n} a_{nn} \,
  \frac{d}{d\tau_2} \left[ \sum_\ell \tau_2^\ell A^{(\ell)}_{nn}
\right]
      \,e^{-\pi \alpha' M_n^2 \tau_2} \right\rbrace\nonumber\\
&& ~~~
-\frac{\pi}{3}  \lim_{\tau_2\to 0} \left[ \sum_\ell 
 \tau_2^\ell \sum_{n} a_{nn} A^{(\ell)}_{nn} \,(\pi \alpha' M_n^2)\,
      e^{-\pi \alpha' M_n^2 \tau_2}
      \right] ~.\nonumber\\
\label{firstline}
\eeqn 
  Critically, we now see that all of our expressions are properly convergent as a result of the exponential damping factors. 
  Indeed, motivated by comparison with our earlier results, we define the {\it regulated}\/ supertrace~\cite{Dienes:1995pm}
\beq
  \Str \, A ~\equiv~   
       \lim_{\tau_2\to 0} \, \sum_n a_{nn} A_{nn} \,e^{- \pi \alpha' M_n^2 \tau_2}~.
\label{supertrace_regulated}
\eeq
Indeed, this is nothing but the supertrace we introduced in Eq.~(\ref{supertracedef2}), with $y$ now identified as $\pi \tau_2$.  Moreover, we now see that Eq.~(\ref{supertrace_regulated}) serves as the correct formal definition of the supertrace previously   
defined below Eq.~(\ref{finalsupertraceresult}).

Written in terms of these supertraces we thus have our general result that expresses our full one-loop string amplitude in terms of supertraces over physical string states: 
\beqn
&& \Bigl\langle\sum_\ell \tau_2^\ell A^{(\ell)}\Bigr\rangle ~=~ 
  \frac{\pi}{3} 
\, \Str \left[
 \frac{d}{d\tau_2} \left( \sum_\ell 
 \tau_2^\ell A^{(\ell)}\right) \right]
 \nonumber\\
 && ~~~~~~~~ -\frac{\pi}{3} 
\, \Str \left[
 \left( \sum_\ell 
 \tau_2^\ell A^{(\ell)}\right) \,(\pi \alpha' M^2) \right] ~.~~~~~~~~~~~~
\eeqn
Equivalently, for any 
modular-invariant operator insertion $\calX$ in four dimensions, we have~\cite{NewSupertraceIdentities}
\beqn
 \Bigl\langle \calX \Bigr\rangle ~&=&~ \frac{\pi}{3} \,
\Str \left(
 \frac{d}{d\tau_2} \calX \right)
 -  \frac{\pi}{3} \,
 \Str \left[ \calX \,(\pi \alpha' M^2) \right] ~\nonumber\\
&=&~ \frac{\pi}{3} \,
\Str \left( D_{\tau_2} \calX \right)~
\label{finalRSresult}
\eeqn
where
\beqn 
D_{\tau_2}~&\equiv&~ \frac{d}{d\tau_2} 
       -  \pi \alpha'M^2 ~\nonumber\\
&=&~ \frac{d}{d\tau_2} 
       -  \frac{1}{4\pi \calM^2} \,M^2~. 
\label{RSderivative}
\eeqn
For modular-invariant insertions $\calX$ that are $\tau_2$-independent (so that we can write $\calX=\mathbbX$ where $\mathbbX$ is itself modular invariant), this result simplifies to
\beq
\Bigl\langle \mathbbX \Bigr\rangle ~=~ - \frac{1}{12\calM^2} \,
 \Str \left( \mathbbX \,M^2 \right)~.
\label{finalRSresult2}
\eeq
Likewise, for operator insertions of the form 
$\calX= \tau_2 \mathbbX_1 + \tau_2^2 \mathbbX_2$, we have
\beqn
&& \Bigl\langle \tau_2 \mathbbX_1 + \tau_2^2\mathbbX_2 \Bigr\rangle ~=~ \frac{\pi}{3} \,
\Str \Bigl(
 \mathbbX_1 + 2 \tau_2 \mathbbX_2 \Bigr)~~~~~~~~~~~~~
 \nonumber\\
 && ~~~~~~~~~~~~~~~~~
 -  \frac{1}{12 \calM^2} \,
 \Str \biggl[ \bigl( \tau_2 \mathbbX_1 + \tau_2^2 \mathbbX_2
 \bigr)\, M^2\biggr]
 %  \Str \left( \calX \, M^2\right)
 ~.~\nonumber\\
\label{finalRSresultagain}
\eeqn
Moreover, under our assumption that the string amplitude is finite, we learn from the first line of Eq.~(\ref{firstline}) that we must have $C_0=0$ as an auxiliary condition.  Given the coefficients in Eq.~(\ref{coeffsevaluated}), this then yields~\cite{NewSupertraceIdentities}
\beq
       \Str \, \calX~=~0~.
\label{singlesupertrace}
\eeq
This auxiliary condition thus accompanies our result in Eq.~(\ref{finalRSresult}) or its simplifications in Eqs.~(\ref{finalRSresult2}) or (\ref{finalRSresultagain}).

Before proceeding further, it is worth emphasizing that 
the quantity in Eq.~(\ref{supertrace_regulated}) is properly viewed as a regulated supertrace only if the inserted operator $A$ is itself $\tau_2$-independent. In such cases, the $\tau_2$-dependent exponential factor in Eq~(\ref{supertrace_regulated}) functions as a true regulator, with $\tau_2$ functioning as a dummy regulator variable whose regulating effects are ultimately removed by taking the $\tau_2\to 0$ limit.  By contrast, in cases in which our inserted operator $A$  has its own $\tau_2$-dependence, we are no longer free to view $\tau_2$ as an independent regulator variable;  rather, the quantity whose limit must be taken in Eq.~(\ref{supertrace_regulated}) becomes inextricably identified with the $\tau_2$ that appears within $A$, so that the $\tau_2\to 0$ limit not only removes the damping exponential but also deforms the operator $A$.
However, in either case, the  operational prescription is clear:   we insert our full operator $A$ within Eq.~(\ref{supertrace_regulated}), along with any $\tau_2$-dependent factors which may appear, and then evaluate the sum and limit accordingly.

The results in Eqs.~(\ref{finalRSresult2}) and (\ref{singlesupertrace})
were originally derived in Ref.~\cite{Dienes:1995pm} for the case $\calX=\mathbbX={\bf 1}$, but we now see~\cite{NewSupertraceIdentities}
that they hold for all modular-invariant $\tau_2$-independent insertions $\mathbbX$ that lead to finite string amplitudes.
Moreover, we now also see that  the result in Eq.~(\ref{finalRSresult2})
is actually only a special case of the more general result in Eq.~(\ref{finalRSresult}).
Indeed, as we have stressed above, our formulation of the Rankin-Selberg transform
in Eq.~(\ref{finalRSresult}) 
is completely general and holds for {\it any}\/ operator insertion $\calX \equiv \sum_\ell \tau_2^\ell A^{(\ell)}$ regardless of the values of $\ell$ involved, so long as the regulated supertrace in Eq.~(\ref{supertrace_regulated}) is used and the inserted operator $\calX$ does not disturb the modular invariance of the amplitude integrand or the finiteness of the resulting amplitude.

Thus far, we have shown that our desired amplitude $\langle \tau_2 \mathbbX_1 + \tau_2^2 \mathbbX_2\rangle$ is given in
Eq.~(\ref{finalRSresultagain}).
However, there is an alternate way in which we might have evaluated this amplitude~\cite{NewSupertraceIdentities}.    Given the sum $S$ in Eq.~(\ref{SSdef}), we immediately recognize that
\beq
      g(\tau_2)~=~ \tau_2^{-1}\, S(\tau_2)~.
\eeq
We therefore directly have
\beqn
&& \Bigl\langle \sum_\ell \tau_2^\ell A^{(\ell)} 
      \Bigr\rangle ~ =~ \frac{\pi}{3}\, \lim_{\tau_2\to 0} \tau_2^{-1} S(\tau_2)
      ~~~~~~\nonumber\\
 && ~~~~~~~~~~~=~  \frac{\pi}{3}\,
 \lim_{\tau_2\to 0} \left\lbrack
     \sum_\ell \tau_2^{\ell-1} \sum_{n} a_{nn} A^{(\ell)}_{nn}\,
      e^{-\pi \alpha' M_n^2 \tau_2} \right\rbrack\nonumber\\
&& ~~~~~~~~~~~=~\frac{\pi}{3}\,\Str \left( \sum_\ell \tau_2^{\ell-1} A^{(\ell)}
       \right)~.
\eeqn
Indeed, for any modular-invariant operator insertion $\calX$, this becomes~\cite{NewSupertraceIdentities}
\beq
\Bigl\langle \calX \Bigr\rangle ~=~ 
  \frac{\pi}{3}\, \Str \left( \tau_2^{-1} \calX\right)~.
\label{alt_RS_result}
\eeq
Comparing this with our result in Eq.~(\ref{finalRSresult}), we thus obtain the remarkable identity~\cite{NewSupertraceIdentities} 
\beq
\Str \left( \tau_2^{-1} \calX\right)~=~ 
\Str \left( D_{\tau_2} \calX \right)
\label{remarkableidentity}
\eeq
where the derivative $D_{\tau_2}$ is given in 
Eq.~(\ref{RSderivative}).
This identity applies to any modular-invariant operator insertion $\calX$ in four dimensions.

Note that this identity 
does {\it not}\/ imply that
$\tau_2^{-1} \calX$ and  $D_{\tau_2} \calX$ are equal. Rather, it implies that both of these quantities have the same {\it supertrace}\/ when this supertrace is evaluated over the entire string spectrum.
For $\tau_2$-independent insertions $\calX=\mathbbX$ this identity takes the simple form~\cite{NewSupertraceIdentities}
\beq
 \Str\left(\tau_2^{-1} \mathbbX\right) ~=~
   - \frac{1}{4\pi \calM^2} \,\Str\left( \mathbbX \,M^2\right)~.
\label{simpleform}
\eeq
Moreover, for the special case in which $\calX=\mathbbX={\bf 1}$, we find
\beq
   \Str \left( \tau_2^{-1} \right) ~=~
     -\frac{1}{4\pi \calM^2} \,\Str\,M^2~.
\label{simpleform2}
\eeq
In conjunction with Eq.~(\ref{Lamresult}), 
this then provides an alternate supertrace expression for the cosmological constant $\Lambda$ in Eq.~(\ref{Lambdanoinsertion}), specifically
\beq
 \Lambda~=~  - \frac{\pi}{6} \,\calM^4 \,
     \Str \left(\tau_2^{-1} \right)~.
\eeq
Finally, for 
$\calX= \tau_2 \mathbbX_1 + \tau_2^2 \mathbbX_2$, the identity in Eq.~(\ref{remarkableidentity}) takes the form
\beq
\Str \Bigl(
  \tau_2 \mathbbX_2 \Bigr) ~=~
  \frac{1}{4\pi\calM^2} \,
 \Str \biggl[ \bigl( \tau_2 \mathbbX_1 + \tau_2^2 \mathbbX_2
 \bigr)\, M^2\biggr] ~.~
\label{tradetau2forM}
\eeq
This then allows us to express the amplitude in Eq.~(\ref{finalRSresultagain}) in the simpler form
\beq
    \Bigl\langle \tau_2 \mathbbX_1 + \tau_2^2\mathbbX_2 \Bigr\rangle ~=~
\frac{\pi}{3}\,\Str \left( \mathbbX_1 + \tau_2 \mathbbX_2\right)~,
\label{simpleamplitudeform}
\eeq
in agreement with Eq.~(\ref{alt_RS_result}).

The identity in Eq.~(\ref{remarkableidentity}) is rather astonishing, leading to results such as those in Eqs.~(\ref{simpleform}), (\ref{simpleform2}), and (\ref{tradetau2forM}) in which a change in the power of  $\tau_2$ 
within a supertrace can be traded for the insertion of an additional factor of the squared {\it mass}\/.    
Moreover, within this context, the condition in Eq.~(\ref{singlesupertrace}) helps us to interpret results such as those in  Eq.~(\ref{simpleform}).  Recall that the definition of the supertrace in Eq.~(\ref{supertrace_regulated}) involves taking the $\tau_2\to 0$ limit.  Thus, when we insert a factor of $\tau_2^{-1}$ into a supertrace --- such as on the left side of Eq.~(\ref{simpleform}) --- or when we equivalently insert a factor of $M^2$ into the supertrace --- such as on the right side of Eq.~(\ref{simpleform}) ---
it would {\it a priori}\/ seem that we are pushing this supertrace towards a divergence, especially since our supertrace involves summing over an infinite tower of states with ever-increasing masses.
 However, these relations come with the additional constraint in Eq.~(\ref{singlesupertrace}) which tells us that the supertrace {\it without}\/ these factors of $\tau_2^{-1}$ or $M^2$ actually vanishes.   These extra factors thus ``lift'' the value of this supertrace away from zero and thereby allow it to have the non-zero result which matches the value of the corresponding amplitude.

Because of its central importance,
it is critical that we understand the implications of Eq.~(\ref{singlesupertrace}).   As we have shown, this relation holds  {\it regardless}\/ of whether $\calX$ contains leading $\tau_2$ factors.
If $\calX$ does not contain any explicit leading $\tau_2$ factors of its own ({\it e.g.}\/, $\calX=\mathbbX= {\bf 1}$), then this constraint is at its most powerful, requiring that
\beq
\sum_n \,a_{nn} \,\mathbbX_{nn} \,e^{-\pi \alpha' M_n^2 \tau_2} ~\to~ 0 ~~~~{\rm as}~~\tau_2\to 0~.
\label{upperboundone}
\eeq
More precisely, in four uncompactified dimensions, we have from Eq.~(\ref{firstline}) that
\beq
\left| \sum_n \,a_{nn}\, \mathbbX_{nn} \,e^{-\pi \alpha' M_n^2 \tau_2}\right| ~\lsim~ \tau_2 ~~~~{\rm as}~~\tau_2\to 0~.
\label{upperboundonep}
\eeq
However, even if $\calX$ has leading $\tau_2$ factors, the constraint in Eq.~(\ref{singlesupertrace})
has teeth.   As an example, let us suppose that $\calX$ has a leading power $\tau_2^\ell$ with some value $\ell>0$, so that we can write $\calX=\tau_2^\ell \mathbbX$.
In such cases the constraints in Eqs.~(\ref{singlesupertrace}) and (\ref{upperboundonep}) 
tell us that 
\beq
\left| \sum_n \,a_{nn} \,\mathbbX_{nn} \,e^{-\pi \alpha' M_n^2 \tau_2}\right| ~\lsim~
    \tau_2^{1-\ell} ~~~~{\rm as}~~\tau_2\to 0~.~~~
\label{upperbound}
\eeq
This constraint continues to provide a significant bound on the spectral sum on the left side of this equation:  as $\tau_2\to 0$, we find that this spectral sum can grow 
no more rapidly than a {\it power}\/ of $1/\tau_2$, with the power given by $\ell-1$.  This constraint is obviously strongest for small values of $\ell$, but nevertheless rules out all exponential growth as $\tau_2\to 0$.   This is a significant exclusion, since exponential growth would be the na\"\i ve expectation in string theory given that the numbers of bosonic and fermionic states in string theory generically grow exponentially with mass.

To illustrate this phenomenon numerically, let us examine a spectrum with alternating, exponentially growing boson/fermion surpluses, as predicted by misaligned supersymmetry, where the growth rates scale as $\sim e^{\sqrt{n}}$ (as required by conformal invariance and the Hagedorn transition).  
For numerical simplicity, we shall model $a_{nn}$ (the net number of bosons minus the number of fermions at a given level $n$) as having the  functional form
\beq
a_{nn} ~=~  (-1)^n e^{\sqrt{n}}~
\label{simplespectrum}
\eeq
where $n$ schematically represents the string level and thus can be associated with the eigenvalue of the corresponding mass $\alpha' M^2$ or  charge $Q^2$.   Here the factor $(-1)^n$ indicates that the even levels are presumed to have surpluses of bosons relative to fermions while the odd levels have surpluses of fermions relative to bosons.
Of course, such a functional form cannot correspond to any actual modular-invariant string theory --- for example, the $a_{nn}$ values in this little exercise are not even integers --- and far more sophisticated functional forms of this general type emerge in actual string models~\cite{Dienes:1994np,Dienes:1995pm, Dienes:2001se}.   However, this simple functional form does capture the essential consequence of misaligned supersymmetry, namely that we have alternating bosonic and fermionic surpluses for which no boson/fermion pairings are possible anywhere in the infinite string spectrum, with the degeneracies $a_{nn}$ lying along  equal but opposite bosonic and fermionic ``envelope functions'' $\sim e^{\sqrt{n}}$~\cite{Dienes:1994np,Dienes:1995pm, Dienes:2001se}.    We shall also imagine that that our insertion $\mathbbX$ has eigenvalues $\mathbbX_{nn}\sim n^\beta$ for some exponent $\beta$, and 
consider the spectral sums
\beq
   f_\beta(\tau_2)~\equiv~ \sum_{n=0}^\infty \,a_{nn}\, n^{\beta} \,e^{-n \tau_2}~.
\label{lookslikesupertrace}
\eeq
For example, the different values of $\beta$ might correspond to different powers of mass/charge insertions, with an insertion of $n^\beta$ corresponding to an insertion of $2\beta$ powers of mass $M$ or charge $Q$.  Indeed, in such cases the fully modular-invariant insertion
would also include an overall factor $\tau_2^\beta$, and thus 
the different values of $\beta$ correspond to different values of $\ell$ in Eq.~(\ref{upperbound}).

In general, we know that for all $\beta>0$ we must have $f_\beta(\tau_2)\to 0$ as $\tau_2\to \infty$, since in this limit only the contributions from the massless ($n=0$) states survive, and these vanish for all $\beta>0$.   However, as we dial $\tau_2$ to smaller values, there will be less and less suppression of the contributions from the heavier states.   Our functions $f_\beta(\tau_2)$ therefore become increasingly sensitive to the exponentially growing  oscillations that exist throughout the massive levels with $n>0$.   Thus, as $\tau_2\to 0$, we expect that $f_\beta(\tau_2)$ will diverge exponentially while simultaneously experiencing rapid oscillations which prevent the extraction of any smooth $\tau_2\to 0$ limit.

 %==========================
\begin{figure}[th]
\centering
\includegraphics[keepaspectratio, width=0.48
\textwidth]{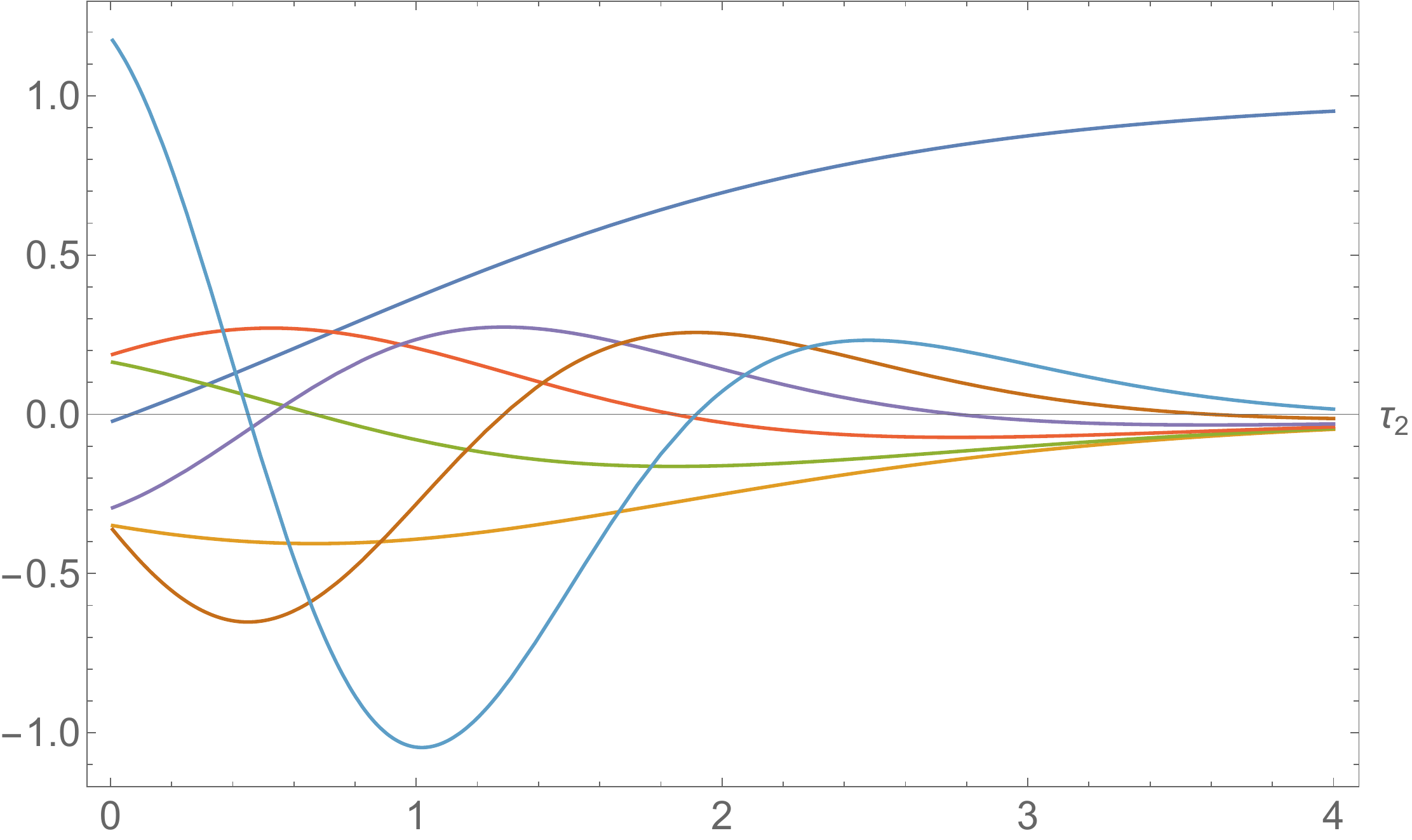}
\caption{
The infinite sums $f_\beta(\tau_2)$ in Eq.~(\ref{lookslikesupertrace}), plotted as functions of $\tau_2$ for $0\leq \beta\leq 6$.  The $\beta=0$ curve is the upper blue curve which asymptotes to $1$ as $\tau_2\to \infty$, while the $\beta=1,2,...,6$ curves all asymptote to $0$ as $\tau_2\to \infty$ and can be identified according to the  increasing values that they have at any fixed large value of $\tau_2$ (with $n=1$ orange, $n=2$ green, $n=3$ red, {\it etc.}\/).  As $\tau_2$ drops to zero from a large value, we see that our functions $f_\beta(\tau_2)$ do not diverge (as would have been expected as we slowly remove the $\tau_2$ cutoff), but instead remain within the bounds indicated in Eq.~(\ref{upperbound}), leading to finite values for $f_\beta(\tau_2)$ as $\tau_2\to 0$.  
}
\label{fig:sums}
\end{figure}
%==========================

In Fig.~\ref{fig:sums} we plot $f_\beta(\tau_2)$ as a function of $\tau_2$ for $0\leq \beta \leq 6$.   As expected, we see that $f_\beta(\tau_2)\to 0$ as $\tau_2\to\infty$ for all $\beta>0$, as discussed above.
However, as $\tau_2$ becomes smaller, 
we find that for each $\beta$ our function $|f_\beta(\tau_2)|$  {\it does not diverge exponentially as $\tau_2\to 0$, but instead remains within the bounds indicated in Eq.~(\ref{upperbound}).}\/  Indeed, this happens even without the ability to realize any boson/fermion pairings within the associated spectrum.
 Moreover, for the simple functional form in Eq.~(\ref{simplespectrum}), we even find that $f_\beta(\tau_2)$ approaches a finite value as $\tau_2\to 0$. Thus, we see that the spectrum in Eq.~(\ref{simplespectrum}) already does a good job of satisfying our supertrace constraints, and even has a $\tau_2\to 0$ limit which comes close to vanishing in the $\beta=0$ case.

 Once again, we stress that the simple spectrum in Eq.~(\ref{simplespectrum}) is only a caricature of an actual fully modular-invariant string spectrum.  This exercise nevertheless illustrates how even the constraint in Eq.~(\ref{upperbound}) has teeth.    

As a final remark, we note that not every oscillating functional form for $a_{nn}$ will exhibit this behavior.   Indeed, the functional form in Eq.~(\ref{simplespectrum}) is particularly ``stringy'':  rather than relying on boson/fermion pairings at any mass level, the controlled behavior as $\tau_2\to 0$  occurs as the result of tight constraints that involve the numbers of states across the entire (infinite) string spectrum.   From this perspective the critical aspect of the spectrum in Eq.~(\ref{simplespectrum}) is that the bosonic and fermionic states share the same exponentially growing degeneracy profile function $e^{\sqrt{n}}$ while nevertheless sampling this function at ``misaligned'' values (in this case, with even $n$ for bosons and odd $n$ for bosons).   This is the underpinning of misaligned supersymmetry, as discussed in Ref.~\cite{Dienes:1994np}.

To see that this is the critical feature,  let us imagine a more ``field-theoretic'' spectrum in which the bosonic states continue to have degeneracies $a_{nn}\sim e^{\sqrt{n}}$  but in which their fermionic would-be superpartners have these same degeneracies but with masses lifted by some small supersymmetry-breaking scale $\delta n$:
\beqn
     f_{\beta}^{\rm (B)}(\tau_2) ~&\equiv&~
      \sum_n  a_{nn}\, e^{-n \tau_2} \nonumber\\
     f_{\beta}^{\rm (F)}(\tau_2) ~&\equiv&~
      \sum_n a_{nn}\, e^{-(n+\delta n) \tau_2}~.
\label{separate}
\eeqn
Here the two distributions in Eq.~(\ref{separate}) indicate bosonic and fermionic states respectively.   Of course, this is a very natural spectrum from a field-theoretic perspective, exhibiting a clear boson/fermion pairing.
However, such cases lack modular invariance, and indeed we find that the corresponding $f_\beta(\tau_2)
\equiv
f_\beta^{\rm (B)}(\tau_2)-
f_\beta^{\rm (F)}(\tau_2)$
functions not only grow exponentially as $\tau_2\to 0$, thereby violating the constraints in Eq.~(\ref{upperbound}), but also exhibit increasingly violent oscillations that preclude numerical extraction of any smooth limiting values as $\tau_2\to 0$.  Similar problematic results arise for other paired bosonic/fermionic splitting patterns as well. 

    Thus, to summarize the results of this section, we find that any four-dimensional string amplitude $\langle \calX \rangle$ with a modular-invariant insertion $\calX$ can be written as~\cite{NewSupertraceIdentities}
\beq
\Bigl\langle \calX \Bigr\rangle~=~
\frac{\pi}{3} \,\Str \left(D_{\tau_2} \calX\right) ~=~
\frac{\pi}{3} \, \Str \left(\tau_2^{-1} \calX\right)~
\label{summaryeq}
\eeq
where $D_{\tau_2}$ is defined in Eq.~(\ref{RSderivative}).
In conjunction with Eq.~(\ref{summaryeq}), we also have the constraint in Eq.~(\ref{singlesupertrace}) which tightly constrains the spectrum of states and renders the quantities in Eq.~(\ref{summaryeq}) finite.
Given the discussion above, we see that supertrace expressions such as these are meaningful and convergent precisely because we are working within a string-theoretic context wherein the corresponding spectra are governed by modular invariance and misaligned supersymmetry, even though bosonic/fermionic pairings are no longer possible.
    Indeed, it is in this way that string theory maintains its finiteness --- even without spacetime supersymmetry, and even in the face of not only  exponentially growing towers of states but also exponentially growing {\it net}\/ (bosonic minus fermionic) numbers of states~\cite{Dienes:1994np,Dienes:1995pm, Dienes:2001se}.

%================================================
\subsection{Entwined amplitudes and entwined supertraces \label{sec:entwined}}

%%%%%%%%%%%%%%%%%%%%%%%%%%%%%%%%%%%%%%%%%%%%%%%

In the previous subsection we derived the general results in Eqs.~(\ref{finalRSresult}) and (\ref{singlesupertrace}) 
for the Rankin-Selberg transform.  
At first glance, our next task would then appear to be to apply these results for the specific operator insertions 
that arise in our calculation of the gauge couplings.
In particular, as evident from Eq.~(\ref{eq:Xs}), there are four specific operator insertions that 
will be required in our gauge-coupling calculation: 
\beq
\overline{Q}_H^2~,~~~
 \overline{Q}_H^2 Q_G^2~,~~~
  \overline{E}_2~,~~{\rm and}~~
   \overline{E}_2 Q_G^2~.
\label{eq:operatorproducts}
\eeq
The first and third of these come from the insertion of $\mathbbX_1$, while the second and fourth come from the insertion of $\mathbbX_2$.

Unfortunately, we cannot yet apply the Rankin-Selberg formalism to all of these operator insertions.   
The insertions $\overline{Q}_H^2$ and $\overline{Q}_H^2 Q_G^2$ behave as assumed above for $A^{(\ell)}$, depositing their corresponding $(m,n)$ eigenvalues  $A^{(\ell)}_{mn}$  
 within the partition-function trace as in Eq.~(\ref{RSdeff1}).
 The same is also true of the factor $Q_G^2$ within the fourth operator in Eq.~(\ref{eq:operatorproducts}).
However the ``operator''
$\overline{E}_2$ is actually a {\it function}\/ of  the complex parameter $\tau$ and thus has its own double power-series expansion in $(q,\qbar)$.  
Moreover, because we are to insert the same $\overline{E}_2$ function for each state within the string spectrum, insertion of this $\overline{E}_2$ function into any partition-function trace yields nothing but a {\it product}\/ of the original partition-function trace and the $\overline{E}_2$ function.
As a result, if $Z_X$ denotes the partition-function trace with an operator $X$ inserted
(\ie, if
 $Z_X \sim \sum_{m,n} a_{mn} X_{mn} \qbar^m q^n$), then
 $Z_{\overline E_2}= Z \cdot \overline E_2$
 and 
 $Z_{ X\overline E_2 } =
 Z_{X} \cdot \overline E_2$.
 The integrand of the  (unregulated) amplitude for our gauge-coupling calculation thus generally takes the form
 \beq
     Z_{\calA+ \calB \overline E_2} ~=~ 
        Z_\calA + Z_\calB \cdot\overline E_2
\label{genform}
\eeq
where ${\cal A}= \sum_\ell \tau_2^\ell \mathbbA^{(\ell)}$ and 
${\cal B}= \sum_\ell \tau_2^\ell \mathbbB^{(\ell)}$.
Here $\mathbbA^{(\ell)}$ and $\mathbbB^{(\ell)}$ are the analogues of $\mathbbX_1$ and $\mathbbX_2$.
Indeed, from Eq.~(\ref{eq:Xs}) we have
\beqn
 && \mathbbA^{(1)} = \frac{\xi}{2\pi} \,\overline Q_H^2~,~~~~
 \mathbbA^{(2)} = -2 \,\overline Q_H^2 Q_G^2~,~\nonumber\\
 &&
 \mathbbB^{(1)} = - \frac{\xi}{24 \pi}~,~~~~~\,
 \mathbbB^{(2)} = \frac{1}{6} \, Q_G^2~.
 \label{ABs}
\eeqn
Note that $\mathbbB^{(1)}$ is just a constant.
Given the form in Eq.~(\ref{genform}), we thus expect the results of our calculations to involve not only the supertraces emerging from $Z_\calA$, as discussed in the previous subsection, but also the supertraces emerging from the {\it product}\/ $Z_\calB \overline E_2$ in Eq.~(\ref{genform}). Unfortunately, as we shall now see, this product structure renders the extraction of the corresponding supertraces more  complicated than before.

To see the implications of this product structure, let us begin by considering a
completely general product of the form $Z_1\cdot Z_2$, where the factors $Z_i$ are arbitrary modular-invariant functions which each have their own $(q,\qbar)$ power-series expansions of the forms
\beqn 
    Z_1 ~&\sim&~  \sum_{m,n} \, b_{mn} \,\qbar^m q^n \nonumber\\
    Z_2 ~&\sim&~  \sum_{r,s} \, c_{rs} \,\qbar^r q^s~.
\label{eq:twoexpansions}
\eeqn
In order to apply the Rankin-Selberg procedure as in the previous subsection, we must first recast the product of these two power-series expansions into the form of a single power-series expansion, as in  Eq.~(\ref{RSdeff1}).  We then wish to consider the corresponding $g$-function, as in Eq.~(\ref{RSdeff3}).

Given Eq.~(\ref{eq:twoexpansions}), we immediately see that
\beq
  Z_1 \cdot Z_2 ~\sim~ \sum_{m,n}\sum_{r,s}\, 
             b_{m,n} \,c_{r,s} \,\qbar^{m+r} q^{n+s}~.
\label{gpreprod}
\eeq
It then follows that
\beq
g_{Z_1 \cdot Z_2} ~\sim~ 
\sum_p \, d_{pp}\, (\qbar q)^p
\label{gprod}
\eeq
where the ``on-shell'' degeneracies $d_{pp}$ associated with the product are given by the discrete convolution
\beq 
        d_{pp} ~\equiv~\sum_{m,n} b_{mn} \,c_{p-m,p-n}~.
\label{dpdef}
\eeq 
This truncation to a single summation in Eq.~(\ref{gprod}) occurs because the $\tau_1$-integration within the definition of $g_{Z_1\cdot Z_2}(\tau_2)$ selects only those terms in Eq.~(\ref{gpreprod}) for which $m+r=n+s= p$.   Thus the new ``degeneracies'' $d_{pp}$ within $g_{Z_1\cdot Z_2}$ depend on the degeneracies of both the level-matched  {\it and}\/ non-level-matched states within the original $Z_1$ and $Z_2$ factors.   In other words, the two factors $Z_1$ and $Z_2$ are now {\it entwined}\/ within the product $Z_1\cdot Z_2$.   
Nevertheless, as evident upon comparing Eq.~(\ref{RSdeff3}) and
(\ref{gprod}), 
the quantities $\lbrace d_{pp}\rbrace$ now play the same role for a product of two modular functions as previously played by the degeneracies $\lbrace a_{nn} A_{nn}\rbrace$ in the case of a single modular function. 

If $Z_2$ is purely anti-holomorphic, we can set $s=0$ above.
We then find that $p=n$, whereupon
\beq 
        d_{pp} 
       ~\equiv~
       \sum_{r=0}^{r_{\rm max}} b_{p-r,p}\, c_{r0}~.
\label{entwin}
\eeq
In general, right-moving worldsheet energies in string theory are bounded from below by the right-moving worldsheet vacuum energy $\Delta$.   For example, we have $\Delta= -1$ for the bosonic string and $\Delta = -1/2$ for the superstring and heterotic string.   In all cases, we must therefore have $p-r\geq \Delta$.  For any value of $p$, as in Eq.~(\ref{entwin}), this therefore imposes an upper bound 
\beq
 r_{\rm max} ~\equiv~ \left\lfloor p-\Delta \right\rfloor
\label{rmaxdef}
\eeq
where $\lfloor x\rfloor$ denotes the greatest integer $\leq x$.
 
 For the special case with $Z_2= \overline{E}_2$, we find $c_{r0} =
\chi_r$ in Eq.~(\ref{entwin}), where $\chi_r$ is defined in Eq.~(\ref{cr0}).
This then yields
\beqn
    d_{pp}~&=&~ b_{pp} - 24 
    \sum_{r=1}^{r_{\rm max}}
    \sigma(r) \,b_{p-r,p}\nonumber\\
    &=&~ \sum_{r=0}^{r_{\rm max}} \,\chi_r \, b_{p-r,p}~.
\label{ddefs}
\eeqn
 We thus see that we obtain the expected result $d_{pp} = b_{pp}$ (for $r=0$) along with a ``correction'' term (for $r\geq 1$) which reflects the entwining of the theories and which is induced by the modular completion. 

There is interesting physics in this entwinement.  For the lowest-lying levels with $p=0$, $p=1$, and so forth, we have
\beqn 
   d_{00} ~&=&~ b_{00}\nonumber\\
   d_{11} ~&=&~ b_{11} -24
    \sigma(1) b_{01} \nonumber\\
    d_{22} ~&=&~ b_{22} -24 \sigma(1) b_{12} - 24\sigma(2) b_{02} \nonumber\\
    d_{33} ~&=&~ b_{33} - 24 \sigma(1) b_{23} -
    24 \sigma(2) b_{13} - 24 \sigma(3) b_{03}~.\nonumber\\
\label{columns}
\eeqn
Of course, if $Z_1$ represents the string partition function itself without any insertions, we then have $b_{mn}=a_{mn}$ where $a_{mn}$ tallies the number of bosonic minus fermionic string states with worldsheet energies $(m,n)$, as in Eq.~(\ref{partfunctbare}).
Otherwise, if $Z_1$ represents the string partition function with an insertion whose $(m,n)$ eigenvalues are given by $A_{mn}$, then 
$b_{mn}=a_{mn}A_{mn}$.   In either case, we see from Eq.~(\ref{columns}) that off-shell (non-level-matched) {\it purely stringy}\/ states are entering into the entwinement.   Moreover, we should also remember that these are not the only states in the theory.  For example, for $p=1/2$, $p=3/2$, {\it etc.}\/, we also have
\beqn 
  d_{\half,\half} ~&=&~ b_{\half,\half} - 24 \sigma(1) b_{-\half,\half}  \nonumber\\
  d_{\threehalf,\threehalf} ~&=&~ b_{\threehalf,\threehalf} - 24 \sigma(1) 
       b_{\half,\threehalf} - 24 \sigma(2) b_{-\half,\threehalf}~.\nonumber\\
\label{dvalues}
\eeqn
Thus even off-shell states with right-moving {\it tachyonic}\/ mass contributions (\ie, with $\alpha' M_R^2<0$) are now entering into the entwinement, even though our theory is presumed to lack physical on-shell tachyons!   (Indeed, if this had been the holomorphic $E_2$-function rather than the anti-holomorphic $\overline{E}_2$-function, even the proto-graviton~\cite{Dienes:1990ij} would have  entered into the entwinement.)  Of course, this analysis pertains to  heterotic strings.   For Type~II theories, by contrast, the proto-graviton states will enter the entwinement even for $\overline E_2$, given that the Type~II string has a holomorphic/anti-holomorphic (or worldsheet left-moving/right-moving) symmetry.  

We also note that in all cases we must always have $p\geq 0$.    This restriction arises for two reasons: 
 because our original theory is presumed tachyon-free (with $b_{pp}=0$ for all $p<0$);  and because $d_{pp}$ with $p<0$ cannot emerge via entwinement (\ie, from some state $b_{p-r,p}$ with $r\geq 1$) because right-moving worldsheet energies in string theory are never smaller than $-1$, even for the bosonic string.   The values of $p$ within the $p$-sum are otherwise unconstrained, and depend on the spectrum of string modes, Kaluza-Klein modes, and winding modes of the particular string under study.   

Given this understanding of the nature of the entwinement induced by the presence of the $\overline E_2$ factor, we can now proceed to derive the general Rankin-Selberg transformation of the amplitude whose integrand is given in Eq.~(\ref{genform}).  Following from Eq.~(\ref{RSdeff2alt}), we have
\beq
\left\langle 
\calA + \calB \overline E_2 \right\rangle ~=~
\frac{\pi}{3}  \,\lim_{\tau_2\to 0} \, g_{\calA+\calB\overline E_2}(\tau_2)
\label{RSgeng}
\eeq
where 
\beq
g_{\calA+\calB \overline E_2}(\tau_2) ~=~
g_\calA(\tau_2) + g_{\calB \overline E_2}(\tau_2)~.
\label{contribs}
\eeq
As before, the first contribution is given by
\beq
   g_\calA(\tau_2) ~=~ 
   \sum_\ell
    \tau_2^{\ell-1} \sum_p  a_{pp} \, \mathbbA^{(\ell)}_{pp} \, 
    e^{-4 \pi p \tau_2}
\label{expon}
\eeq
where in the exponential we identify $p$ with $\alpha M^2/4$, as usual, with
$M$ denoting the total mass of the given string state within the $p$-sum.   
By contrast, for the second contribution in Eq.~(\ref{contribs}), we see from Eqs.~(\ref{gprod}) and (\ref{entwin}) that
\beq
   g_{\calB\overline E_2} (\tau_2) ~=~
  \sum_\ell \tau_2^{\ell-1} 
  \left[ \sum_p 
  \sum_{r=0}^{r_{\rm max}}
        \chi_r \, a_{p-r,p} \, \mathbbB^{(\ell)}_{p-r,p} \right]  e^{-4\pi p\tau_2}
\label{firstexp}
\eeq
where $\chi_r$ is given in Eq.~(\ref{cr0}) and where
the quantity in square brackets 
is nothing but $d_{pp}$ in Eq.~(\ref{gprod}).
Indeed, the double $(p,r)$-sum in Eq.~(\ref{firstexp}) is essentially a sum over all of the string states with right- and left-moving worldsheet energies $(m,n)$ with $m\leq n$, even those with $m\not=n$.   Of course, all string states have $m-n\in\mathbbZ$ as a result of the invariance of the string partition function under $\tau\to \tau+1$.

The expression in Eq.~(\ref{firstexp}) 
is properly convergent:  the $r$-summation has only a finite range $0\leq r\leq r_{\rm max}$ for any fixed value of $p$, while the subsequent $p$-sum --- although infinite --- is kept convergent by the $p$-suppressed exponential.  However, it is possible to rearrange the terms of this double $(p,r)$ sum in a manner which continues to include all $(m,n)$ string states with $m\leq n$ but which renders this convergence more explicit and is closer to the fundamental string symmetries.

\begin{figure}[t]
\centering
\includegraphics[keepaspectratio, width=0.49\textwidth]{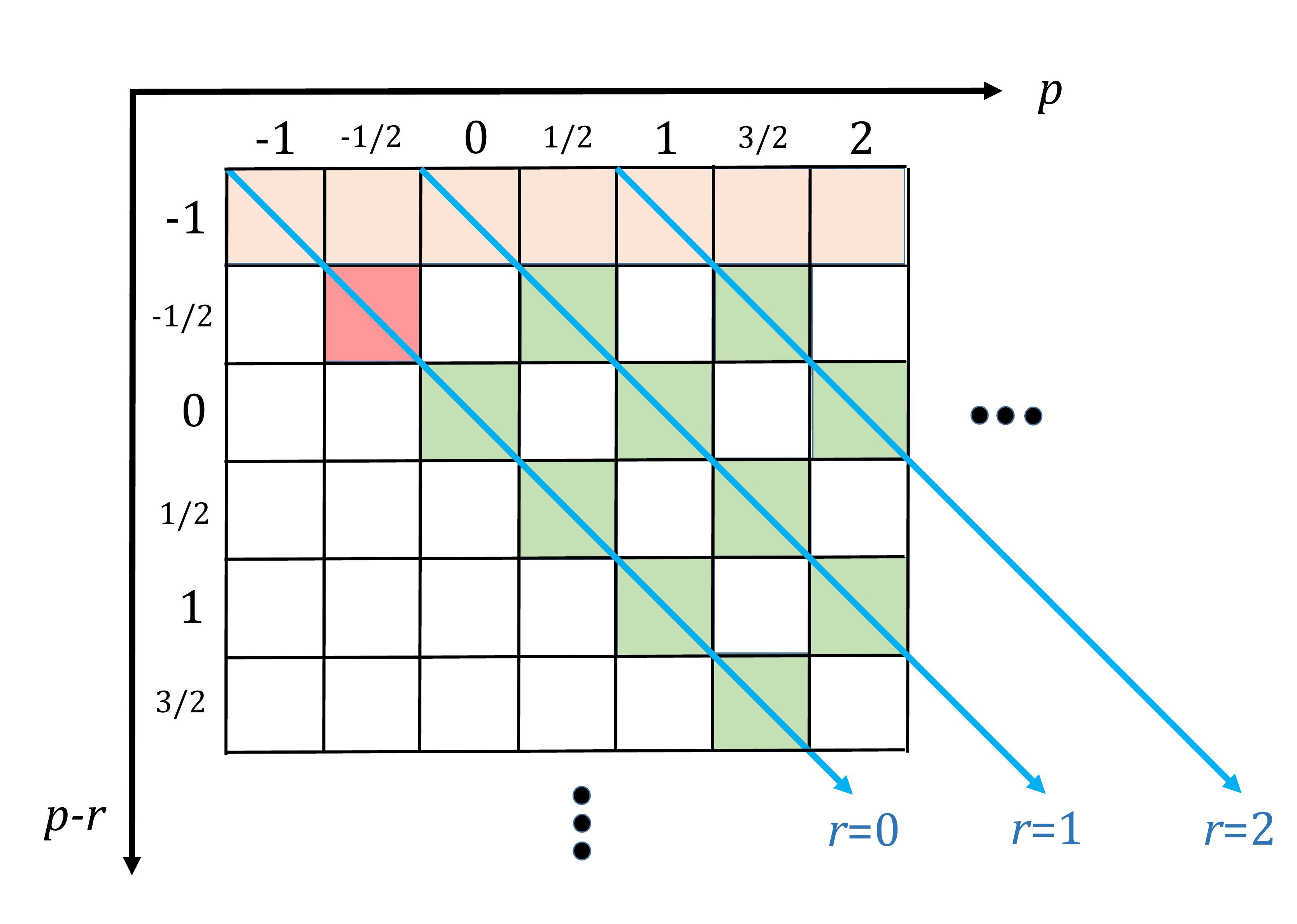}
\caption{String states arranged as a matrix according to their right-moving (vertical) and left-moving (horizontal) worldsheet energies $(m,n)\equiv (p-r,p)$, respectively. The  states with $r=0$ lie along the principal diagonal, while the states with $r=1,2,...$ lie along successive shifted diagonals.  The requirement $r\geq 0$ selects only those string states along or above the principal diagonal, and the $\tau\to \tau+1$ invariance of the partition function ensures that only those ``squares'' shaded in green with $m-n\in\mathbbZ$ can be populated. The pink square is necessarily empty in any tachyon-free theory, and the row shaded in orange is excluded for heterotic strings because such strings have right-moving worldsheet energies $\geq -1/2$.  In drawing this figure we have assumed that states populate only integer or half-integer mass levels, but in general the spectrum of states can be far denser and may even approach a continuum in $(m,n)$ [or equivalently in $p$] for exceedingly large or small compactification radii.
}
\label{fig:matrix}
\end{figure}

To see this, let us imagine the different $(m,n)\equiv (p-r,p)$ string states as populating a matrix, with $m\equiv p-r$ and $n\equiv p$ indicating the corresponding row and column, respectively. Such an arrangement is illustrated in Fig.~\ref{fig:matrix}.  Indeed, we see that for any fixed $p$ there is a maximum corresponding value of $r$, as already noted in Eq.~(\ref{rmaxdef}).  The $(p,r)$ double sum as written in Eq.~(\ref{firstexp}) then corresponds to tallying our states vertical column by vertical column within the upper triangular $m\leq n$ portion of the matrix ({\it i.e.}\/, the portion on or above the diagonal).  However, as evident from Fig.~\ref{fig:matrix}, there is another way in which we might sum these states.
First, we can sum the states which lie along the diagonal (which we will now call the {\it principal diagonal}\/) --- these are the physical string states, all of which have $r=0$.
 Next, we can sum  the states along the ``first shifted diagonal'' which is one column/row displaced (or shifted) from the principal diagonal.   These entries tally the contributions from the unphysical string states whose left- and right-moving worldsheet energies differ by one unit, {\it i.e.}\/, states with $r=1$. Next, we can sum over the ``second shifted diagonal'' which is two units removed from the principal diagonal, and so forth.  In this way, we can equivalently reach all of our $(m,n)$ states with $m\leq n$ and $m-n\in\mathbbZ$.   However, organizing our states according to the diagonals on which they lie is tantamount to organizing our states according to their $L_0-\overline{L}_0$ eigenvalues, where $L_0$ is the zero-mode Virasoro operator.  
 Indeed, this method of summing along diagonals  
 is even suggested by the $p$ and $r$ variables themselves, since $r$ essentially specifies the  diagonal on which a given state lies while $p$ then indicates the location along this diagonal.  
 
 Performing our summation along successive diagonals rather than column-by-column is tantamount to replacing
\beq
 \sum_{p}^\infty ~ \sum_{r=0}^{r_{\rm max}}
 ~\longrightarrow~ 
 \sum_{r=0}^\infty~ \sum_{p}^\infty
 \label{shufflesums} 
 \eeq
 within Eq.~(\ref{firstexp}),
 where as always the $p$-sum is an infinite one which includes all of the values relevant for the particular string model in question.
 These values include not only integers [for states such as those in Eq.~(\ref{columns})] but also half-integers [for states such as those in Eq.~(\ref{dvalues})], {\it etc.}\/   
Of course, while the $p$-sum on the left side of Eq.~(\ref{shufflesums}) nominally begins at $p=0$, the $p$-sum on the right side of Eq.~(\ref{shufflesums}) begins at higher values of $p$ so that $p-r\geq \Delta$.
Implementing Eq.~(\ref{shufflesums}) within Eq.~(\ref{firstexp})
is thus equivalent to reshuffling the contributions from the different terms within Eqs.~(\ref{columns}) 
 and (\ref{dvalues}) so that we sum along the vertical columns (rather than horizontal rows) within these equations.
 We therefore obtain
\beq
   g_{\calB\overline E_2} (\tau_2) ~=~
  \sum_\ell \tau_2^{\ell-1} 
  \sum_{r=0}^\infty \sum_p \chi_r\,
        a_{p-r,p} \, \mathbbB^{(\ell)}_{p-r,p} \,
       e^{-4\pi p\tau_2}\, .
\label{secondexp}
\eeq

Our next step is to understand the exponential suppression factor $e^{-4\pi p\tau_2}$ that appears in Eq.~(\ref{secondexp}).  In Eq.~(\ref{expon}), a similar exponential factor appeared, and we identified $p$ in the exponential with the total squared mass $\alpha' M^2/4$
of the corresponding string state, where
$M^2 \equiv (M_L^2+M_R^2)/2$.
However, this identification is no longer appropriate for Eq.~(\ref{secondexp}).  Because of the entwinement that appears in Eq.~(\ref{entwin}), we see that
our exponential factor $(\qbar q)^p$ actually represents the product
\beq
  (\qbar q)^p ~=~ 
   \left( \qbar^{p-r} q^p\right) \cdot 
    \qbar^r
\eeq
where only the parenthesized first factor encapsulates the mass contributions  from the states in the original theory.  By contrast, the second factor emerges purely from the entwinement function.  Indeed, for $r\not=0$ this is precisely why our results are sensitive to the off-shell string states.
However, this means that our identification of $p$ and $r$ with the masses of our underlying string states must now take the form
\beq
 \alpha' M_R^2 = 4 (p-r) ~,~~~~~
 \alpha' M_L^2 = 4 p~,
\label{massids}
\eeq
implying that $r = \alpha' (M_L^2 - M_R^2 )/4 \equiv \alpha' (\Delta M^2)/4$.   This is consistent with our identification of $r$ as indicating how far from the principal diagonal a given state lies.   However, we now see directly from Eq.~(\ref{massids})  that $p$ represents only the {\it left-moving}\/ contribution $\alpha' M_L^2/4$ to the total mass of the state.   Thus our exponential suppression factor is given by $e^{-\pi \alpha' M_L^2 \tau_2}$.

This result may seem strange, especially given that the exponential damping factor within the $g$-function has the level-matched form $(\qbar q)^p$, as it must.  However, in the present circumstance we identify
\beqn
(\qbar q)^p ~&=&~
 (\qbar^{p-r} q^p )\cdot \qbar^r
 \nonumber\\
 &=&~ \left( \qbar^{\alpha' M_R^2/4}\,
      q^{\alpha' M_L^2 /4}\right)\, 
      \qbar^r\nonumber\\
&=&~ e^{-2\pi \tau_2 \left(
   \half \alpha' M^2 + r\right) }
    \,
    e^{2\pi i \tau_1
     \left[ \quarter \alpha' (M_L^2-M_R^2) - r\right] }~.
     \nonumber\\
\label{pedantic}
\eeqn
Moreover, we know that $r= \alpha'(M_L^2-M_R^2)/4$.   Inserting this result for $r$ then eliminates the second exponential in Eq.~(\ref{pedantic}) --- as it must, given that $g(\tau_2)$ cannot have any residual $\tau_1$-dependence ---  and the first exponential becomes
\beqn
   e^{-2\pi \tau_2 \left(
    \alpha' M^2 + r\right) }
~&=&~
   e^{-2\pi \tau_2 \left[
   \quarter \alpha' (M_L^2+M_R^2) + \quarter\alpha' (M_L^2 - M_R^2) \right] }\nonumber\\
&=&~
   e^{-\pi \alpha' M_L^2 \tau_2}~,
\label{MLsuppression}
\eeqn
thereby reproducing the same exponential suppression as given above.

Thus, putting the pieces together, we find that the $E$-entwined portion of our $g$-function takes the form 
\beqn
   && g_{\calB\overline E_2} (\tau_2) ~=~ \label{thirdexp} \\
   && ~~~~~
  \sum_\ell \tau_2^{\ell-1} 
  \sum_{r=0}^\infty \sum_p
       \chi_r\, a_{p-r,p} \, \mathbbB^{(\ell)}_{p-r,p} \,
         e^{-\pi \alpha' M_L^2 \tau_2}~,\nonumber
\eeqn
whereupon we have
\beqn
&& \left\langle 
\calA + \calB \overline E_2 \right\rangle ~=~
\frac{\pi}{3}  \,\lim_{\tau_2\to 0} \,\tau_2^{-1} \, \Bigl\lbrack\nonumber\\ 
&& ~~~~~~~\phantom{+}  \sum_\ell
    \tau_2^{\ell} \sum_p  a_{pp} \,\mathbbA^{(\ell)}_{pp} \, 
    e^{-\pi \alpha' M^2 \tau_2} 
    \nonumber\\
&& ~~~~~~+
\sum_\ell \tau_2^{\ell} 
  \sum_{r=0}^\infty \sum_p \chi_r\,
        a_{p-r,p} \,\mathbbB^{(\ell)}_{p-r,p} \, e^{-\pi \alpha' M_L^2 \tau_2}\Bigr\rbrack~.~~~~
        \nonumber\\
\label{piecestogether}
\eeqn
The rest of our analysis proceeds precisely as for the case without entwinement.
Following Eq.~(\ref{assumedtau2dep}),
we can assume that the total quantity within square brackets in Eq.~(\ref{piecestogether}) has an overall $\tau_2$-dependence of the
form $\sum_{j=0}^\infty C_j \tau_2^j$.
We then have 
\beqn
  C_0~&=&~ \lim_{\tau_2\to 0} \,[...]~,\nonumber\\
  C_1~&=&~ \lim_{\tau_2\to 0}\,
  \frac{d}{d\tau_2}\, [...]~
  \label{preentwinedRSresults}
\eeqn
where $[...]$ represents the quantity in square brackets in Eq.~(\ref{piecestogether}). 
The presumed finiteness of $\bigl\langle \calA + \calB\overline E_2\bigr\rangle$ then leads us to conclude that
\beqn
   C_0 ~&=&~ 0~, \nonumber\\
   \left\langle \calA + \calB \overline E_2\right\rangle &~=~& 
\frac{\pi}{3} \,C_1~.
\label{entwinedRSresults}
\eeqn

Just as in the case without entwinement, these results can be given a direct interpretation in terms of supertraces over our string states.   We have already remarked that our un-entwined supertrace in Eq.~(\ref{supertrace_regulated}) is nothing but an operator eigenvalue-weighted sum over the states that lie along the principal diagonal.
Given this, let us also define an analogous {\it shifted supertrace}\/ as the sum over the states that lie along  the $r^{\rm th}$ {\it shifted}\/ diagonal:
\beq 
    {\rm Str}^{(r)} X ~\equiv~ 
    \lim_{\tau_2\to 0} \,\sum_{p}\, a_{p-r,p} \,X_{p-r,p} \, e^{-\pi \alpha' M_L^2 \tau_2}~
\label{shifted_supertrace}
\eeq
where $\alpha' M_L^2=4p$ and where the $p$-sum, as always, is over all of the states in the spectrum of the string model under consideration.
Note that for $r=0$, level-matching implies that $M_L=M_R=M$.   We thus find that
the $r=0$ shifted supertrace is nothing but our ordinary supertrace:
\beq
\Str^{(r=0)} X ~=~ \Str\,X~.
\label{requal0}
\eeq
The shifted supertraces with $r>0$ may thus be considered to be the generalizations of the ordinary supertrace to off-shell states --- \ie, states that lie along  non-principal diagonals.  As noted above, the $p$-sums along non-principal diagonals typically begin with non-zero values of $p$, so that $p-r$ continues to exceed the right-moving vacuum energy $\Delta$ of the string model under consideration (with $\Delta= -1/2$ for the heterotic string).  However, this restriction merely characterizes the existing states in the theory. 
  No states are excluded by these observations, and indeed these sums continue to tally
all of the (off-shell) states that exist in the theory.

We can also define an {\it $E$-entwined supertrace}\/ $\Str_E$ as the $\chi_r$-weighted sum of all of these shifted supertraces:
\beq
     {\rm Str}_E \,X ~\equiv ~
               \sum_{r=0}^\infty \,
               \chi_r \, \Str^{(r)} X~.
\label{Esupertrace}
\eeq
As such, this entwined supertrace $\Str_E X$ not only tallies both the physical and the unphysical string states, but also organizes the latter naturally according to how non-level-matched they are.  In this way the $E$-entwined supertraces elegantly capture the string-theoretic nature of our full string spectrum, where the $\overline E_2$ function determines the $\chi_r$ coefficients and thereby determines the precise nature of the entwinement (motivating us to refer to this as an $E_2$-entwinement, or $E$-entwinement for short). 

The definition in Eq.~(\ref{Esupertrace}) has an important simplification if our supertrace is restricted to only those states with $M_L=0$.  Such states have $p=0$, but for heterotic strings this in turn implies that we can only have $r=0$, since any greater value of $r$ would result in a right-moving worldsheet energy $\overline L_0$ less than $-1/2$.  We therefore have
\beq
\zLStrE\, X
~=~
\chi_0 \,\zStr^{\!\!\!(0)}  X
 ~=~ 
    \zStr\,X~
\label{nutheridentity}
\eeq
where the first equality is a consequence of the restriction to $r=0$ while the second equality is a consequence of Eq.~(\ref{requal0}) in conjunction with the fact that $\chi_0=1$.  Indeed, the quantity in Eq.~(\ref{nutheridentity})
is nothing but $a_{00} X_{00}$. 

Our results in Eqs.~(\ref{preentwinedRSresults}) and (\ref{entwinedRSresults}) can be easily expressed in terms of these shifted and entwined supertraces.   In particular, we see that
\beqn 
     C_0    ~&=&~ \Str\, \calA + \Str_E \,\calB~,\nonumber\\ 
    C_1 ~&=&~ 
     \Str\left( \frac{d\calA}{d\tau_2}\right)
     - \Str\left\lbrack \calA (\pi \alpha' M^2)\right\rbrack     \nonumber\\
       &&~~+~  
        \Str_E\left( \frac{d\calB}{d\tau_2}\right)
     - \Str_E\left\lbrack \calB (\pi \alpha' M_L^2)\right\rbrack \nonumber\\
&=&~ 
    \Str \Bigl[
 D_{\tau_2} \calA \Bigr]
+ \Str_E \left[ 
D^{(L)}_{\tau_2} \calB \right]           
\eeqn
where, in analogy to Eq.~(\ref{RSderivative}), we have
\beq 
D^{(L)}_{\tau_2}~\equiv~ \frac{d}{d\tau_2} 
       -  \pi \alpha'M_L^2 ~.
\label{RSderivativeL}
\eeq
Our Rankin-Selberg relations for the entwined case then become
\beqn
      \Str\, \calA ~&=&~ -\Str_E \calB   \nonumber\\
    \left\langle \calA + \calB \overline E_2\right\rangle ~&=&~\frac{\pi}{3} \left\lbrace \Str \left[
 D_{\tau_2} \,\calA \right]
+ \Str_E \left[ 
D^{(L)}_{\tau_2} \calB \right]\right\rbrace ~.
\nonumber\\
\label{RSresull}
\eeqn
In cases for which the operator insertions $\calA$ and $\calB$ are $\tau_2$-independent, this last result reduces to
\beq
\left\langle \mathbbA + \mathbbB \overline E_2\right\rangle ~=~ -\frac{1}{12{\cal M}^2} \biggl\lbrack \Str 
\left( \mathbbA M^2
\right) + \Str_E \left(
\mathbbB M_L^2\right) \biggr\rbrack ~.~\label{eq:abE2}
\eeq
We also note that the results in Eq.~(\ref{RSresull}) reduce to those in Eqs.~(\ref{singlesupertrace}) and (\ref{finalRSresult}) for the case in which the entwinement is removed, \ie, the case in which $\overline E_2 \to 0$, or $\chi_r\to 0$ for all $r\geq 0$.

We have already seen in Eq.~(\ref{strone}) that $\Str\,{\bf 1}=0$ in any closed-string theory which is free of physical (on-shell) tachyons.  Indeed, this is a completely general result which was first obtained in Ref.~\cite{Dienes:1995pm} and which holds even if the string model in question lacks spacetime supersymmetry.   Indeed, this result is one of the fundamental predictions of {\it misaligned supersymmetry}~\cite{Dienes:1994np,Dienes:1995pm}, a hidden symmetry which must always exist in any string spectrum and which plays a critical role in ensuring the finiteness of closed-string amplitudes.   Given this result, it is natural to wonder whether the corresponding {\it entwined}\/ supertrace $\Str_E\,{\bf 1}$ vanishes as well.  From Eq.~(\ref{Esupertrace}) we see that this would require either that the shifted supertraces $\Str^{(r)}{\bf 1}$ each vanish individually or take values which cancel in the sum over $r$ in Eq.~(\ref{Esupertrace}). 
However, just as $\Str\,{\bf 1}$ vanishes when evaluated along the principal diagonal for any self-consistent tachyon-free closed string theory,  it can be shown~\cite{offshellMisSUSY} that 
$\Str^{(r)}{\bf 1}=0$ as well --- \ie, that this supertrace also necessarily vanishes along each of the shifted diagonals.  This result is ultimately the result of an {\it off-shell}\/ misaligned supersymmetry that exists within the {\it off-shell}\/ structure of any tachyon-free string theory~\cite{Dienes:1994np,offshellMisSUSY}.
We thus find the general result
\beqn
   \Str^{(r)}{\bf 1}~& =&~ 0~~ {\rm for~all~} r ~~~~~~\nonumber\\
   \Longrightarrow~~~ 
     \Str_E\,{\bf 1}~&=&~0~.
\label{eq:StrE10}
\eeqn

Given these results, it is now straightforward to evaluate the supertraces of the overall operator insertions $\mathbbX_1$ and $\mathbbX_2$ in Eq.~(\ref{eq:Xs}).  We thus obtain the relations
\beqn
  {\rm Str}\, \mathbbX_1 ~&=&~ \frac{\xi}{2\pi} \,
      {\rm Str}\,\overline Q_H^2 
      % - \frac{\xi}{24\pi} \, {\rm Str}_E \,{\bf 1} 
      \nonumber\\
      {\rm Str}\, \mathbbX_2 ~&=&~ -2 \,{\rm Str}\, \overline Q_H^2 Q_G^2
        + \frac{1}{6}\, {\rm Str}_E \,Q_G^2~~~~
\label{presupertraceresults}
\eeqn
where ${\rm Str}_E$ is defined as in Eq.~\eqref{Esupertrace} in terms of principal and shifted supertraces and where we have eliminated a term proportional to $\Str_E\, {\bf 1}$ that would otherwise have appeared in the first line of Eq.~(\ref{presupertraceresults}). Indeed, with these results we have succeeded in writing the supertraces of our operator insertions $\mathbbX_1$ and $\mathbbX_2$ in terms of the supertraces of our physical charges $\overline Q_H^2$ and $Q_G^2$ across the states in the string spectrum.

We see, then, that the entwinement has had a profound effect on our theory, leading to supertraces over more than merely the physical string states.
At first glance, this might seem to violate our claim --- as expressed in Sect.~II --- that the Rankin-Selberg procedure leads to supertraces over only the physical string states.   However, the Rankin-Selberg procedure always involves performing the $\tau_1$-integral when defining $g(\tau_2)$, and thus it always projects onto those overall $\qbar^m q^n$ contributions for which $m=n$.   By contrast, what has occurred is that the presence of the $\overline E_2$ insertion --- which has its own intrinsic $\taubar$-dependence ---  has effectively {\it deformed}\/ how the different $(m,n)$ states ultimately contribute within the relevant modular integrands, ultimately allowing off-shell states within the original theory to become ``physical'' (\ie, level-matched) within the modular-completed calculation of the gauge couplings.  This then allows such states to  contribute in the  large-$\tau_2$ limit (\ie, in the deep IR), just as we expect for physical states. 
Thus, in this sense, it is the {\it modular completion}\/ --- along with the appearance of the  $\overline E_2$ factor ---  which has  deformed the notion of ``physicality'' insofar as the Rankin-Selberg procedure is concerned, allowing string states which were originally non-level-matched to behave as physical states in our gauge-coupling calculation.   Entwinement thereby widens the class of states which can ultimately contribute to the supertraces when calculating string amplitudes.

\subsection{Generic picture of running gauge couplings in
string theory}

\label{subsec:eval}

Having assembled all of the relevant conceptual ingredients, we are now ready to tackle our main task: to utilize the Rankin-Selberg procedure in order to evaluate  
the string amplitude that yields the regulated one-loop contribution to the gauge coupling,
and to recast this amplitude in terms of spectral supertraces. 
As we have seen in Eqs.~(\ref{eq:Ctilde})
and (\ref{genform}),
this amplitude is given by
\beqn
   \widehat \Delta_G ~&=&~ 
   \Bigl\langle 
   \left(\tau_2 \mathbbX_1 + \tau_2^2 \mathbbX_2\right) 
   \,\widehat\calG \Bigr \rangle~\nonumber\\
~&=&~
\Bigl\langle 
   \left(\calA + \calB \overline{E}_2\right) 
   \,\widehat\calG \Bigr \rangle~~~~~\nonumber\\   
   ~&=&~ 
   \Bigl\langle 
   \Bigl[\left(\tau_2 \mathbbA^{(1)} + \tau_2^2 \mathbbA^{(2)} \right) 
   \nonumber\\
     && ~~~~~~~+
   \left(\tau_2 \mathbbB^{(1)} + \tau_2^2 \mathbbB^{(2)} \right) \overline E_2
  \Bigr] \,\widehat\calG \Bigr \rangle~~~~~~
\label{eq:stringamp}
\eeqn
where the $\mathbbX_\ell$ are given in Eq.~(\ref{eq:Xs}), where $\calA$, $\calB$, $\mathbbA^{(\ell)}$, and $\mathbbB^{(\ell)}$  are given in Eq.~(\ref{ABs}), and where the regulator function $\widehat \calG_\rho(a,\tau)$ is given in Eq.~(\ref{regG}).   

The first thing we notice 
from the last term on the final line of Eq.~(\ref{eq:stringamp}) is that we  are now dealing not with a mere entwinement between two modular functions $Z_\calB(\tau,\taubar)$ and $\overline E_2(\taubar)$, but rather a {\it triple entwinement}\/ between $Z_\calB(\tau,\taubar)$, $\overline E_2(\taubar)$, and our regulator function $\widehat \calG_{\rho}(a,\tau,\taubar)$.
Thus, in principle, we should first develop a formalism for handling such a triple entwinement.   Moreover, as we have discussed in Sect.~\ref{sec:regulator}, we would like to evaluate this amplitude as a function of $a$, since $\rho a^2$ will eventually be identified with our running scale $\mu$ in units of $M_s$.   However, the value of $a$ affects the nature of the triple entwinement in a highly non-trivial way.

For these reasons, we shall adopt a different approach.  In particular, we shall follow the methodology first established in Appendix~A of Ref.~\cite{Abel:2021tyt} for calculations of the Higgs mass, only suitably adapted for our gauge-coupling calculation.  
To do this, we observe upon comparing Eqs.~(\ref{Higgsstep1}) and (\ref{eq:stringamp}) that the one-loop Higgs mass has the same algebraic structure as the one-loop contribution the gauge coupling.   Indeed, the only difference between the two expressions is a change in the particular operator insertions $\mathbbX_\ell$.
However, the calculation in Appendix~A 
of Ref.~\cite{Abel:2021tyt}  does not rely on the precise operator insertions as long as they have the general modular structure $\tau_2 \mathbbX_1 + \tau_2^2 \mathbbX_2$.   For this reason, we can borrow the results from Appendix~A of Ref.~\cite{Abel:2021tyt} and then simply update these results using the new operator insertions appropriate for our gauge-coupling calculation.

This procedure is greatly facilitated by first observing that 
the form of the regulator function $\widehat\calG_\rho(a,\tau)$ in Eq.~(\ref{regG}) allows us to reduce the calculation of $\widehat \Delta_G$ to a calculation of the ``reduced'' amplitude
\beq
       P(a) ~= ~ \Bigl\langle  \left( \tau_2 \mathbbX_1+\tau_2^2 \mathbbX_2 \right)
      \, Z_{\rm circ}(a,\tau)\Bigr\rangle ~
\label{mainint}
\eeq
where $Z_{\rm circ}(a,\tau)$ is the circle-compactification function in  Eq.~(\ref{Zcircdef}).  Indeed, once we have evaluated $P(a)$, it follows from Eq.~(\ref{regG}) that we can then easily evaluate $\widehat \Delta_G$ through the relation
\beq
\widehat\Delta_G(\rho,a)  ~=~
\frac{a^2}{1+\rho a^2} 
\frac{\rho}{\rho-1} 
\frac{\partial}{\partial a} \,
          \biggl[   P(\rho a) - P(a) \biggr]~.
\label{operator}
\eeq

Our task is therefore to evaluate $P(a)$.   However, this is precisely what is done in Ref.~\cite{Abel:2021tyt} for cases in which the operators $\mathbbX_\ell$ do not contain any entwinements.  Indeed, in such cases it is shown that $P(a)$ is given by
\beqn
P(a) ~&=&  ~~ \zStr \mathbbX_1 \left\lbrack f_1(a) + f_2(a) \right]  \nonumber\\
     ~&& +  \zStr \mathbbX_2 \left\lbrack f_3(a) \right\rbrack \nonumber\\
     ~&& +  \pStr \mathbbX_1 \left\lbrack f_2(a) +  f_4(M,a)
  \right\rbrack ~~\nonumber\\
     ~&& + \pStr \mathbbX_2 \left\lbrack f_5(M,a) \right\rbrack ~
\label{finalPafromhiggs}
\eeqn
where
\beqn
f_1(a) ~&=&~ \frac{\pi a}{3}\nonumber\\
f_2(a) ~&=&~ \frac{\pi}{3a} \nonumber\\
f_3(a) ~&=&~ - \frac{2}{ a} \log\,a\, \nonumber\\
f_4(M,a) ~&=&~ \frac{2}{\pi} \,\sum_{r=1}^\infty \left(\frac{M}{r\calM}\right) \,K_1\left( \frac{r  M}{a\calM} \right) 
  \nonumber\\
f_5(M,a) ~&=&~  \frac{4}{a}\, \sum_{r=1}^\infty  
            K_0\left(  \frac{ r  M}{a\calM} \right) ~.
\label{ffunctions}
\eeqn
Here $\calM= M_s/(2\pi)$ is the reduced string scale and $K_\nu(z)$ denotes the modified Bessel function of the second kind. 
Likewise, the notations $\zStr \mathbbX f(M)$ and $\pStr \mathbbX f(M)$ respectively indicate contributions from purely massless and massive string states.  
The  result in Eq.~(\ref{finalPafromhiggs}) is exact for all $a$ and holds for general 
un-entwined modular-invariant operator insertions of the form $\calX=\tau_2 \mathbbX_1 +\tau_2^2 \mathbbX_2$. 
Thus we  see from Eq.~(\ref{finalPafromhiggs}) that
our reduced amplitude in Eq.~(\ref{mainint}) for {\it un-entwined}\/ operators $\mathbbX_1$ and $\mathbbX_2$ 
can ultimately be expressed in terms of combinations of supertraces of the form $\Str \,\bigl[\mathbbX f(M)\bigr]$ for various combinations of $\mathbbX_1$ and $\mathbbX_2$ and for various functions $f(M)$.
We further note that the functions $f_1$ through $f_3$ are wholly independent of any aspect of the spectrum of the string theory under study, and thus these functions can be taken outside their respective supertraces.
By contrast, the functions $f_4$ and $f_5$ depend not only on the regulator parameter $a$ but also on the mass $M$ of the contributing state.   Such functions are thus intrinsically part of the supertrace and cannot be factored out.

Although the result in Eq.~(\ref{finalPafromhiggs}) was derived for operators $\mathbbX_\ell$ that do not involve any entwining, it is not difficult to generalize this result to the case in which a given operator $\mathbbX$ is entwined, \ie, takes the form $\mathbbX=\mathbbA + \mathbbB\cdot \overline{E}_2$ with $\mathbbB\not=0$.   Indeed, tracing through the derivations that originally led to Eq.~(\ref{finalPafromhiggs}), we find that in such cases we can simply replace
\beqn
    && \Str\, \Bigl[\mathbbX\, f(M)\Bigr] 
    \nonumber\\
    &&~~~~ \longrightarrow~
    \Str\,\Bigl[ \mathbbA\, f(M)\Bigr] + \Str_E\, \Bigl[\mathbbB \,f(M_L)\Bigr]~~~~~~~~~
\label{replacement}
\eeqn
within Eq.~(\ref{finalPafromhiggs}). This illustrates the power of the entwined-supertrace
formalism we have developed. 
Moreover, when restricting to massive states [\ie, states whose contributions to $g(\tau_2)$ have an exponential $\tau_2$-dependent suppression], we have
\beqn
    && \pStr\, \Bigl[\mathbbX\, f(M)\Bigr] 
    \nonumber\\
    &&~~~~ \longrightarrow~
    \pStr\,\Bigl[ \mathbbA\, f(M)\Bigr] + \pStrE\, \Bigl[\mathbbB \,f(M_L)\Bigr]~~~~~~~~~~~~~
\label{replacement3}
\eeqn
The critical point here is that the restriction to massive states for $\mathbbX$ becomes a restriction to states with positive {\it left}\/-moving mass $M_L$ for the $E$-entwined supertrace.  This makes sense since the exponential suppression within $g(\tau_2)$ for the entwined supertrace depends on $M_L^2$ rather than $M^2$.
By contrast, when restricting to {\it massless}\/ states [\ie, states that contribute to $g(\tau_2)$ without exponential suppression], the $f(M)$ function becomes a constant which can be pulled outside the trace.  We can then push this one step further and write
\beqn
  \zStr\, \mathbbX
   ~&\longrightarrow& ~
    \zStr\,\mathbbA
    + \zLStrE\, \mathbbB \nonumber\\
    ~&=&~
    \zStr\,\mathbbA
    + \zStr \mathbbB~.
\label{replacement2}
\eeqn
The top line is of course analogous to what occurs in Eq.~(\ref{replacement3}), but the additional step --- the passage to the second line --- 
follows from Eq.~(\ref{nutheridentity}).

As an example of Eq.~(\ref{replacement2}), let us consider the cases when $\mathbbX=\mathbbX_1$ and $\mathbbX=\mathbbX_2$. 
In these cases we have
\beqn
\zStr  \mathbbX_1 ~&=&~
\frac{\xi}{2\pi} \,\zStr \overline Q_H^2 - \frac{\xi}{24\pi} \,\zLStrE {\bf 1} ~~~~\nonumber\\
~&=&~
\frac{\xi}{2\pi} \,\zStr  \left( \overline Q_H^2 - \frac{1}{12}\right)~
\eeqn
and 
\beqn
\zStr \mathbbX_2 ~&=&~ -2\,\zStr\,
\overline Q_H^2 Q_G^2 + \frac{1}{6} \,\zLStrE Q_G^2 ~~~~\nonumber\\
&=&~ -2 \zStr \left( 
\overline Q_H^2 - \frac{1}{12} \right) Q_G^2 ~.~~~~
\label{otherresult}
\eeqn
Interestingly, we see that in both cases the restriction to massless states has completely eliminated the modular completion that we originally performed in Eq.~(\ref{Eisen_helicity}). 
We further note the identity
\beqn
 \pLStrE {\bf 1} ~&=&~
   \Str_E {\bf 1} -
   \zLStrE {\bf 1} \nonumber\\
  ~&&=~  - \zStr {\bf 1}
   ~=~ \pStr {\bf 1}~.~~~
\label{oneidentities}
\eeqn
Here the first equality follows
from the observation originally made in the paragraph below that containing Eq.~(\ref{dvalues}), namely that $p\geq 0$ (implying that $M_L\geq 0$).  Likewise, 
the second equality follows from 
Eq.~(\ref{eq:StrE10}) and the final equality follows from Eq.~(\ref{strone}).

Let us now start with Eq.~(\ref{finalPafromhiggs}) and insert our expressions for $\mathbbX_1$ and $\mathbbX_2$, where these expressions are given in Eq.~(\ref{eq:Xs}).  Bearing in mind the substitutions and simplifications in Eqs.~(\ref{replacement}) through (\ref{oneidentities}), we obtain
\beqn
P(a) ~&=&  ~~ 
% line 1
\zStr \left\lbrace \frac{\xi}{2\pi}  \left(\overline Q_H^2 - \frac{1}{12}\right) \bigl[f_1(a) + f_2(a)\bigr] \right\rbrace    \nonumber\\
%  line 2
     ~&& -  \zStr \left\lbrack  2\left(\overline Q_H^2 - \frac{1}{12}\right) Q_G^2 
    \,  f_3(a)   \right\rbrack            \nonumber\\    
%  line 3
     ~&& + 
     \pStr  \left\lbrack \frac{\xi}{2\pi } \left(
      \overline Q_H^2 - \frac{1}{12} \right)  
       f_2(a)\right]
      \nonumber\\
%  line 4
  ~&& + \pStr  \left[  \frac{\xi}{2\pi} \,\overline Q_H^2 \,  f_4(M,a)
           \right]  
          ~\nonumber\\
%  line 5 
  ~&& - 
  \pLStrE   \left[  \frac{\xi}{24\pi} 
  f_4(M_L,a) 
       \right] 
          ~\nonumber\\
%   line 6
     ~&&  -
     \pStr \left\lbrack 2 \,\overline Q_H^2 Q_G^2  \,f_5(M,a) \right\rbrack 
     \nonumber\\
%   line 7
      ~&& + 
      \pLStrE \left\lbrack
   \frac{1}{6}\, Q_G^2 \,
            f_5(M_L,a) \right\rbrack  ~.~~~~~~
\label{finalPafromhiggs2}
\eeqn
This expression represents the line-by-line result of substituting the values of the $\mathbbX_i$ from Eq.~(\ref{eq:Xs}) into Eq.~(\ref{finalPafromhiggs}) and implementing the identities in Eqs.~(\ref{replacement}) through (\ref{oneidentities}).   

Additional manipulations can further simplify this expression and render it more compact while also simultaneously elucidating its algebraic structure. This rewriting will also be useful for understanding certain properties of the resulting running of the gauge couplings.  In particular, we note that the final $f_2(a)$ term on the first line of Eq.~(\ref{finalPafromhiggs2}) can be joined with the third line of this equation in order to remove the $M>0$ restriction on the latter.  The removal of this restriction then further allows us to eliminate the $-1/12$ term as a result of the identity $\Str\, {\bf 1}=0$. Likewise, the $f_1(a)$ term on the first line can be combined with the expressions on the fourth and fifth lines in order to remove their $M>0$ and $M_L>0$ restrictions as well;  indeed, these latter observations follow as a result of the fact that $K_1(z)\sim z^{-1}$ as $z\to 0$, whereupon we see that 
\beq
\lim_{M\rightarrow 0} \,
   \sum_{r=1}^\infty \,
     \left(\frac{ M}{r\calM}\right)\, 
             K_1\left( \frac{r  M}{a\calM} \right) ~=~
 \sum_{r=1}^\infty \frac{a}{r^2} ~=~ \frac{\pi^2 a}{6}~,
\eeq
or equivalently
\beq
\lim_{M\to 0} \,f_4(M,a) ~=~f_1(a)~.
 \label{Mzerolimit}
\eeq
Interpreting the quantity in Eq.~(\ref{Mzerolimit}) as $f_4(0,a)$, 
we thus find that Eq.~(\ref{finalPafromhiggs2}) simplifies to
\beqn
P(a) ~&=&  ~- 
     \zStr \left[ 2 \left(\overline Q_H^2 - \frac{1}{12}\right) Q_G^2 \,
     f_3(a) \right] \, \nonumber\\    
%  line 3
     ~&& + \,
     \Str  \left[ \frac{\xi}{2\pi } \,
      \overline Q_H^2 
       \,f_2(a)\right]
      \nonumber\\
%  line 4
  ~&& + \,\Str\left[  \frac{\xi}{2\pi}\, \overline Q_H^2 \,  f_4(M,a)
           \right] 
          ~\nonumber\\
%  line 5 
  ~&& - \,
  \Str_E   \left[ \frac{\xi}{24\pi} \,
  f_4(M_L,a) 
       \right] 
          ~\nonumber\\
%   line 6
     ~&&  -\,
     \pStr \left\lbrack 2 \,\overline Q_H^2 Q_G^2  \,f_5(M,a) \right\rbrack 
     \nonumber\\
%   line 7
      ~&& + \,
      \pLStrE \left\lbrack
   \frac{1}{6}\, Q_G^2 \,
            f_5(M_L,a) \right\rbrack  ~.~~~~~~
\label{finalPafromhiggs3}
\eeqn
At first glance it might seem that the first line of Eq.~(\ref{finalPafromhiggs3}) could be rewritten in an analogous manner as the $M\to 0$ limit of the terms in the final two lines.  This would require that $f_3(a)$ somehow emerge as the $M\to 0$ limit of $f_5(M,a)$.  However, this is ultimately not the case:   we instead find
that
\beq
f_5(M,a ) ~\sim~ f_3(a) +\frac{M_s}{M} + \frac{2}{a}\log \left( e^\gamma \frac{M}{2M_s}\right)  ~~{\rm as}\,M\to 0~.
\label{f5limit}
\eeq
While the extra $a$-independent constant $M_s/M$ would ultimately prove irrelevant under the operation in Eq.~(\ref{operator}), the logarithmic divergence in Eq.~(\ref{f5limit}) spoils the uniform convergence of the Bessel-function sum.
This issue has important  implications and is discussed in detail in Sect.~V of Ref.~\cite{Abel:2021tyt}.

There is also a third way of rewriting these expressions which can be useful for understanding the ramifications of the entwinement in these theories.   From Eqs.~(\ref{Esupertrace}), (\ref{requal0}), and (\ref{cr0}) we can write
\beq
 \Str_E X ~=~ \Str\, X + \sum_{r=1}^\infty \chi_r \,\Str^{(r)} X~.
 \eeq
 However, as we discussed below Eq.~(\ref{requal0}), we must have $p-r\geq \Delta$ where $\Delta$ is the right-moving vacuum energy within the type of string theory under study.  We thus find that $p\geq \Delta +r$, which implies that we cannot have $r\geq 1$ unless $p\geq \Delta +1$.  This last constraint is evident in Fig.~\ref{fig:matrix} for the heterotic case in which $\Delta = -1/2$.  From Eq.~(\ref{massids}) this last constraint corresponds to $\alpha' M_L^2\geq 4(\Delta+1)$.  
 We can thus write
 \beq
\Str_E X ~=~  \underset{M <M_E} {\Str} \,X
  ~+~
  \underset{M_L\geq M_E}{\Str_E}\, X~
  \label{newidentity}
\eeq
where $M_E$ denotes the {\it entwinement scale}
\beq
   M_E ~\equiv~ 2\sqrt{\Delta+1} \,M_s~
\label{MEdef}
\eeq
at which the entwinement first appears.
For heterotic strings we have $M_E= \sqrt{2}M_s$.
Note that Eq.~(\ref{newidentity}) is a general result, valid for all strings.  
Despite the $M<M_E$ upper limit
on the first of the sums in Eq.~(\ref{newidentity}), this sum can nevertheless involve a large number of states; this is especially true for cases in which compactification radii are far from the string scale.  

Using Eq.~(\ref{newidentity}), we can finally rewrite Eq.~(\ref{finalPafromhiggs3}) in the form
\beqn
&& P(a) ~= 
    ~ - \zStr \left[ 2 \left(\overline Q_H^2 - \frac{1}{12}\right) Q_G^2 
    \,  f_3(a) \right]               \nonumber\\
% line 3
     ~&& + \,
     \Str  \left[ \frac{\xi}{2\pi } \,
      \overline Q_H^2 
       \,f_2(a)\right]
      \nonumber\\  
%  line 4
  ~&& +\,\underset{ 
  M<M_E}{\Str} ~\left[ \frac{\xi}{2\pi}  \left(\overline Q_H^2 -\frac{1}{12}\right) \,    f_4(M,a) \right]  \,
          ~\nonumber\\
  ~&& +\,\underset{M\geq M_E}{\Str} 
        ~\left[ \frac{\xi}{2\pi} \, \overline Q_H^2 \,
          f_4(M,a) \right] \,
          ~\nonumber\\
%  line 5 
  ~&& - \,\underset{M_L\geq M_E}{\Str_E} ~
  \left[ \frac{\xi}{24\pi} \,
        f_4(M_L,a) \right] 
          ~\nonumber\\
%   line 6
     ~&&  -\,\underset{ 0< M<M_E}{\Str} 
     \left\lbrack 2\left( \overline  Q_H^2 -\frac{1}{12}\right) Q_G^2  \,f_5(M,a) \right\rbrack  ~\nonumber\\
     ~&&  -\,\underset{M\geq M_E}{\Str}
     ~ \left\lbrack 2 \,\overline Q_H^2 Q_G^2  \,f_5(M,a)\right\rbrack ~\nonumber\\
%   line 7
      ~&& + \, \underset{M_L\geq M_E}{\Str_E}~
    \left\lbrack \frac{1}{6}\,
      Q_G^2 \,f_5(M_L,a) \right\rbrack ~.~~~~~~~~~~~~~
\label{finalPafromhiggs4}
\eeqn
Although this expression has more individual terms than its two predecessors, it explicitly shows that the entwinement is wholly restricted to string states with $M_L\geq M_E$.   Indeed, all terms that receive contributions from states with masses $M< M_E$ depend not on $\overline Q_H^2$ but rather on the explicit un-entwined combination $\overline Q_H^2 - 1/12$.   This statement includes the contribution on the second line of Eq.~(\ref{finalPafromhiggs4}) once we realize that the $1/12$ term that would otherwise appear there has vanished as a result of the identity $\Str\, {\bf 1}=0$.

Eqs.~(\ref{finalPafromhiggs}), 
(\ref{finalPafromhiggs2}), (\ref{finalPafromhiggs3}), and (\ref{finalPafromhiggs4})
represent fully modular-invariant evaluations of the reduced string amplitude $P(a)$ in Eq.~(\ref{mainint}), expressed purely in terms of supertraces over our string states. 
Given that these supertraces are to be evaluated over the states within the spectrum of whatever the string model happens to be, these results are completely general and model-agnostic, applicable to any four-dimensional string model --- with or without spacetime supersymmetry --- so long as the model lacks physical tachyons.   Although the modular invariance of $P(a)$ in each of these expressions is not manifest, it is hidden in supertrace identities that relate the various terms in these expressions to each other.   

Using these expressions in conjunction with Eq.~(\ref{operator}) we can then trivially evaluate our full desired amplitude $\widehat \Delta_G(\mu)$ for the running of the gauge couplings.  Indeed, we see from Eq.~(\ref{operator}) that 
we can turn $P(a)$ into $\widehat \Delta_G(\mu)$ simply by replacing each term within $P(a)$ according to the schematic substitution $\Str [\mathbbX  f_i(a)]\to  \Str[ \mathbbX \phi_i(\mu)]$ where the operators $\mathbbX$ are unchanged and where the new functions $\phi_i(\mu)$ for each $f_i(a)$ are given by
\beq
 \phi_i(\mu)  \,\equiv\,
 \frac{1}{1+\rho a^2} \,\frac{\rho}{\rho-1} a^2 \frac{\partial}{\partial a} \biggl\lbrack f_i(\rho a) - f_i(a)\biggr\rbrack~ 
\label{regGf}
\eeq 
where we first evaluate the right side of Eq.~(\ref{regGf}) as a function of $\rho$ and $a$, and then identify $\mu^2 \equiv \rho a^2 M_s^2$ with $\rho=2$ chosen as a benchmark value.  Indeed, given the $f_i$-functions given in Eq.~(\ref{ffunctions}), we find
\beqn
\phi_1(\mu) \,&=&\, \frac{\pi}{3} \,\frac {\mu^2/M_s^2}{1+ \mu^2/M_s^2}
\nonumber\\
\phi_2(\mu) \,&=&\, \frac{\pi}{3}\,\frac {1}{1+ \mu^2/M_s^2}
\nonumber\\
\phi_3(\mu) \,&=&\, \frac{2}{1+\mu^2/M_s^2}\,\log\left( \frac{2\sqrt{2} e M_s}{  \mu}\right) 
\nonumber\\
\phi_4(M,\mu) \,&=&\, \frac {1}{1+ \mu^2/M_s^2} \, \frac{1}{\pi} \!
\left(\frac{M}{\calM}\right)^2 
 \! \left\lbrack    
        \calK_0^{(0,1)}\!(z) \! + \!
        \calK_2^{(0,1)}\!(z)  \right\rbrack
\nonumber\\
\phi_5(M,\mu) \,&=&\, \frac {2}{1+ \mu^2/M_s^2}\,
\left \lbrack 
\calK_1^{(1,2)}\!\left( z\right) -
        2\calK_0^{(0,1)}\!\left( z \right)         
             \right\rbrack ~~~~ \nonumber\\
\label{phifunctions}
\eeqn
where $z\equiv 2\sqrt{2}\pi M/\mu$ and where we have defined the Bessel-function combinations~\cite{Abel:2021tyt}
\beq
     \calK_\nu^{(n,p)} (z) ~\equiv~ \sum_{r=1}^\infty ~ (rz)^{n} \Bigl\lbrack 
       K_\nu(rz/\rho) - \rho^p K_\nu(rz)  \Bigr\rbrack~.~~~
\label{Besselcombos}
\eeq
Note that $\phi_1(\mu) + \phi_2(\mu)= \pi/3$.

These $\phi_i$-functions are extremely important and have direct physical interpretations.  While the specific charges that enter into the $\mathbbX_i$ expressions tell us which  specific quantity is under study (such as the Higgs mass versus the gauge coupling), and while the particular numerical values of these charges tell us about the particular string model under study, the $\phi_i$ functions are essentially universal and tell us how these phenomenological quantities {\it run}\/ as functions of the scale $\mu$ in the corresponding EFT.~  As we have seen, these running functions are universal for all quantities (such as the one-loop Higgs mass or gauge couplings) which have at most a logarithmic divergence in string theory prior to regularization.
More specifically, substituting
$P(a) \to \widehat \Delta(\mu)$ and $f_i(a)\to \phi_i(\mu)$ within
our previous expression for $P(a)$
in Eq.~(\ref{finalPafromhiggs2}),
we find 
that
\begin{itemize}
\item  $\phi_1(\mu)$ is the contribution to $\widehat \Delta_G(\mu)$ per unit ($\mathbbA^{(1)}+\mathbbB^{(1)}$) charge from each massless physical state;
\item $\phi_2(\mu)$ is the additional contribution to $\widehat \Delta_G(\mu)$ per unit
($\mathbbA^{(1)}+\mathbbB^{(1)}$) charge from each physical state, {\it regardless}\/ of mass;
\item $\phi_3(\mu)$ is the additional contribution to $\widehat \Delta_G(\mu)$ per unit ($\mathbbA^{(2)} +  \mathbbB^{(2)}$) charge from each massless physical state;
\item $\phi_4(M,\mu)$ is the additional contribution to $\widehat \Delta_G(\mu)$ per unit $\mathbbA^{(1)}$ charge for each physical state of non-zero mass $M$, while $\phi_4 (M_L,\mu)$ is $\chi_r^{-1}$ times the additional contribution to $\widehat \Delta_G(\mu)$ per unit $\mathbbB^{(1)}$ charge for each physical {\it or unphysical}\/ string $(m,n)$ state with left-moving mass $M_L$ for which $n-m\equiv r$ with $r\geq 0$;  and
\item $\phi_5(M,\mu)$ is the additional contribution to $\widehat \Delta_G(\mu)$ per unit 
$\mathbbA^{(2)}$ charge for each physical state of non-zero mass $M$, while 
$\phi_5 (M_L,\mu)$ is $\chi_r^{-1}$ times the additional contribution to $\widehat \Delta_G(\mu)$ per unit $\mathbbB^{(2)}$ charge for each physical {\it or unphysical}\/ string $(m,n)$ state with left-moving mass $M_L$ for which $n-m\equiv r$ with $r\geq 0$.
\end{itemize}
Here the $\mathbbA^{(i)}$ and $\mathbbB^{(i)}$ charges are given in Eq.~(\ref{ABs}),
and the above results are quoted for bosonic states;
fermionic states of course contribute with opposite signs.
Once again, we stress that these results are completely general for all phenomenological quantities which diverge at most logarithmically when unregulated;  it is only when we substitute the particular forms of $\mathbbA^{(i)}$ and $\mathbbB^{(i)}$ in Eq.~(\ref{ABs}) that 
we limit our attention to $\widehat \Delta_G(\mu)$ of the gauge couplings.   Indeed, in the case of the Higgs mass in Ref.~\cite{Abel:2021tyt}, no entwinement occurs and we have $\mathbbB^{(1)} = 
\mathbbB^{(2)} =0$.

As we have seen, the results quoted above for $P(a)$ in Eqs.~(\ref{finalPafromhiggs2}), (\ref{finalPafromhiggs3}), and (\ref{finalPafromhiggs4}) come directly from the result in  
Eq.~(\ref{finalPafromhiggs}).
This in turn is taken directly from
Eq.~(A15) of Ref.~\cite{Abel:2021tyt}.
Although the derivation given in Ref.~\cite{Abel:2021tyt} is sufficient for the Higgs mass, it makes the implicit assumption that supertraces of the general form $\Str\, [\tau_2 \mathbbX_2 f(M)]$ are all vanishing as a result of the explicit factor of $\tau_2$ within the supertrace. Otherwise, such terms would also have appeared in Eq.~(\ref{finalPafromhiggs}).  At first glance, the absence of such terms from {\it all}\/ calculations might appear to be justified, given that our supertraces are defined in Eq.~(\ref{supertrace_regulated})
in terms of a limiting procedure that involves taking the $\tau_2\to 0$ limit.   Indeed, in most circumstances (including those considered in Ref.~\cite{Abel:2021tyt}, where the Higgs mass was calculated), the extra factor of $\tau_2$ inside the supertrace would drive the overall supertrace to vanish, as assumed. 
However, as discussed earlier, it is possible (especially near the borders of moduli space) that our spectrum of string states can become extremely dense.   In such cases, the accumulated ``pile-up'' of states can cause quantities such as $\Str\,\mathbbX_2 f(M)$ to diverge, thereby allowing supertraces such as $\Str\,[\tau_2 \mathbbX_2 f(M)]$ to have non-zero values.
This ``pile-up'' phenomenon will be discussed in more detail in Ref.~\cite{NonRenormalizationTheorems}.   

In the present calculation of gauge couplings, we would like to maintain complete generality 
and allow our results to remain valid even as we approach the boundaries of moduli space.   For this reason, we must amend our results for $P(a)$ quoted above.   However, it turns out that this is relatively straightforward and amounts to introducing only one additional contribution
\beq
   \pStr \, \tau_2\mathbbX_2  \left[f_2(a)\right]~
\label{pileup}
\eeq
within Eq.~(\ref{finalPafromhiggs}).
This extra term will then propagate into
Eqs.~(\ref{finalPafromhiggs2}), (\ref{finalPafromhiggs3}), and (\ref{finalPafromhiggs4}).   
In fact, given that the ``pile-up'' phenomenon that gives rise to this term involves the infinite towers of massive states, we note that
$ \zStr \, \tau_2\mathbbX_2  \left[f_2(a)\right]=0$.
The contribution in Eq.~(\ref{pileup}) can thus be equivalently written as
 \beq
 \Str \, \tau_2\mathbbX_2  [f_2(a)]~,
 \label{pileup2}
 \eeq
 with no restriction on the masses of the states in the supertrace.

At this stage, we have now completed Step~8, as outlined in Sect.~II.~
This enables us to extract a considerable amount of information about the running of $\widehat \Delta_G(\mu)$. For example, let us consider the behavior of $\widehat \Delta_G(\mu)$ in the deep-IR limit, \ie, as $\mu\to 0$.
As $\mu\to 0$, we find that $\phi_1(\mu)$, $\phi_4(\mu)$, and $\phi_5(\mu)$ all vanish;  in the latter two cases this happens because the Bessel functions $K_2(z)$ in Eq.~(\ref{phifunctions}) all vanish exponentially as $z\to \infty$.  Thus, only $\phi_2(\mu)$ and $\phi_3(\mu)$ survive in the deep-IR limit.  Of course, $\phi_3(\mu)$ actually diverges in this limit.  This divergence is not a surprise, however, since the deep-IR limit corresponds to the limit $a\to 0$ in which our regulator is removed.   {\it Thus this divergence corresponds to the logarithmic divergence of our original unregulated quantity.}\/    
As anticipated in Sect.~\ref{sec:regulator}, and as apparent from each of our above expressions for $P(a)$,
this divergence is proportional to
\beq
\zStr \left( \mathbbA^{(2)} + \mathbbB^{(2)} \right) ~=~ 
 -2\, \zStr \,\left(
  \overline Q_H^2 -\frac{1}{12}\right)
  Q_G^2~.
  \label{masslessX2}
\eeq
However, in theories for which this quantity vanishes, we find that only $\phi_2(\mu)$ survives, with $\lim_{\mu\to0} \phi_2(\mu) = \pi/3$.   In such cases we find from Eqs.~(\ref{finalPafromhiggs2}) and 
(\ref{pileup2}) that
\beqn
&& \lim_{\mu\to 0} \widehat \Delta_G(\mu) 
~=~ \frac{\pi}{3}\,
\Str\left( \mathbbX_1 +  \tau_2 \mathbbX_2\right)  \nonumber \\ 
&& ~= ~  \frac{\xi}{6}\, \Str\, \overline Q_H^2 
- \frac{2\pi}{3}  \left( \Str\,   \tau_2 \overline Q_H^2 Q_G^2 - \frac{1}{12}  \Str_E  \, \tau_2Q_G^2 \right) .\nonumber\\
\label{deepIRlimit}  
\eeqn
 The fact that $\widehat \Delta_G(\mu)$ asymptotes to a 
 constant as $\mu\to 0$ is not particularly surprising, given the assumed vanishing of
 the quantity in Eq.~(\ref{masslessX2}).
However, what {\it is}\/ surprising is that the asymptotic value in Eq.~(\ref{deepIRlimit})  receives contributions not only from the light or massless string states, but from the {\it entire tower of string states}\/, all the way up into the UV!~   Indeed, {\it all}\/ of the string states contribute to the unrestricted supertraces in Eq.~(\ref{deepIRlimit}).  This is a graphic example of the UV/IR mixing inherent in a modular-invariant theory such as string theory.

Let us now consider the behavior of $\widehat \Delta_G(\mu)$  as we proceed upwards in energy scale $\mu$ away from the deep-IR limit.  Indeed, much of the following discussion mirrors the discussion for the Higgs mass in Ref.~\cite{Abel:2021tyt}, to which we refer the reader for details not provided here.   Let us first focus on energy scales for which $\mu\ll M_s$.   
In this regime, we find that $\phi_1$ and $\phi_4$ continue to remain vanishingly small.   However, whether $\phi_5$ remains small as well for a particular state of mass $M$ depends on the value of $z\sim M/\mu$ --- \ie, on whether the state whose contribution we are assessing is heavier or lighter than $\mu$.
In this connection it is important to realize that
our supertraces receive contributions from the entire string spectrum.
This necessarily includes states with masses {\mbox{$M\gsim M_s$}}, but may also include
potentially light states with non-zero masses far below $M_s$.
The existence of such light states depends on our string construction and
on the specific string model in question.  Thus, even though we are considering situations in which $\mu\ll M_s$, there need not be any fixed hierarchical relationship between $\mu$ and $M$.

%=================================================================
\begin{figure}
\centering
\hspace{-0.75cm}
\includegraphics[width=0.5
\textwidth]{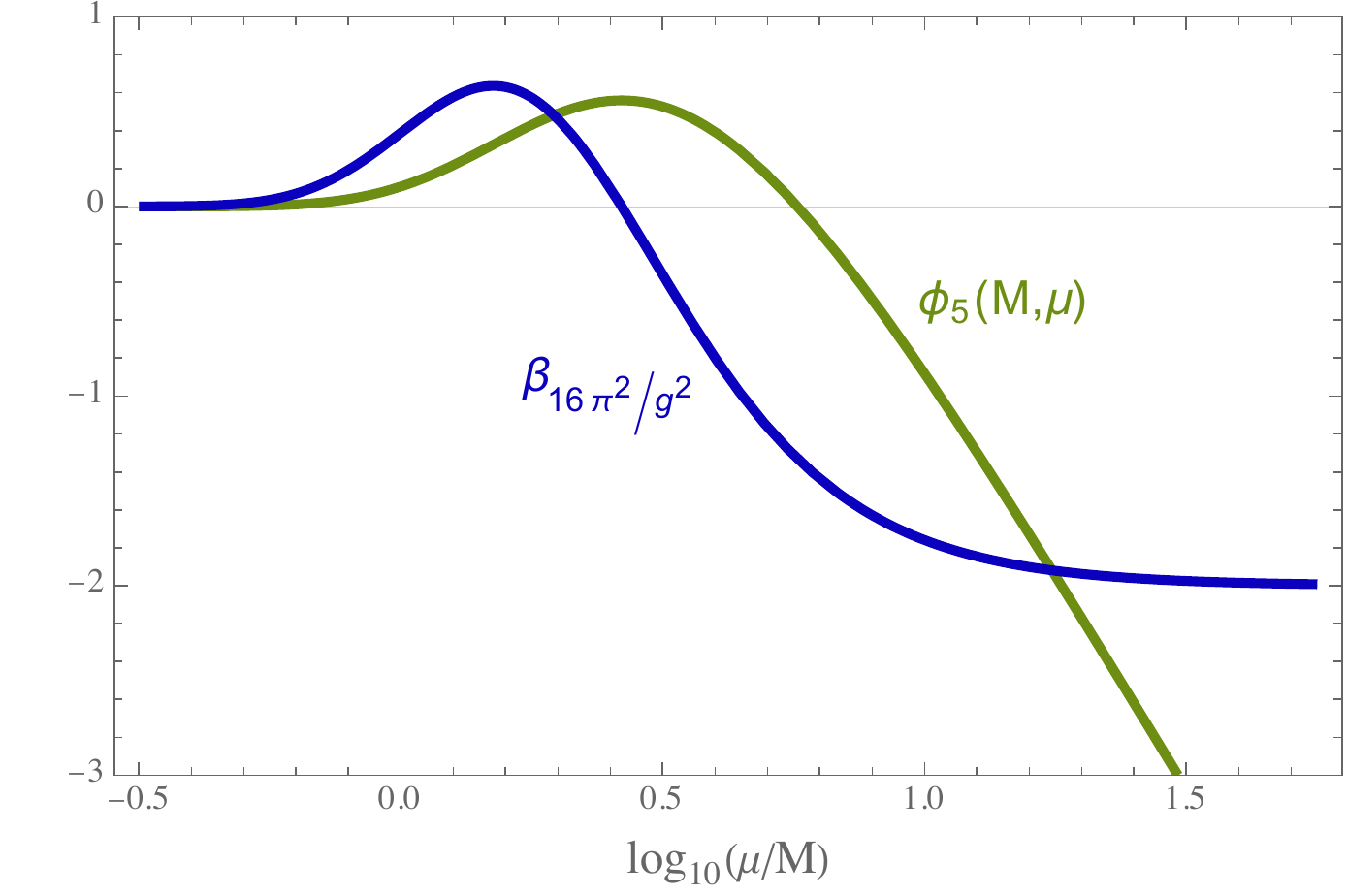}
\caption{ 
The function $\phi_5(M,\mu)$
in Eq.~(\ref{phifunctions}) and the corresponding contribution to the beta function $\beta_\Delta$ in Eq.~(\ref{beta_Delta}), plotted as functions of $\log_{10}(\mu/M)$  [green and blue, respectively].  Note that $\phi_5(M,\mu)$ is the Bessel-function contribution to $\widehat \Delta_G(\mu)$ per unit $\mathbbA^{(2)}$ charge from a given physical bosonic string state of non-zero mass $M$.
When $\mu\ll M$, the state is effectively integrated out of the theory, whereupon the running contribution $\phi_5(M,\mu)$ asymptotes to a constant. However, at larger energy scales $\mu\gg M$, the state is fully dynamical and produces a running which is effectively logarithmic.  Finally, within the intermediate $\mu\sim M$
region, the Bessel-function expression for $\phi_5(M,\mu)$ in Eq.~\ref{phifunctions}) provides
a smooth connection between these two asymptotic behaviors
and even gives rise to a transient ``hump'' in the value of $\widehat \Delta_G(\mu)$, or equivalently a ``dip'' in the value of the running coupling $g_G(\mu)$.}   
\label{fig:particlebeta}
\end{figure} 
%=================================================================

In Fig.~\ref{fig:particlebeta}, we have plotted $\phi_5(M,\mu)$ as a function of $\mu/M$.   Recalling that this is the contribution to $\widehat \Delta_G(\mu)$ per unit $\mathbbA^{(2)}$ charge from a given physical bosonic string state of mass $M$, certain aspects of this behavior are easy to understand.   For example, when $\mu\ll M$ the state is much heavier than the relevant energy scale $\mu$ and is effectively ``integrated out'' of our theory.   Thus all running stops, and $\phi_5(M,\mu)$ becomes flat.  Mathematically, this occurs because
\beq
       \calK_\nu^{(n,p)}(z) ~\sim~ \sqrt{\frac{\pi \rho}{2}} \,z^{n-1/2} \,e^{-z/\rho} ~~~~{\rm as}~ z\to \infty~.
\label{asymptoticform}
\eeq
Thus all running is exponentially suppressed as $z\sim M/\mu\to \infty$, leaving behind only an exponential tail.
By contrast, for energy scales $\mu\gg M$,  
our state is still dynamical.
We then see from Fig.~\ref{fig:particlebeta} that the effective contribution to the running from this state is effectively {\it logarithmic}\/.
Indeed, as {\mbox{$z\to 0$}}, we find that~\cite{Paris}
\beqn
          \calK_0^{(0,1)}(z)~&\sim &~ - \half \log\,z  + \half\left[ \log\,(2\pi) - \gamma\right]  \nonumber\\        
          \calK_1^{(1,2)}(z)~&\sim &~ 1~
\label{Kasymp}
\eeqn
where $\gamma$ is the Euler-Mascheroni constant.
For $\mu\gg M$, this leads to an 
asymptotic logarithmic running for $\phi_5(M,\mu)$ of the form 
\beq
  \phi_5(M,\mu) ~\approx~ -2  \log\left[ \frac{1}{\sqrt{2}}\,e^{-(\gamma+1)} \frac{\mu}{M}\right]~.
\label{loglimit}
\eeq
Finally, between these two extremes, we see that $\phi_5(M,\mu)$
interpolates smoothly and even gives rise to a transient ``hump'' in $\widehat \Delta_G(\mu)$, or equivalently a ``dip'' in $g_G$.
This behavior results from the 
specific combination of Bessel functions  within $\phi_5(M,\mu)$.
Of course, the statistics factor $(-1)^F$ within the supertrace
flips the sign of this contribution for degrees of freedom which are fermionic.  

Likewise,
for any fixed scale $\mu$,
adjusting the mass $M$ upwards or downwards simply corresponds
to shifting this curve rigidly to the right or left, respectively.
In this way one can imagine summing over all such contributions to the running (while also weighting each contribution by the appropriate net numbers of states at each mass level) as one takes the supertrace over the
entire string string spectrum.
Of course, for any energy scale $\mu$, the contributions from states with $M\gg \mu$ are
exponentially suppressed, as discussed above.   Thus, at any energy scale $\mu$, the only states which contribute meaningfully to
$\widehat\Delta_G(\mu)$  are those with $M\lsim \mu$. 

Thus, combining these Bessel-function contributions with those from Eq.~(\ref{deepIRlimit})
and keeping only those (leading) terms which dominate when $M\lsim \mu\ll M_s$,
we see that we can approximate the exact result in Eq.~(\ref{finalPafromhiggs4}) as
\beqn
&&  \widehat \Delta_G(\mu) ~\approx~
 \frac{\pi}{3}\,\Str\, \left( \mathbbX_1 
+ \tau_2 \mathbbX_2\right) 
\nonumber\\
&& ~~~~~
+ 2\, \zStr\,\left(  \mathbbA^{(2)} + \mathbbB^{(2)} \right) \,\log\left(
         \frac{2\sqrt{2} e M_s}{\mu} \right)
    \nonumber\\
&& ~~~~~ - 2 \,\effStr 
\left (
\mathbbA^{(2)} + \mathbbB^{(2)} \right)
\log \left\lbrack \frac{1}{\sqrt{2}} e^{-(\gamma+1)} \frac{\mu}{M} \right\rbrack 
\nonumber\\
\label{approxDeltapre}
\eeqn
or equivalently
\beqn
&&  \widehat \Delta_G(\mu) ~\approx~
 \frac{\xi}{6}\, \Str\, \overline Q_H^2 
\nonumber \\
&&  ~~~~~ -
\, \frac{2\pi}{3}  \left( \Str\,   \tau_2 \overline Q_H^2 Q_G^2 - \frac{1}{12}  \Str_E   \, \tau_2Q_G^2 \right) ~
\nonumber\\
&& ~~~~~ -4\,\zStr \left( \overline Q_H^2 - \frac{1}{12}\right) Q_G^2 \,\log\left(
     \frac{2 \sqrt{2} e M_s}{\mu} \right)
      \nonumber\\
&& ~~~~~ 
+ 4 \,\effStr \!\!\left( \overline Q_H^2 - \frac{1}{12}\right) Q_G^2\, \log \left\lbrack \frac{1}{\sqrt{2}} e^{-(\gamma+1)} \frac{\mu}{M} \right\rbrack .\nonumber\\
\label{approxDelta}
\eeqn

Given these results, our gauge couplings $g_G(\mu)$ can exhibit a variety of running behaviors.
These will ultimately depend on the spectrum of states associated with the string model under study.  Of course, the final terms in 
Eqs.~(\ref{approxDeltapre}) and (\ref{approxDelta}) do
not exhibit any running 
until we reach $\mu \sim M_{\rm lightest}$,
where $M_{\rm lightest}$ is the mass of the lightest massive string state
carrying a non-zero $(\mathbbA^{(2)} + \mathbbB^{(2)})$ charge.
Therefore, if we first restrict our attention to energy scales $\mu\lsim M_{\rm lightest}$,
the only running that arises is due to the massless $(\mathbbA^{(2)} + \mathbbB^{(2)})$-charged states.   These are the contributions that appear on the second and third lines of
Eqs.~(\ref{approxDeltapre}) and (\ref{approxDelta}), respectively.

This running can be expressed in a manner which is more traditional for describing the running of gauge couplings in string theory, namely in terms of an RGE that relates the couplings $g_G(\mu)$ to their values {\it at the string scale $M_s$}\/
(see, \eg, Refs.~\cite{Dienes:1995sv,Dienes:1995bx,Dienes:1995sq,Dienes:1996du}). 
From Eq.~(\ref{oneloopcontribution}), we obtain the general running equation for the total gauge couplings $g_G(\mu)$:
\beq 
 \frac{16\pi^2}{g^2(\mu)} 
    ~-~
    \frac{16\pi^2}{g^2_{{\rm tree}}}
  ~=~\widehat\Delta_G(\mu)
\label{oneloopcontribution2}
\eeq
where from Eq.~(\ref{approxDelta}) can write
\beqn
\widehat\Delta_G(\mu) ~&=&~ \widehat \Delta_G(M_s) -2\,
     \zStr\left(  \mathbbA^{(2)} + \mathbbB^{(2)} \right) \,\log\left( \frac{\mu}{M_s} \right)~\nonumber\\
&\equiv&~ \widehat \Delta_G(M_s) +\beta_\Delta(0)\, 
\log\left( \frac{\mu}{M_s} \right)~.
\label{intermedi1}
\eeqn
Here the quantity $\beta_\Delta(0)$ 
may be regarded as the $\mu=0$ value of the
 general beta function $\beta_\Delta(\mu)$ for $\widehat \Delta_G(\mu)$, which in turn is defined as
\beq
\beta_\Delta(\mu) ~\equiv~
\frac{\partial \widehat \Delta_G(\mu)}
{\partial \log\,\mu}~=~
    -\frac{32\pi^2}{g_G^3} \,\beta_g(\mu)~
\label{beta_Delta}
\eeq
where $\beta_g(\mu)\equiv \partial g_G/\partial \log\mu$ is the usual beta function for the gauge coupling $g_G$.
Indeed, we see from Eq.~(\ref{intermedi1}) that $\beta_\Delta(0)$ is 
precisely $-2$ times the quantity given in Eq.~(\ref{masslessX2}), \ie,
\beqn
     \beta_\Delta(0) ~&=& ~ 
     -2 \,\zStr\left(  \mathbbA^{(2)} + \mathbbB^{(2)} \right) \nonumber\\
         &=&~ 
         4\, \zStr \,\left(
  \overline Q_H^2 -\frac{1}{12}\right)
  Q_G^2~.~~~~~~
\eeqn
Likewise, using an asterisk `$\ast$' to  indicate the couplings that would have arisen in our theory if $\beta_\Delta(0)$ had vanished, we can write
\beq
\widehat\Delta_G(M_s) ~=~ \widehat \Delta_G^\ast(0) + \kappa
\label{kappaimplicit}
\eeq
where $\widehat\Delta_G^\ast(0)$ is 
given in Eq.~(\ref{deepIRlimit})
and where [within our regulator scheme defined by our regulator function $\widehat \calG_\rho(\mu)$
with $\rho=2$] we have
\beq  
  \kappa ~=~ -\beta_\Delta(0) \left[ 1+ \log\left( 2\sqrt{2} \right)\right]~.
\eeq
Thus, putting the pieces together, we have
the RGE
\beq 
 \frac{16\pi^2}{g^2(\mu)} 
    ~-~
    \frac{16\pi^2}{g^2_{{\rm tree}}}
  ~=~ 
   \widehat \Delta_G(M_s) ~+~ 
    \beta_\Delta(0)\, 
\log\left( \frac{\mu}{M_s} \right)~\\
\label{gaugerunning}
\eeq
where $\widehat \Delta_G(M_s)$ is given in
Eq.~(\ref{kappaimplicit}).

Thus far, Eq.~(\ref{gaugerunning}) captures the effects of the massless $(\mathbbA^{(2)} + \mathbbB^{(2)})$-charged string states.
However, as $\mu$ increases still further,
additional $(\mathbbA^{(2)}+\mathbbB^{(2)})$-charged string states 
enter the EFT and contribute their own individual logarithmic contributions. 
Of course, if these additional states 
have masses {\mbox{$M\gg M_{\rm lightest}$}},
the logarithmic nature of the running shown in Fig.~\ref{fig:particlebeta}
from the state with mass $M_{\rm lightest}$
will survive intact until $\mu \sim M$.
However, if the spectrum of states is relatively dense 
beyond $M_{\rm lightest}$, the logarithmic contributions from each of these states
must be added together, leading to a far richer behavior.

One interesting possibility arises in cases of string theories with large compactification radii $R\gg M_s^{-1}$.   In such cases, our theory will have Kaluza-Klein (KK) modes with masses $m_n\sim n/R$ that appear well below $M_s$.   Thus, as $\mu$ increases towards $M_s$, increasingly many KK states enter the EFT.~  Although each KK state contributes the same logarithmic running, our natural field-theoretic expectation is that the full supertrace over the string spectrum will begin to experience an accumulated effective {\it power-law}\/ growth, with $\widehat \Delta_G(\mu) \sim \mu^\delta$ where $\delta$ is the number of spacetime dimensions whose inverse compactification radii $R^{-1}$ lie below $\mu$.  Indeed, this is precisely the field-theoretic behavior discussed in Refs.~\cite{Dienes:1998vh, Dienes:1998vg}, which can algebraically be interpreted as resulting from a beta function $\beta_\Delta(\mu)$ which itself is growing polynomially with $\mu$. However, as we shall shortly see, in a string context we also have a scale-duality symmetry under $\mu \to M_s^2/\mu$.   This means that even at energy scales $\mu\ll M_s$ the {\it winding}\/ modes associated with such compactifications can also contribute.   Remarkably, these have the generic effect of {\it cancelling}\/ this power-law running (and even the original logarithmic running), thereby producing a situation in which there can be no running at all!  We will refer to the region in which such running terminates as a ``fixed-point region''.    This non-renormalization phenomenon, surprising as it is, is actually quite general and will be discussed in detail in Ref.~\cite{NonRenormalizationTheorems}.

Thus far our results are valid for energy scales below the string scale.
However, as mentioned above, it turns out that we also have information about what happens in the opposite region, namely that with
$\mu > M_s$:  {\it we simply enter a ``dual'' infrared 
region in which this same behavior again emerges, but in reverse.}\/
This is a direct consequence of the modular invariance which we have been careful to maintain throughout our calculations.
Indeed, modular invariance ensures that the running is symmetric under
the {\it scale-inversion}\/ duality transformation
\beq
  \mu ~\to~ \frac{M_s^2}{\mu}~.
\label{muduality}
\eeq
As a result, when plotted as a function of $\log(\mu/M_s)$, 
the behavior of $\widehat \Delta_G(\mu)$ for $\mu\ll M_s$ is reflected symmetrically
through the self-dual point $\mu_\ast=M_s$ to yield the reverse behavior for $\mu\gg M_s$.

As discussed in Ref.~\cite{Abel:2021tyt}, the origins of this scale-duality symmetry are easily understood.
To see this, we note that in general
the contribution of a string states of mass $M$ to the one-loop partition function
experiences a Boltzmann-like suppression factor  $\sim e^{-\pi \alpha' M^2 \tau_2}$.
Thus, for any particular benchmark value $\tau_2= t$, 
we can separate our string spectrum into two groups:   ``heavy'' states whose Boltzmann suppressions
at $\tau_2=t$ are significant according to some convention, and whose contributions therefore do not require regularization, 
and ``light'' states whose Boltzmann suppressions are not significant, and whose contributions
therefore require regularization.  
On this basis, with an eye towards interpreting these results in terms of an EFT with a running scale $\mu$,
we are directly led to identify $\mu^2$ inversely with $t$.
However, modular invariance tells us that any physical quantities which
depend on $\tau$ must be invariant under $\tau\to -1/\tau$.
Along the $\tau_1=0$ axis, this becomes an invariance under $\tau_2\to 1/\tau_2$.   This then immediately
implies an invariance 
under $t \to 1/t$, or equivalently 
under $\mu \to \mu_\ast^2 /\mu$  where $\mu_\ast$ is an arbitrary self-dual mass scale.
Of course, the choice of normalization for $\mu$ in relation to $t$ is purely a matter of convention, since the former is a dimensionful spacetime quantity while the latter is a dimensionless worldsheet quantity.  In keeping with the traditional string-theoretic conventions relating worldsheet and spacetime physics, we take this conversion factor to be given by $\alpha'$.    This then tells us that $\mu_\ast= M_s$.

Although this scale-duality symmetry follows directly from modular invariance, 
its implications are profound.
{\it Ultimately, the existence of such a symmetry implies the existence of a fundamental limit 
on the extent to which 
an EFT perspective can possibly remain valid in string theory.}
We have already noted in the Introduction that  string theory is rife with duality symmetries which defy EFT notions:  an immediate example of this
is T-duality, under which the physics associated with a closed string 
propagating on a spacetime with a compactified dimension of radius $R$
is indistinguishable from the physics associated with a closed string propagating
on a spacetime with a compactified dimension of radius $R'\equiv \alpha'/R$.  This is true as
an exact symmetry not only for the string spectrum but also for all interactions.
Thus such strings cannot distinguish between large and small compactification geometries,
thereby preventing us from establishing a Wilsonian EFT-like ordering of length scales from large to small, 
or equivalently from IR to UV.~
Phrased somewhat differently, the existence of a T-duality symmetry tells us that there is a fundamental limit to which we may consider a spacetime compactification radius to be ``small''.
However, what we are seeing now is that a somewhat different phenomenon --- namely the scale-duality symmetry under $\mu\to M_s^2/\mu$ which is guaranteed by modular invariance --- implies a fundamental limit on the extent to which our EFT can exhibit UV behavior.   
Indeed, pushing $\mu$ beyond  $M_s$ only serves to reintroduce the original IR behavior of our theory ---
a behavior which we may now associate with the {\it dual}\/ energy scale 
{\mbox{$\mu'\equiv M_s^2/\mu$}} associated with a ``dual'' EFT.~
In this sense, the energy scales near $M_s$ exhibit the ``most possible UV'' behavior
that can be realized.

At first glance, it may be tempting to associate this scale-duality symmetry with T-duality, since both tend to place limits on UV behavior and have similar algebraic forms.   We stress, however, that T-duality is a {\it spacetime}\/ symmetry, and as such comes with certain assumptions about the spacetime geometry.  By contrast, modular invariance is a fundamental {\it worldsheet}\/ symmetry which is required for the self-consistency of the theory itself.   As such, modular invariance and T-duality are unrelated.    Indeed, T-duality relates  two {\it a priori}\/ different string theories to each other, one with a large compactification volume and the other with a small compactification volume, and maps a given state with KK- and winding-numbers $(m,n)$ in the first theory to the {\it equally massive}\/ $(n,m)$ state in the other.  By contrast, modular invariance is a symmetry that operates {\it within}\/ a single string theory and  involves Poisson resummations across the {\it entire}\/ string spectrum simultaneously.  As such, no two string states in the theory are directly related to each other under modular transformations.  Indeed, only through such non-trivial resummations involving the {\it 
 entire}\/ string spectrum --- including the oscillator states as well --- could we ever hope to obtain features such as misaligned supersymmetry and supertrace constraints (such as $\Str\, {\bf 1}=0$) that simultaneously balance all of our string states at all energy scales against each other within a single string theory, even without supersymmetry.

%======================== BEGIN FIGURE ======================================== 
\begin{figure*}[t!]
\centering
\includegraphics[keepaspectratio, width=0.99\textwidth]{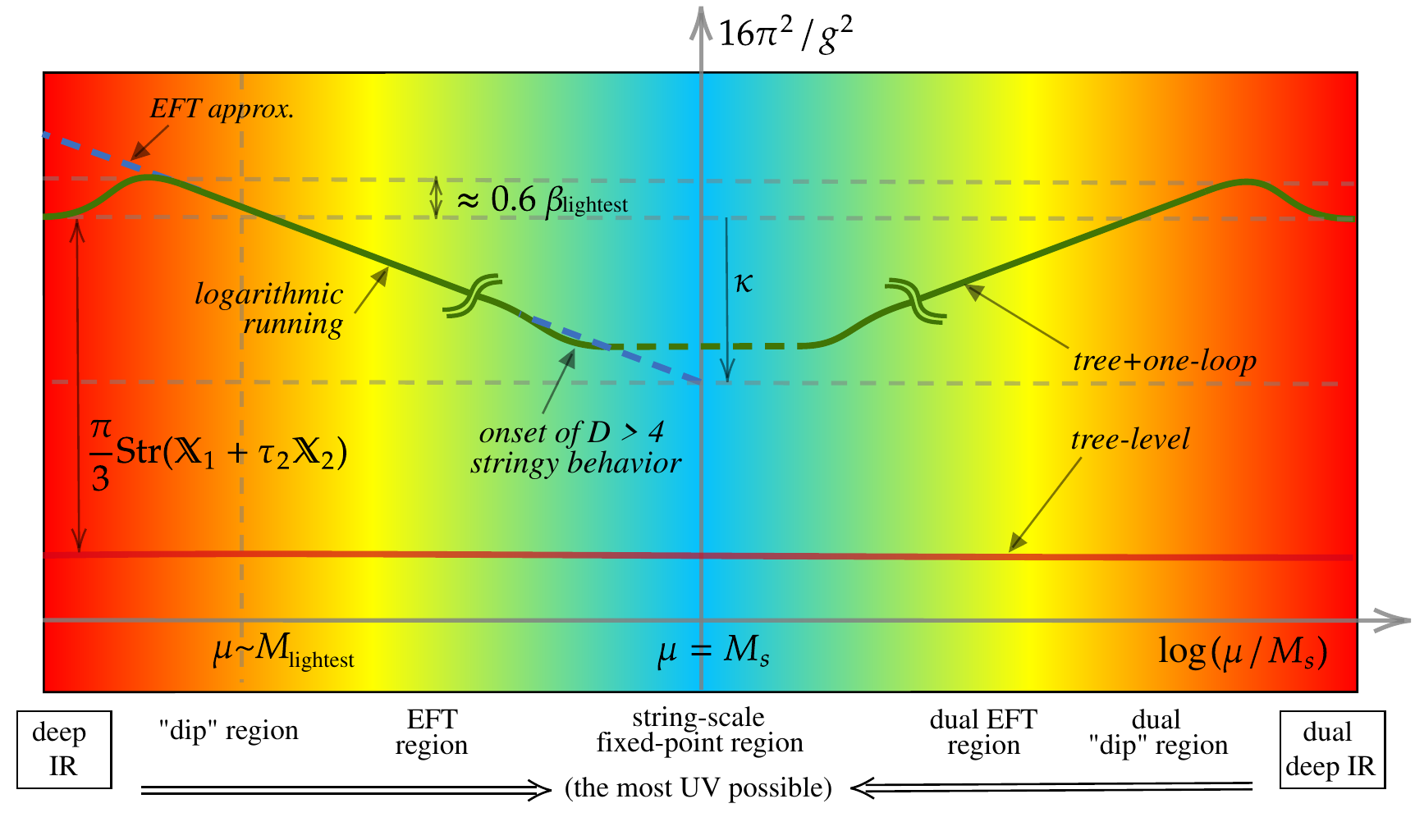}
\caption{The one-loop running of the inverse gauge coupling $\widehat \Delta_G = 16 \pi^2/g_G^2$ for any gauge group $G$, as calculated
from first principles in a fully modular-invariant string framework. The tree-level contribution is sketched in red, and the total one-loop coupling is sketched in green.
In the deep IR, the coupling approaches an asymptotic
value  which receives contributions from {\it all}\/ of the states in the string spectrum which carry non-trivial helicity $\mathbbA^{(1)}\sim\overline Q_H^2$ charges.  This assumes that our theory contains no net massless states charged under
$\mathbbA^{(2)}+\mathbbB^{(2)}\sim ( \overline Q_H^2-1/12) Q_G^2$ where $Q_G^2$ 
 is the sum of the squares of the charges in the Cartan subalgebra of $G$;
  otherwise $\widehat \Delta_G(\mu)$ diverges in the IR limit.
Moving towards higher values of $\mu$, we see that a non-trivial scale-dependence does not emerge until $\mu\sim M_{\rm lightest}$,
where $M_{\rm lightest}$ collectively represents the masses of the lightest massive states which are charged under $\mathbbA^{(2)}+\mathbbB^{(2)}$.
The non-monotonic ``dip'' in $g_G$ (or ``hump'' in $\widehat \Delta_G$) that is observed in this region is a transient effect
which smoothly connects the asymptotic deep-IR region $\mu\ll M_{\rm lightest}$ to
an EFT-like region $M_{\rm lightest}\lsim \mu \ll M_s$.
Beyond the dip region, the theory then enters an EFT-like region
in which the gauge coupling experiences a logarithmic running.
As $\mu\to M_s$, it is possible that we might cross the energy threshold $R^{-1}$ associated with large compactification radii.  In such cases, this logarithmic running can be modified by the appearance of Kaluza-Klein and winding states which might appear at mass scales significantly below $M_s$ and which
might tend to {\it cancel}\/ this logarithmic running, leading to the existence of a higher-dimensional fixed-point regime, as shown.  
The subtleties involved in this behavior will be discussed further in Ref.~\cite{NonRenormalizationTheorems}.
However, as a general principle, modular invariance requires that the running of $\widehat\Delta_G$ exhibit an invariance under $\mu \to M_s^2/\mu$.   Thus, 
as $\mu$ increases beyond $M_s$,
the theory inevitably begins to re-enter an IR-like regime which we may associate with a ``dual'' EFT, followed by a dual dip region and then a dual deep-IR region.  The background colors of this sketch indicate the transition from the deep IR (red) to the UV (blue) and
then back to IR (red).
As such, there is a maximum degree to which our theory can approach the UV:   
once the energy scale $\mu$ passes the self-dual 
point $\mu\sim M_s$, further increases in $\mu$ only push us towards increasingly IR behavior.   The quantity $\kappa$ is defined in Eq.~(\protect\ref{kappaimplicit}).
 }
\label{anatomy}
\end{figure*}
% =========================================

Taken together, all of these observations lead to a running for $\widehat\Delta_G(\mu)$ as sketched in Fig.~\ref{anatomy}.   
In the deep IR, $\widehat\Delta_G(\mu)$ approaches a constant unless the quantity in Eq.~(\ref{masslessX2}) is non-zero.
As $\mu$ increases, our theory then passes through a ``dip'' region and a subsequent EFT region characterized by logarithmic running.
If our theory has large compactification radii $R\gg M_s^{-1}$, the contributions from the corresponding Kaluza-Klein and winding states can then conspire to eliminate this running, leading to the existence of a higher-dimensional string-scale fixed-point region. Beyond $M_s$, our theory enters a ``dual'' regime in which further increases in $\mu$ only reproduce the IR behavior we have already seen, only in reverse.

We conclude this section with three comments.
First, we observe that the running of the gauge coupling is essentially the same as the running of the Higgs mass in Ref.~\cite{Abel:2021tyt} --- indeed for $\mu \lsim M_E$  the only differences are the {\it coefficients}\/ of the different running terms.   These coefficients change because they tally the appropriate charges of our states across the string spectrum, and the charges that are appropriate or relevant change when we switch our attention from the Higgs mass to the gauge couplings.

Our second comment 
concerns the running of the gauge couplings themselves.  Within our calculations we have implicitly assumed that these couplings remain perturbative throughout the running shown in Fig.~\ref{anatomy};  otherwise our one-loop calculation is no longer applicable and higher-loop (and even non-perturbative) calculations would be needed.
Depending on the relative signs and magnitudes of the various supertraces involved, these couplings could be in danger of becoming non-perturbative either as $\mu\to M_S$ (which represents one extremum of the gauge-function plotted in Fig.~\ref{anatomy}) or within the ``dip'' region.

Most importantly, however, there is a deep and fundamental difference between the running of the gauge couplings $\widehat\Delta_G$ and the running of the Higgs mass in Ref.~\cite{Abel:2021tyt}.
As we see directly from Eqs.~(\ref{finalPafromhiggs2}), (\ref{finalPafromhiggs3}), and (\ref{finalPafromhiggs4}), the gauge-coupling calculation now includes contributions from off-shell string states for which $M_L\not = M_R$.   This is a strange but not entirely unexpected feature:  states which are not {\it physical}\/ in the underlying string theory, and which therefore can only contribute in string loop diagrams, also contribute to the running of the gauge couplings in the corresponding low-energy EFT!~
This feature did not appear in the running  of the Higgs mass in Ref.~\cite{Abel:2021tyt}.   However, as we have seen, this feature ultimately stems from the fact that the contributions to the Higgs mass are proportional not to the square of the helicity charge $\overline Q_H^2$, but rather to this quantity {\it minus}\/ 1/12.   In field theory, this extra $-1/12$ is not problematic.  However, in string theory it has deep repercussions because a pure number
such as $-1/12$ cannot be subtracted from a squared-charge operator such as $\overline Q_H^2$ because a pure number has modular weight $k=0$ while the squared-charge operator has modular weight $k=2$.   Modular invariance thus requires that the $-1/12$ term be ``completed'' to the weight-two modular function $\overline E_2/12$, and this in turn has reverberations throughout the string spectrum, shifting left-moving string masses $M_L$ relative to right-moving string masses $M_R$.   This is why the non-level-matched string states now survive the Rankin-Selberg procedure and appear in our running calculation.

That said, these states do not contribute to the low-energy running in a standard way.   Normally, we would expect a string state to contribute in the low-energy theory according to its mass $M^2= (M_L^2+M_R^2)/2$.    Indeed, this quantity in some sense tells us how much worldsheet energy (as measured by eigenvalues of $\overline L_0$ and $L_0$ respectively) has been ``invested'' in creating that state as an excitation in the underlying worldsheet theory.   However, what we are now learning from Eqs.~(\ref{finalPafromhiggs2}), (\ref{finalPafromhiggs3}), and (\ref{finalPafromhiggs4}) is that although a given entwined string state may have a string-theoretic mass given by $M$,  it contributes to the low-energy EFT precisely as if it had a mass simply given by $M_L$!   In other words, the combined {\it string-theoretic mass}\/ $M$ is irrelevant;   what matters --- and what we may therefore consider to be the effective  {\it EFT mass}\/ in such theories, at least as far as the gauge couplings are concerned  --- is determined by $M_L$ alone.  This, of course, is the effect of the shift in left-moving masses relative to right-moving masses induced by $\overline E_2$.  

We also observe that the entwined resonances all have left-moving masses that exceed the string scale: $M_L\geq M_E = \sqrt{2} M_s$.   Thus, one might be tempted to argue that these states have no effects at energy scales below $M_s$.
However, this would not be correct.
Thanks to scale-inversion duality, any state that affects the running of quantities {\it above}\/ the string scale will also affect the running of these quantities {\it below}\/ the string scale.  This is not a new phenomenon unique to the entwined states.   After all, we have already seen that the behavior of our amplitudes in the deep IR is in part determined by the extremely heavy string states in the deep UV.~   In a similar way, the entwined states also have effects below the string scale and thereby also have an indirect role in affecting the low-energy EFT below the string scale.  

Ultimately, this can be understood from the perspective of modular invariance and misaligned supersymmetry.   In the two cases studied in Ref.~\cite{Abel:2021tyt} (namely the cosmological constant and the Higgs mass), the relevant supertraces involved only physical states and closed under modular transformations into themselves.  Thus the corresponding amplitudes in each case were fully modular invariant and yet could be written purely in terms of supertraces over only physical string states. However, for the present gauge-coupling calculation, slightly different supertraces are involved.   Amongst these supertraces, those involving only physical string states do not close into each other under modular transformations.  Rather, closure for these supertraces also involves certain entwined supertraces as well.
This is ultimately why the entwinement occurs in these theories.  Through the entwinement, the off-shell string states continue to make explicit contributions to the relevant amplitudes.

This last observation is in fact part of a more general lesson.   In ordinary quantum field theory, one can meaningfully seek to identify the physical effects that arise due to the existence of specific states in the spectrum.   For example, we might attempt to determine the 
energy scales at which a given state of mass $M$ contributes to the running of a given quantity.    However, in a modular-invariant theory, this question has no meaning because of UV/IR mixing.  Every state within the spectrum is deeply connected to the states at all other energy scales.   It is therefore impossible to uniquely isolate the contributions of a single state within the spectrum because there is no modular-invariant way to perform such an analysis.

As a dramatic example of this phenomenon, let us consider the unphysical $(m,n)=(0,-1)$ proto-graviton states~\cite{Dienes:1990ij} that arise within all string models, and ask whether these states contribute to the corresponding one-loop cosmological constant $\Lambda$. 
 On the one hand, we might claim that the proto-graviton states do not contribute to $\Lambda$ because we know that we can write $\Lambda$ as the supertrace of $M^2$ over the purely physical states in the spectrum, as in Eq.~(\ref{Lamresult}).  However, a direct calculation of the one-loop torus integral associated with $\Lambda$ demonstrates that these states not only contribute, but actually provide contributions that dominate over those of all other states (see, \eg, Table~2 of Ref.~\cite{Dienes:2006ut}).
 The underlying reason for this apparent contradiction is that our question about which states contribute does not have a modular-invariant answer. 
 Modular transformations allow us to reshuffle our contributions so that the effects of one state can be re-interpreted as the resummed effects of other states instead.   Indeed, as we have already asserted, the un-entwined supertraces that contribute to the gauge-coupling running do not close into themselves under modular transformations;  they also involve the entwined supertraces.   All supertraces --- both entwined and un-entwined --- therefore contribute together in a modular-invariant way.
 
In this connection, we observe that the lightest entwined states actually have vanishing string-theoretic masses!   These are the states that populate the $(m,n)=(-1/2,+1/2)$ square in Fig.~\ref{fig:matrix}.   Such states require only a minimal amount of energy to produce on the worldsheet --- indeed, exactly the same amount of energy as required to produce the physical massless states that populate the $(0,0)$ square and presumably include the Standard-Model states.   It will be interesting to explore the ramifications of this observation~\cite{offshellMisSUSY}.

\bigskip
\bigskip

\section{Conclusions, discussion, and future directions}
\label{sec:discussion}

In this paper we developed a general framework for analyzing the running of gauge couplings within closed string theories.  Unlike previous discussions in the literature, our calculation fully incorporates the underlying modular invariance of the string and includes the contributions from the infinite towers of string states which are ultimately responsible for many of the properties for which string theory is famous, including its enhanced degree of finiteness and UV/IR mixing. In order to perform our calculations, we adopted a formalism~\cite{Abel:2021tyt} that was recently developed for calculations of the Higgs mass within such theories.

In general, this formalism --- which builds upon the Rankin-Selberg technique~\cite{rankin1a,rankin1b} but which also includes additional critical features such as an identification between worldsheet parameters and an effective spacetime 
energy scale $\mu$ ---
gives rise to an ``on-shell'' EFT description in which the final results are expressed in terms of supertraces over the physical string states, and in which these quantities exhibit an EFT-like ``running'' as a function of the scale $\mu$.  We found, however, that the calculation of the gauge couplings differs in one deep way from that of the Higgs mass:  while the latter results depend on purely on-shell supertraces, the former results have a different modular structure which causes them to depend on supertraces over {\it off-shell}\/ string states as well.    Indeed, as explained at the end of Sect.~\ref{sec:entwined}, the entwinement induced by the modular completion of the helicity operator needed for calculating the gauge couplings has ``deformed'' the notion of physicality for the string states, allowing states which are not level-matched to nevertheless act as physical states which contribute to the physical supertraces describing the values of physical string amplitudes.   We have also seen that although our results yield the expected logarithmic running of the gauge couplings within certain energy scales, they also yield a number of intrinsically stringy behaviors that transcend what might be expected within an effective field theory approach.

A central feature of our treatment is our use of a modular-invariant regulator $\widehat {\cal G}_\rho$ to define a physical energy scale $\mu$ in the system and simultaneously eliminate any logarithmic divergences that might arise from the contributions of certain massless states in the theory.  Since this regulator is modular invariant, it suppresses the contributions from the lighter string states in a smooth manner which is consistent with UV/IR mixing and which therefore naturally incorporates the contributions from the infinite towers of string states in a modular-invariant way.  Use of this regulator not only eliminates logarithmic IR divergences but more importantly but also allows us to study how the gauge couplings ``run'' as a function of the spacetime energy scale $\mu$.  Indeed this procedure can be thought of as a ``functional renormalization group'' (FRG) approach to scaling~\cite{Polchinski:1983gv,WETTERICH199390,Morris:1993qb}  for UV-complete modular-invariant theories.
Notably, the use of such a regulator allows us to  sidestep the need to introduce a sharp cutoff which would be very difficult if not impossible to reconcile with our fundamental UV/IR-mixed string symmetries.
This also allows us to avoid the need to designate which states are ``light'' with respect to $\mu$ (and which therefore contribute to the running), and which are ``heavy'' (and therefore do not).   Through such modular-invariant regulator functions, we can develop a notion of ``running'' gauge couplings and beta functions, with our modular-invariant regulator allowing us to extract this apparent EFT-like behavior as a function of the spacetime energy scale.

The final global picture that emerges is shown in Fig.~\ref{anatomy}.  Perhaps the most prominent feature is the existence of a scale-duality symmetry, \ie, an invariance under $\mu \to M_s^2/\mu$.   As we have discussed, this is an inevitable consequence of the modular invariance that underlies our calculations.    However, the impact of this scale-duality symmetry is felt even at energy scales below the string scale.   For example, we have seen that the {\it IR value} of the one-loop contribution to the gauge coupling is given by
$\frac{\pi}{3} \Str(\mathbbX_1 +\tau_2\mathbbX_2)$
where the supertrace is over {\it all}\/ of the states in the theory, regardless of their masses.  From a na\"ive field-theory perspective, such a supertrace would appear to control a quadratic {\it UV divergence}\/.
However, the deep IR in such theories is also equivalent to the deep UV, where one would expect all of the states to play a role.
Indeed, these IR predictions are RG invariants, in the sense that they define fixed-point values.  These predictions are also independent of our choice of regulator function. 

As we move away from the $\mu\to 0$ limit and proceed towards higher energy scales, the system evolves away from this asymptotic IR behavior.    Once $\mu$ exceeds the masses of the lowest-lying states, the one-loop contribution to the gauge couplings passes through a localized ``dip''  
and then 
begins to experience a non-trivial logarithmic running that can be associated with an EFT-like description.
This running then continues towards higher energy scales, possibly passing through a sequence of EFT-like descriptions.   If our string compactification geometry has effective radii $R_i^{-1}\ll M_s$, then this running continues until $\mu \sim R_i^{-1}$.   Above this scale we find a surprising new behavior:  the running stops, with the gauge coupling entering
a string-scale ``fixed-point'' region. This surprising behavior will be discussed in more detail in 
Ref.~\cite{NonRenormalizationTheorems}
and ultimately results from the combined effects of both KK states and winding states.  It may seem strange that both kinds of states should be playing a role at scales $\mu\ll M_s$, but this is a direct consequence of the scale-duality symmetry under $\mu\to M_s^2/\mu$.
The fact that both KK {\it and}\/ winding states are simultaneously playing important roles further implies that the behavior in this region is not only fully higher-dimensional but also intrinsically {\it stringy}\/. 
This transition to entirely stringy behavior is an inevitable and profound consequence of an RG procedure which is consistent with modular invariance:  a modular-invariant regulator cannot  distinguish between KK and winding modes and therefore can only act to suppress the contributions of both or neither. 

Beyond $\mu\sim M_s$, we enter the ``dual'' phase in which the running of the gauge coupling is inverted.   This inversion of the running is quite remarkable.  Indeed, from a na\"ive field theoretic perspective, this kind of complete reversal would be disallowed (by \eg, the $a$-theorem).  Of course, we do not expect such a theorem to hold in a UV/IR-mixed theory such as string theory.

Indeed this inversion of the running of the gauge couplings can best be understood by recognizing that the fundamental degrees of freedom within the  dual phase of the theory are not those of the original low-energy theory.  The original theory and its dual are nothing but modular transformations of each other --- indeed, the relationship between these two ``phases'' of the theory is outlined in Fig.~4 of Ref.~\cite{Abel:2021tyt}, where they are shown to lie along different but equivalent ``spokes'' of the same fundamental diagram.  Thus each phase carries the same information and can be viewed as representing the same fundamental theory, consistent with the idea that modular symmetries (like gauge symmetries) do not relate physically different theories to each other, but rather represent redundancies of description.
However, under modular transformations, the states within our string theory at all mass levels are non-trivially mixed with each other.   Thus, a given state in the original theory is mapped to a highly non-trivial combination of states in the dual theory, while each state in the dual theory is likewise mapped to a highly non-trivial combination of states in the original theory.  Demanding that this mapping nevertheless produce the same theory is the essence of what it means for a theory to be modular invariant.  More explicitly, as discussed in Ref.~\cite{Abel:2021tyt}, the scale-duality map $\mu\to M_s^2/\mu$ is intimately related to the $\tau\to -1/\tau$ modular transformation evaluated along the $\tau_1=0$ line.   This modular transformation induces a Poisson resummation amongst the states of the original theory, so that
 the degrees of freedom in the dual phase of the theory with $\mu \geq M_s$ are Poisson-resummed versions of the degrees of freedom in the original phase of the theory with $\mu< M_s$.
It is ultimately this Poisson resummation which
is responsible for the inversion in the running of the gauge couplings once we cross between the $\mu<M_s$ and $\mu>M_s$ regions in any modular-invariant theory.

Although this physical picture is relatively simple,
it actually encapsulates a considerable amount of non-trivial physics.  As we would expect in any UV/IR-mixed theory, string modes that are light are being mixed with those which are heavy.  However, such heavy string states are super-Planckian, and may (depending on the string coupling) include black holes.  They are also likely to include so-called ``long'' strings, \ie, strings with large numbers of oscillator excitations.   
However, as long as we maintain a constant definition of the physical spacetime energy scale $\mu$, modular invariance requires that these states all conspire (through Poisson resummations) to achieve this apparent reversal in the directionality of the gauge-coupling running at $\mu=M_s$. 

It would be an interesting exercise to develop an understanding of the dual running directly in terms of these dual degrees of freedom.  Moreover, although we have concentrated in this paper on the one-loop running of the gauge couplings, we expect similar results to apply to other one-loop amplitudes, such as might be involved in string {\it scattering}\/.   Here too one must identify a physical string scale $\mu$ in terms of certain renormalization conditions and then study how such amplitudes depend on $\mu$.  

Above the string scale any alternative  RG prescription derived from such amplitudes --- \eg, a prescription based on suppressing the contributions of certain momentum modes -- should therefore be defined in terms of {\it dual momenta} for those asymptotic eigenstates that can be prepared in this regime.
These would comprise the long-string modes discussed above. In this way the $\mu\rightarrow M_s^2/\mu$ symmetry would be faithfully  respected.  It is for this reason that our worldsheet regulator prescription gives a correct physical picture of renormalization, one which is applicable for all values of $\mu$. 
 Indeed, such a regulator prescription is also correctly aligned with both
cosmological~\cite{Brandenberger:1988aj} and thermal~\cite{Dienes:2003du,Dienes:2003dv} dualities.   We leave such investigations for future work.

Given these observations, we now discuss possible new approaches to  hierarchy problems.  We begin by recalling from Sect.~\ref{sec:preliminaries} that within the string context there is no notion of a spacetime energy scale $\mu$ before we insert a regulator and identify $\mu$ in terms of the parameters of this regulator.  Thus, the unregulated modular integrals that govern the couplings of the theory  --- integrals such as $\Delta_G$ or equivalently $\widehat\Delta_G(\mu=0)$ --- are simply one-loop contributions to the effective action.   Such integrals are either logarithmically divergent (if the string spectrum contains a non-zero net number of exactly massless $\mathbbX_2$-charged states), or finite otherwise.
Thus, upon introducing the modular-invariant regulator $\widehat \calG_\rho(a,\tau)$,
it is inevitable that the values of the couplings in the deep IR (\ie, as $\mu \to 0$) are also the ``bare'' couplings that we would expect to obtain as $\mu\to \infty$.     To see this, we note that
within the FRG approach the ``average effective action'' is normally taken to interpolate between the effective action at $\mu=0$ and the ``bare'' action at $\mu\rightarrow \infty$. 
However, within our UV/IR-mixed context, the effective actions that one obtains in the $\mu\to 0$ and $\mu\to \infty$ limits are one and the same.    This is because the modular invariance of our regulator implies that when we are regulating the IR, we are also equivalently regulating the UV.~
Indeed, any divergence that would arise within the $\tau\to i\infty$ region of the modular integral (and which therefore would normally be interpreted as an IR divergence) is the same as the divergence that would arise within the $\tau\to 0$ region of the modular integral (and which would therefore be interpreted as a UV divergence). 

This relation between UV and IR divergences has an important implication.
By locking these two types of divergences together, our theory cannot exhibit any UV divergence that is not also present as an IR divergence.
However, as we have seen, the $\tau\to i \infty$ limit of our theory can support at most a logarithmic IR divergence arising from the contributions of certain massless states.   This is an expected divergence that indicates that in the low-energy regime the theory will behave like a quantum field theory. However, modular invariance then implies that we cannot have quadratic divergences as $\tau\to 0$ (\ie, in the UV).~   Indeed,
in any modular-invariant theory, stringy ``miracles'' (such as the cancellation of certain supertraces) have no choice but to eliminate the quadratic divergences because there is simply no place left for them.
In this connection, we note that this argument relies directly on modular invariance alone, and is not specific to the specific form chosen for our regulator so long as it is modular invariant.  

This suggests that there is a fundamental difference  between hierarchy problems in field theory and hierarchy problems in string theory (and in UV/IR mixed theories more generally).  In order to  analyze hierarchy problems within a Wilsonian field theory,  one starts by separating operators into ``relevant'' ones that grow in the IR and ``irrelevant'' ones that grow in the UV.~ Relevant operators typically begin at (possibly Gaussian) fixed points in the UV.~
 Their RG trajectories are then determined by a set of ``unpredicted'' free parameters that are chosen by hand in order to make the associated couplings agree with the desired (presumably measured) IR values. Meanwhile the irrelevant operators flow to attractive fixed points in the IR, thereby becoming ``predictions'' of the theory.  In this language, hierarchy problems arise whenever there is an extreme sensitivity of the relevant operators (which control the RG trajectory) to the intermediate physics.  Such sensitivity confronts us with what is essentially a ``shooting problem'' because it requires us to keep fine-tuning our RG trajectory in order to hit the desired IR values.  However the UV/IR mixing in string theory (and its attendant $\mu\rightarrow M_s^2/\mu$ symmetry) removes the underpinnings of this entire picture, as
operators cannot even be designated as ``relevant'' or ``irrelevant'' until we decide which direction corresponds to `UV' and which corresponds to `IR'  within our definition of the energy scale. 
In this context we refer the reader to Fig.~4 in Ref.~\cite{Abel:2021tyt}, which graphically demonstrates the different possibilities.
Indeed, the only reliable quantities before we make this choice are the values of supertraces over the infinite towers of states. As we have mentioned, these are by definition invariant under our choice of regulator and also invariant under changes in the relevant energy scale. As such, they remain invariant within the emergent EFT,  simultaneously determining from the outset both the bare action and the effective action to which the theory must flow in the deep IR regardless of what intermediate physics may exist. 

 \interfootnotelinepenalty=10000  
  % --- insert this to prevent footnotes from being split across pages.

We find these observations to be  compelling foundations  for future, more general, phenomenological studies. 
In particular, the transformation of apparent hierarchy problems into statements about the properties of supertraces over infinite towers of physical string states suggests that seemingly miraculous cancellations and ``magic zeros'' are unavoidable features of the  effective field theory stemming from a natural UV/IR-mixed modular-invariant theory. 
  In such a framework, the solutions to hierarchy problems such as the gauge-hierarchy problem and the cosmological-constant problem rely on conspiracies between physics at all energy scales simultaneously, and would thus be essentially invisible to low-energy observers.  In this sense, they might be considered to exhibit what has recently been dubbed ``neutral naturalness'', except in a form that does not involve pairwise cancellation mechanisms that operate scale-by-scale but rather through seemingly miraculous cancellations that operate at all scales simultaneously.
  Within such frameworks, retaining the full spectrum of states within our calculations is therefore critical for obtaining a full understanding of naturalness.\footnote{
  This point of view regarding modular symmetries addressing hierarchy problems was first advocated in Ref.~\cite{Dienes:2001se}.~~More recent discussions along these lines can be found in Ref.~\cite{Abel:2021tyt} and in the  talk delivered by KRD at the CERN Theory Workshop ``Exotic Approaches to Naturalness'', Jan.-Feb.\ 2023, slides available at 
{\tt https://indico.cern.ch/event/1204192/contributions/
5217557/attachments/2584637/4458574/dienes$\_$cern.pdf}. }

Given these comments, there exist many promising avenues for future research targetting the development of a fuller understanding of naturalness in UV/IR-mixed theories.   Indeed, 
one of these is to study the manner in which the running of the gauge couplings is deformed in the presence of large extra dimensions.   For example, following Ref.~\cite{Dixon:1990pc},
we might study the running of the gauge couplings when our theory has a compactification geometry of the form $\mathbbK \times \mathbbT^2$ where $\mathbbK$ has a characteristic volume near the string scale and where $\mathbbT^2$ indicates a two-torus with radii $R_i\gg M_s^{-1}$.  
The results of this analysis will be presented in Ref.~\cite{NonRenormalizationTheorems}, where we shall find that the promotion of logarithmic running to power-law running --- as expected from a field-theoretic analysis~\cite{Dienes:1998vh, Dienes:1998vg}
--- does not occur in string theory.  
Indeed, as far as power-law running is concerned, this is a kind of ``non-renormalization'' theorem for string theory.
As we shall see in Ref.~\cite{NonRenormalizationTheorems},
this cancellation of power-law running is the result of a delicate conspiracy between modular invariance and the manner in which extra spacetime dimensions emerge in string theory as their radii become large.   Indeed, this cancellation ultimately reflects a subtle entanglement between the properties of renormalization in higher dimensions and the requirements of modular invariance.
Moreover, as we shall demonstrate in Refs.~\cite{NonRenormalizationTheorems, NewSupertraceIdentities}, there are also many additional supertrace relations which govern the spectra of modular-invariant string theories with large compactification radii --- supertrace relations whose role is to enforce these remarkable cancellations.  Such supertrace relations are thus also responsible for the finiteness properties of these theories.

The results shown in Fig.~\ref{anatomy} also call for a reappraisal of the possibilities for gauge-coupling unification in modular-invariant theories.  Although the gauge couplings continue to exhibit logarithmic running --- thereby suggesting that a traditional logarithmic unification may continue to be viable --- the existence of a possible fixed-point regime near the string scale has the potential to alter this situation.   This is especially true given that the scale of unification typically assumed for heterotic strings is only one or two orders of magnitude below the string scale~\cite{Kaplunovsky:1987rp, Dienes:1996du}.   It is even possible that the existence of this fixed-point regime may serve to reconcile the long-standing discrepancy~\cite{Dienes:1996du} between the string-predicted perturbative heterotic unification scale and the traditional GUT scale extrapolated from experimental measurements of the Standard-Model gauge couplings at low energies.

Overall, the message seems clear.
Hierarchy problems (and even issues related to gauge-coupling unification) assume traditional field-theory relationships 
between UV and IR physics.  By contrast, string theory tells us that we have UV/IR 
mixing, misaligned supersymmetry, softened divergences (even finiteness), scale duality, and so forth.
Thus, within the context of string theory, hierarchy 
problems may not be fundamental or survive in the manner we normally assume.

\begin{acknowledgments}
The research activities of SAA were supported by the STFC grant ST/P001246/1, and the Institut Pascal at
Université Paris-Saclay with the support of the P2IO Laboratory of Excellence, the P2I axis of the Graduate School Physics of Université Paris-Saclay, and the IN2P3 master projet UCMN. 
The research activities of KRD were supported in part by the U.S.\ Department of Energy
under Grant DE-FG02-13ER41976 / DE-SC0009913, and also 
by the U.S.\ National Science Foundation through its employee IR/D program.
The opinions and conclusions
expressed herein are those of the authors, and do not represent any funding agencies.

\end{acknowledgments}

\bigskip\bigskip

\appendix 

\vfill\eject

\bibliography{TheLiterature}
\end{document}